\documentclass[fleqn,usenatbib]{mnras}

\usepackage{newtxtext,newtxmath}

\usepackage[T1]{fontenc}
\usepackage{ae,aecompl}
\usepackage[flushleft]{threeparttable}
\usepackage{rotating}

\usepackage{graphicx}	
\usepackage{amsmath}	
\usepackage{amssymb}	

\title[Hosts of high-redshift obscured quasars]{Host galaxies of high-redshift extremely red and obscured quasars}

\author[Zakamska et al.]{
Nadia L. Zakamska$^{1,2}$\thanks{e-mail: zakamska@jhu.edu},
Ai-Lei Sun$^{1}$,
Michael A. Strauss$^{3}$,
\newauthor
Rachael M. Alexandroff$^{4}$,
W. N. Brandt$^{5}$,
Marco Chiaberge$^{6,1}$,
\newauthor
Jenny E. Greene$^{3}$,
Fred Hamann$^{7}$,
Guilin Liu$^{8}$,
\newauthor
Serena Perrotta$^{7}$,
Nicholas P. Ross$^{9}$, and
Dominika Wylezalek$^{10}$
\\
$^{1}$Department of Physics and Astronomy, Bloomberg Center, Johns Hopkins University, Baltimore MD 21218, USA \\
$^{2}$Institute for Advanced Study, Einstein Dr., Princeton NJ 08540, USA \\
$^{3}$Department of Astrophysical Sciences, Princeton University, Princeton NJ 08544, USA\\
$^{4}$Canadian Institute for Theoretical Astrophysics, The University of Toronto, Toronto, ON M5S 3H8, Canada; \\
Dunlap Institute for Astronomy and Astrophysics, The University of Toronto, Toronto, ON M5S 3H4, Canada\\
$^{5}$Department of Astronomy \& Astrophysics, 525 Davey Lab, The Pennsylvania State University, University Park, PA 16802, USA; \\
Institute for Gravitation and the Cosmos, The Pennsylvania State University, University Park, PA 16802, USA; \\
Department of Physics, The Pennsylvania State University, University Park, PA 16802, USA\\
$^{6}$AURA for ESA, Space Telescope Science Institute, 3700 San Martin Dr., Baltimore MD 21210, USA\\
$^{7}$Department of Physics and Astronomy, University of California, 900 University Avenue, Riverside CA 92521, USA\\
$^{8}$CAS Key Laboratory for Research in Galaxies and Cosmology, Department of Astronomy, University of Science and \\
Technology of China, Hefei, Anhui 230026, China;\\
School of Astronomy and Space Sciences, University of Science and Technology of China, Hefei 230026, China\\
$^{9}$Institute for Astronomy, University of Edinburgh, Royal Observatory, Blackford Hill, Edinburgh EH9 3HJ, United Kingdom\\
$^{10}$European Southern Observatory, Karl-Schwarzschildstr. 2, 85748 Garching bei M\"unchen, Germany
}

\date{Accepted XXX. Received YYY; in original form ZZZ}

\pubyear{2018}

\begin{document}
\label{firstpage}
\pagerange{\pageref{firstpage}--\pageref{lastpage}}
\maketitle

\begin{abstract}
We present Hubble Space Telescope 1.4$-1.6$\micron\ images of the hosts of ten extremely red quasars (ERQs) and six type 2 quasar candidates at $z=2-3$. ERQs, whose bolometric luminosities range between $10^{47}$ and $10^{48}$ erg/sec, show spectroscopic signs of powerful ionized winds, whereas type 2 quasar candidates are less luminous and show only mild outflows. After performing careful subtraction of the quasar light, we clearly detect almost all host galaxies. The median rest-frame $B$-band luminosity of the ERQ hosts in our sample is $10^{11.2}L_{\odot}$, or $\sim 4L^*$ at this redshift. Two of the ten hosts of ERQs are in ongoing mergers. The hosts of the type 2 quasar candidates are 0.6 dex less luminous, with 2/6 in likely ongoing mergers. Intriguingly, despite some signs of interaction and presence of low-mass companions, our objects do not show nearly as much major merger activity as do high-redshift radio-loud galaxies and quasars. In the absence of an overt connection to major ongoing gas-rich merger activity, our observations are consistent with a model in which the near-Eddington accretion and strong feedback of ERQs are associated with relatively late stages of mergers resulting in early-type remnants. These results are in some tension with theoretical expectations of galaxy formation models, in which rapid black hole growth occurs within a short time of a major merger. Type 2 quasar candidates are less luminous, so they may instead be powered by internal galactic processes. 
\end{abstract}




\section{Introduction}

Models for galaxy formation increasingly invoke feedback from active galactic nuclei (AGN) to explain the cessation of star formation in ellipticals and the shape of the galaxy luminosity function (e.g., \citealt{silk98,crot06}). A popular evolutionary paradigm that naturally incorporates AGN into galaxy evolution models invokes mergers and interactions of gas-rich galaxies to trigger both intense bursts of star formation and black hole growth (e.g., \citealt{sand88,hopk06}). In these scenarios, the black holes grow largely in an obscured phase inside the dusty starbursts until a blowout of gas and dust, fueled at least partly by the central AGN, halts the star formation and reveals a visibly luminous quasar in the galactic nucleus. Quasars that are obscured and reddened by dust are valuable for testing this evolution scheme because they should be in the brief blowout phase, close in time to the initial merger/starburst event. 

Due to the need for high spatial resolution and high contrast observations in order to detect the faint host galaxies surrounding the bright central quasars, studies of quasar hosts have long been a key science topic for the Hubble Space Telescope (HST). As observational techniques were being developed, pioneering HST studies of quasar hosts naturally concentrated at low redshift ($z<1$, e.g., \citealt{bahc97, kirh99, dunl03, floy04}). However, even at low redshifts there remains a significant controversy as to whether quasars are associated with enhanced or decreased star formation in their hosts, whether there are signs of merger activity beyond those expected in similar inactive galaxies, and whether black hole growth is primarily related to stellar mass or star formation rate \citep{yang17}. Furthermore, it is unclear how these relationships depend on circumnuclear obscuration, radio loudness or quasar luminosity. For example, \citet{wyle16a} compared type 2 quasar hosts at $z\sim 0.6$ with a stellar-mass matched sample of inactive galaxies, and found that type 2 quasars were hosted by massive galaxies undergoing minor mergers. \citet{cana01} similarly found a connection between quasars in a particular ``transition'' phase and interactions and star formation in their hosts. \citet{goul18b} found that mergers are particularly important triggers of luminous type 2 quasar activity. But other low-redshift studies, especially of type 1 quasars, have often failed to find a relationship between quasar and enhanced merger activity \citep{vill17}. 

Studies of quasar hosts at low redshifts probe the relationship between black holes and their hosts at the epoch when most of the stellar and black hole mass is already in place. In contrast, the epoch $2<z<3$ is particularly interesting for probing models of joint black-hole and host evolution because this is the period when black holes and their hosts were evolving most rapidly \citep{boyl98, rich06b} and when the quasars must have had the strongest impact on the formation of massive galaxies \citep{hopk06}. But observations of high-redshift active galaxies are even more ambiguous than those at low redshifts, complicated both by cosmological surface brightness dimming and by the relative compactness of high-redshift massive galaxies \citep{vand10}. 

At $z>2$, the host galaxies of low-luminosity active galaxies in the CANDELS survey \citep{grog11} appear to be mostly undisturbed and indistinguishable from inactive galaxies of comparable luminosities \citep{scha11, koce12}. These findings echo those of the GOODS survey for active galaxies at $z\sim 1$ \citep{grog05, vill12}. Perhaps low- and moderate-luminosity sources are not fueled by the merger process and are instead triggered by secular processes within galaxies \citep{hopk09b}, in which case the true tests of merger-driven galaxy and black hole formation scenarios must come from studies of high-luminosity sources. 

Until recently, very little was known about the hosts of high-luminosity quasars at $z \sim 2.5$. Attempts to image the hosts of high-redshift luminous quasars were at the limit of detectability \citep{kuku01, croo04, mcle09}, with an occasional aid from gravitational lensing \citep{peng06b}. Observations of the molecular gas content of powerful quasars using sub-millimeter continuum and molecular line emission has also yielded conflicting results. Some objects appear gas-rich and strongly star-forming \citep{beel04, pitc19}, while others are devoid of cool gas \citep{tsai15}. A slew of recent studies take advantage of the improved resolution and sensitivity of Wide Field Camera 3 (WFC3) on HST, as well as new methods to select obscured quasars, enabling new tests of the merger hypothesis \citep{glik15, fan16, hilb16, mech16, farr17}. However, disentangling the emission of the bright nuclear source from the surrounding faint and compact host galaxy remains a significant challenge.

In this paper we present the results of an HST study of luminous red and obscured quasars at $z\sim 2.5$, some of which show signatures of extreme quasar feedback. We describe sample selection, ancillary data and HST observations in Section \ref{sec:data}. We present host galaxy measurements in Section \ref{sec:host}. We discuss our results in comparison with other samples in Section \ref{sec:discussion}, concluding in Section \ref{sec:conclusions}. We use a $h=0.7$, $\Omega_{\rm m}=0.3$ and $\Omega_{\Lambda}=0.7$ cosmology throughout this paper. Following long-standing convention, emission lines are identified by their wavelength in air (e.g., [OIII]$\lambda$5007\AA), but all wavelength measurements are performed on the vacuum wavelength scale. W1$-$4 refer to magnitudes observed by the Wide-field Infrared Survey Explorer (WISE; \citealt{wrig10}); we use unWISE flux measurements \citep{lang16}. We define the $B$-band luminosity as $\nu L_\nu$ in the rest-frame $B$-band and when necessary we apply the conversion for the Sun, $\nu L_{\nu,\odot}$[$B$]$=0.57 L_{\odot}$, where $L_{\odot}=3.83\times 10^{33}$ erg/sec is used as a luminosity unit.

\section{Sample, observations and data reductions} 
\label{sec:data}

In this paper, we present results of HST observations of two somewhat different, but related populations of quasars at $2<z<3$. In Section \ref{sec:data:erq} we describe selection and ancillary data for a sample of extremely red quasars (ERQs), observed in HST programme GO-14608 (PI: Zakamska). In Section \ref{sec:data:t2} we similarly describe selection and ancillary data for a sample of type 2 quasar candidates, observed in HST programme GO-13014 (PI: Strauss). In Section \ref{sec:data:hst} we describe our HST observations and data reduction, and in Section \ref{sec:data:psf} we describe our modeling of the point spread function (PSF). The targets are listed in Table \ref{tab:measure}, with spectral energy distributions (SEDs) in Figure \ref{pic:sed}, separated into the two target categories discussed throughout the paper.

\subsection{Sample of extremely red quasars} 
\label{sec:data:erq}

ERQs were first defined by \citet{ross15} who selected spectroscopically confirmed quasars \citep{pari14, pari17} with $g<22$ mag in the Baryon Oscillation Spectroscopic Survey \citep{daws13} of the Sloan Digital Sky Survey-III \citep{eise11} based on very red infrared-to-optical colours, as measured from WISE and SDSS photometry. A subset of these objects showed spectroscopic properties not commonly seen in type 1 quasars, such as high ratios of NV$\lambda$1240\AA\ to Ly $\alpha$, high rest equivalent widths (REW) of CIV$\lambda$1550\AA\ and unusual ``stubby'' (wingless) emission line profile shapes.  

Extremely red quasars (ERQs) studied in this paper are drawn from the parent sample by \citet{hama17} who used a combination of selection criteria based on broad-band colours ($i_{\rm AB}-W3_{\rm Vega}>9.8$ mag) and emission-line properties (REW of CIV$>$100\AA) to define a sample of 97 ERQs at redshifts $2<z<3.4$. We have been conducting extensive follow-up of these objects across the electromagnetic range. Among the most striking properties of these objects are their extremely high bolometric luminosities, typically $L_{\rm bol}\simeq 10^{47}$ erg/sec, but reaching $10^{48}$ erg/sec in some cases \citep{hama17}, which is an order of magnitude greater than the Eddington limit for a $10^9M_{\odot}$ black hole.  

Near-infrared (rest-frame optical) spectra of ERQs routinely show forbidden emission lines such as [OIII]$\lambda$5007\AA\ with unprecedented velocity widths, reaching full width at half maximum (FWHM) of 6000 km/sec and showing clear signatures of outflows \citep{zaka16b,perr19}. Due to the relatively low critical density, this transition is thought to arise on relatively large scales, at least $\sim 100$ pc \citep{hama11} based on theoretical expectations and out to $\sim 10$ kpc scales \citep{liu13a, liu13b} as seen in low-redshift quasars, although the spatial extents of [OIII] winds in ERQs have not yet been measured. 

As a population, ERQs are radio-quiet but not radio-silent \citep{hwan18}. Their radio luminosities and [OIII] velocity widths lie on the extreme end of the radio/[OIII] relationship found in low redshift quasars \citep{zaka14}. These observations are consistent with the picture in which their radio emission is a by-product of quasar-driven winds, produced when relativistic particles are accelerated in the resulting shocks. Radio data reinforce our hypothesis that ERQs are hosting some of the most extreme galactic winds known \citep{alex16, hwan18}. 

X-ray observations demonstrate that ERQs are an X-ray luminous, highly obscured population ($N_{\rm H}\simeq 10^{24}$ cm $^{-2}$; \citealt{goul18a}) with an X-ray-to-infrared ratio similar to that of other luminous quasars, indicating that most of the infrared emission is likely due to the quasar, not to the star formation in the host galaxy. Although their optical-to-infrared colours are red, their optical (rest-frame UV) continua are relatively bright and flat and their UV emission lines are broad (FWHM of CIV$>$2000 km/sec), inconsistent with those of type 2 quasars \citep{zaka03}. This contradiction between optical and X-ray classifications is resolved by optical spectropolarimetric observations \citep{alex18}. These data show that the rest-frame UV continua of ERQs are highly polarized ($\ga 10\%$) and are therefore likely dominated by the light from the obscured quasar scattered into our line of sight by the surrounding medium rather than by direct emission from the nucleus. Furthermore, polarization signatures within the UV emission lines (Ly $\alpha$, CIV) indicate outflow activity on scales of several parsecs. Therefore, outflow activity in ERQs is apparent on a wide range of spatial scales. 

\begin{figure*}
\includegraphics[scale=0.85, trim=0cm 10cm 0cm 0cm, clip=true]{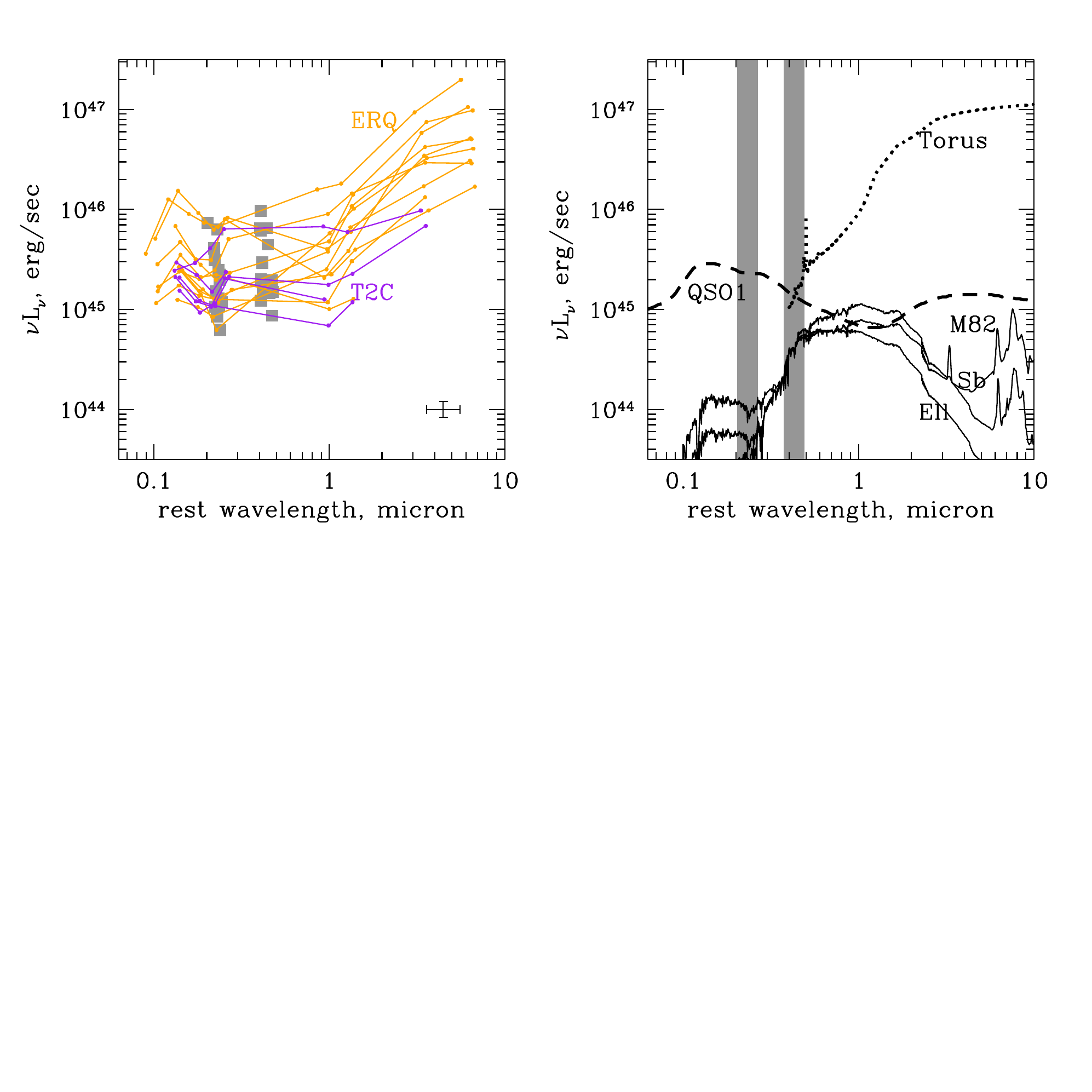}
\caption{{\it Left:} SEDs of ERQs (orange) and type 2 quasar candidates (purple) from SDSS and WISE data, with wavelengths of HST observations presented in this paper shown in grey squares corrected to the rest frame. An error bar illustrating the median filter width and the flux uncertainty required for detection is shown in the bottom right corner. {\it Right:} For comparison, to represent the infrared emission of our targets we show the `torus' template of \citet{poll07} normalized to $\nu L_{\nu}$[5\micron]$=10^{47}$ erg/sec, to represent the scattered light we show the type 1 quasar template of \citet{rich06} normalized to a $10^{46}$ erg/sec bolometric luminosity, and we display three galaxy templates \citep{poll07} normalized to $\nu L_{\nu}=10^{11.0} L_{\odot}$ in the B band. The grey bands mark the approximate wavelengths of our HST observations.}
\label{pic:sed}
\end{figure*}

For the HST follow-up we prioritized eleven ERQs with then-ongoing follow-up spectroscopic observations covering [OIII]. As near-infrared observations arrived, [OIII] outflow activity in ERQs appeared ubiquitous \citep{perr19}. Compared to the parent sample of ERQs from \citet{hama17}, ERQs targeted for follow-up NIR spectroscopy (and therefore for the HST follow-up) are restricted both in redshift, as [OIII] must fall within the $H$-band or the $K$-band, and in position on the sky for ground-based observability. We have verified that the 11 ERQs selected for the HST study are consistent (in the sense of the Kolmogorov-Smirnov test) with the parent population from \citet{hama17} in all other parameters: infrared luminosity and optical-to-infrared colours, as well as CIV emission line width, equivalent width and kurtosis. The optical-to-infrared SEDs of ERQs selected for the HST observations are shown in Figure \ref{pic:sed}. 

\subsection{Sample of type 2 quasar candidates}
\label{sec:data:t2}

In the classical geometric unification model of AGN \citep{anto93}, obscured (type 2) and unobscured (type 1) objects are distinguished primarily on the basis of the width of the permitted optical emission lines such as H$\alpha$ and H$\beta$ \citep{khac74}. In this model, the high velocity widths seen in type 1 AGN are appropriate for the gas seen close to the black hole, whereas the low velocity widths seen in type 2 AGN are commensurate with the much shallower potential of the extended host galaxy. A FWHM of 2000 km/sec is typically used to separate type 1 and type 2 quasars \citep{zaka03, reye08}.

\citet{alex13} selected 145 type 2 candidates from the BOSS survey in the redshift range $2<z<4.3$ based on the velocity width of Ly $\alpha$ and CIV. WISE fluxes of these objects indicate bolometric luminosities of up to a few $\times 10^{46}$ erg/sec \citep{alex13}. Near-infrared (rest-frame optical) spectroscopy of a subset of these objects \citep{gree14b} demonstrates that most of them have a broad H$\alpha$ component and that they therefore have moderate extinction toward the nucleus, $A_{\rm V}\la 2$ mag, although several quasars with type 2 rest-frame optical spectra (i.e., no broad H$\alpha$ or H$\beta$) have also been identified \citep{alex18}. Their optical (rest-frame UV) spectra show polarization of a few per cent, suggesting that at least some of the continuum is dominated by scattered light. 

[OIII] outflow signatures are present in some of these type 2 quasar candidates \citep{alex18}, although they are not nearly as extreme as those seen in ERQs. Given the correlation between line width and radio emission among the radio-quiet quasars \citep{zaka14}, it is not surprising that the radio luminosities of high-redshift type 2 quasar candidates are also appreciably lower than those of ERQs \citep{alex16}. We conclude that the objects identified by \citet{alex13} are in some ways similar to the classical type 2 quasars at low redshifts \citep{zaka03, reye08, yuan16}, although with a range of extinctions to the nucleus extending to lower values, as appropriate for the selection based on rest-frame UV (instead of rest-frame optical) properties. 
 
As the WFC3/F160W filter fits neatly between the Balmer break and the [OIII]$\lambda$5007\AA\ line for redshifts between 2.4 and 2.6, this was our primary selection criterion of type 2 quasar candidates for HST follow-up. There are 16 objects in this redshift range in the parent sample by \citet{alex13}. We chose the six objects with the strongest and narrowest emission lines to observe with HST. The FWHM of the CIV line in these objects range from 900 to 1300 km/s, with line luminosities ranging from $8\times 10^{42}$ to $8\times 10^{43}$ erg/sec. Their SEDs from SDSS and WISE photometry are shown in Figure \ref{pic:sed}. Since they were selected from the SDSS survey without regard for their infrared luminosity, they are not always detected in WISE. 

\begin{figure}
\includegraphics[scale=0.85, trim=0cm 10cm 10cm 0cm, clip=true]{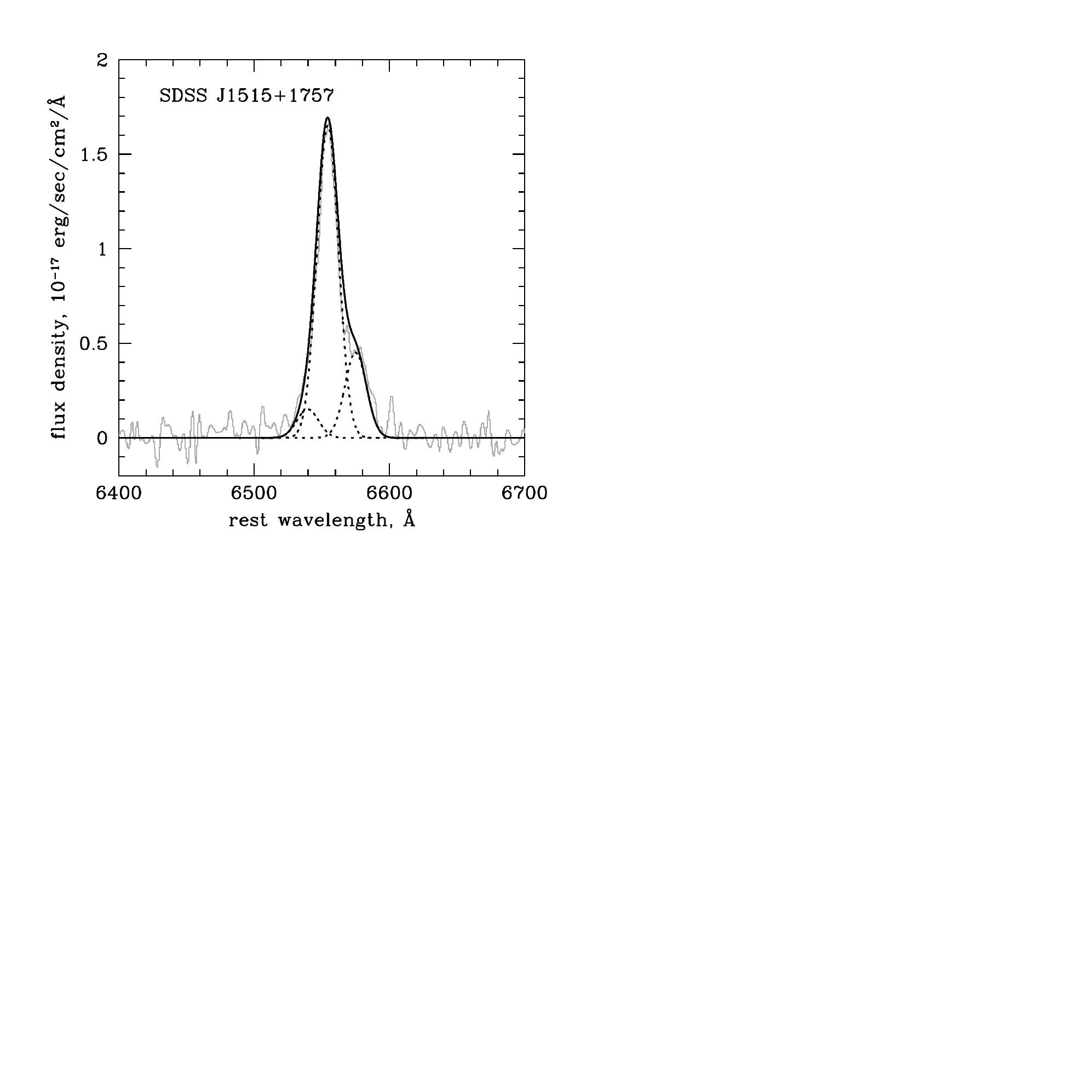}
\caption{Previously unpublished Gemini near-infrared spectrum of the H$\alpha$+[NII]$\lambda$6548,6563\AA\ line blend of SDSS~J1515+1757, fitted with the line profile derived from the [OIII]$\lambda$5007\AA\ fit \citep{alex18}. No additional broad component in H$\alpha$ is required, and therefore the target is consistent with the classical type 2 definition.}
\label{pic:gemini}
\end{figure}

Of the six objects, only two have available near-infrared (rest-frame optical) spectroscopy. One, SDSS~J1444$-$0013, shows narrow H$\beta$ and high [OIII]/H$\beta$ ratio characteristic of the narrow-line region, but a strong broad H$\alpha$ component \citep{gree14b}, so it would be classified as a type 1.9 source based on its rest-frame optical spectrum. The other, SDSS~J1515+1757, shows the same type of H$\beta$+[OIII] spectrum \citep{alex18}, but has no sign of a broad H$\alpha$ component (Figure \ref{pic:gemini}) and is therefore a spectroscopically confirmed type 2 quasar. In the absence of a spectroscopic type for every object, throughout this paper we continue to refer to this subsample as `type 2 quasar candidates'. 

\subsection{HST images and data reduction}
\label{sec:data:hst}

In both HST programmes, we observe each object in the rest-frame ultraviolet with the F814W filter of ACS and rest-frame optical with the F140W or F160W filter of WFC3. At $z\sim 2.5$, the ACS/F814W filter spans rest-frame wavelengths 2000$-$2700\AA, an essentially line-free region, thus sampling the near-ultraviolet continuum as a probe both of active star formation and scattered quasar light. The exposure times are one orbit per object for the ERQs and two orbits per object for the type 2 quasar candidates. 

The purpose of our near-infrared HST WFC3 observations is to observe starlight from the host galaxy, so for each object we choose the filter to sample longward of the 4000\AA\ break. All type 2 quasar candidates are observed with the F160W filter. In the three lowest redshift objects the [OIII]$\lambda$5007\AA\ emission line falls within the filter and may potentially contaminate the extended emission. For ERQs, the filter choice -- F140W or F160W -- is adjusted based on the redshift (Table \ref{tab:measure}), so only in one target (SDSS~J2215$-$0056) do we expect the [OIII] line to fall within the filter. In Section \ref{sec:host:mass} we discuss the possible contamination of our host observations by emission lines and find it to be negligible. Each target is observed for one orbit with WFC3. 

In both programmes, ACS observations are set up with default \texttt{acs-wfc-dither-box} patterns. WFC3 observations of type 2 quasar candidates use both \texttt{wfc3-ir-dither-box-min} and \texttt{wfc3-ir-dither-blob} dithers, while the ERQ observations use only the former. One observation, of ERQ SDSS~J0936+1019, suffered from guide star acquisition failure in the WFC3 image (the ACS image is not affected), rendering the image unusable for most of our analyses. Thus the final sample observed with HST consists of 10 ERQs and 6 type 2 quasar candidates, listed in Table \ref{tab:measure}. 

In the following analysis we focus on the WFC3 data to constrain the host galaxy properties, leaving ACS images for future work. We combine the calibrated WFC3 images (\texttt{flt}) of each target into a single drizzled image (\texttt{drz}) using \texttt{DrizzlePac} version 2.1.22. There are 4-6 exposures (\texttt{flt}) per target. In this process cosmic rays and the background are removed and the images are corrected for image distortion and rebinned (``drizzled'') onto a final grid \citep{fruc02}. The pixel size of the original \texttt{flt} image for WFC3 IR is 0.13\arcsec, but improved resolution can be achieved by taking advantage of the sub-pixel dithers we use in the observations. During final drizzling, we shrink the input pixel (drop size or \verb|final_pixfrac|) by a factor of 0.8 and subsample the output image with a factor of two finer pixels (0.065\arcsec). The final \texttt{drz} images are registered on a coordinate grid with North at the top. We visually inspect the drizzled images and find the \texttt{drz} images to be sharper than the input \texttt{flt} images. In particular, the first Airy ring of the PSF and its substructure are clearly visible in the final images. 

\subsection{PSF modeling}  
\label{sec:data:psf}

In Section \ref{sec:host}, we present the analysis of the quasar host emission using two different methods -- by subtracting the quasar PSF and conducting aperture photometry on the residuals, resulting in 1D surface brightness profiles of the host galaxies, and by jointly fitting PSF+parametric host models to the 2D images. In both cases a PSF model is required. To model the PSF we use \texttt{Tiny Tim} version 7.5, a software package for simulating the HST PSF \citep{hook08, kris11}. The fitting is performed on each of the flat-fielded (\texttt{flt}) images that are then combined to make the final drizzled (\texttt{drz}) image. Before the fit, we subtract a constant background level estimated by \texttt{AstroDrizzle} in the original reduction procedure from the flat-fielded image. Then we generate PSF models for the flat-fielded images using \texttt{Tiny Tim}. The PSF model depends on the filter, the focus, the object spectrum, and the position of the object on the detector. We determine the focus of each exposure by fitting the PSF to a reference star in the same field as the quasar, and we determine the spectral type of the reference star from its SDSS colours; the variations in the focus among different images are a negligible source of error. We approximate the spectrum of the quasar by power-law interpolation between the SDSS $z$-band and WISE W1 fluxes. 

The PSF model generated has a diameter of 6\arcsec\ to cover extended spikes and is oversampled by a factor of six to allow sub-pixel shifts. This oversampling factor is chosen as a compromise between being able to capture the structure within the PSF while keeping the computational time manageable. The free parameters are the peak position of the PSF, its normalization, and the width of a symmetric Gaussian kernel that is convolved to the PSF model to simulate pointing jitter. During the fitting process we also apply a charge diffusion kernel provided by \texttt{Tiny Tim}. The best-fit PSF of each exposure is then geometrically corrected and dither-combined with \texttt{AstroDrizzle} in the same way as the images, which accounts for the sub-pixel dithering between constituent exposures. The combined PSF model is then subtracted from the drizzled image (with extension \texttt{drz}). 

We find that fits that minimize $\chi^2$ often over-subtract the PSF at the nucleus, leaving strong negative residuals in a small number of pixels. This is likely due to the PSF model trying to match the extended features (the host galaxy) by increasing the normalization of the PSF. To reduce the over-subtraction, we penalize the negative residuals by increasing their weights in the $\chi^2$ calculation by a factor of 9 relative to the positive residuals.

\begin{figure*}
    \centering
    \hbox{
    \includegraphics[width=4in]{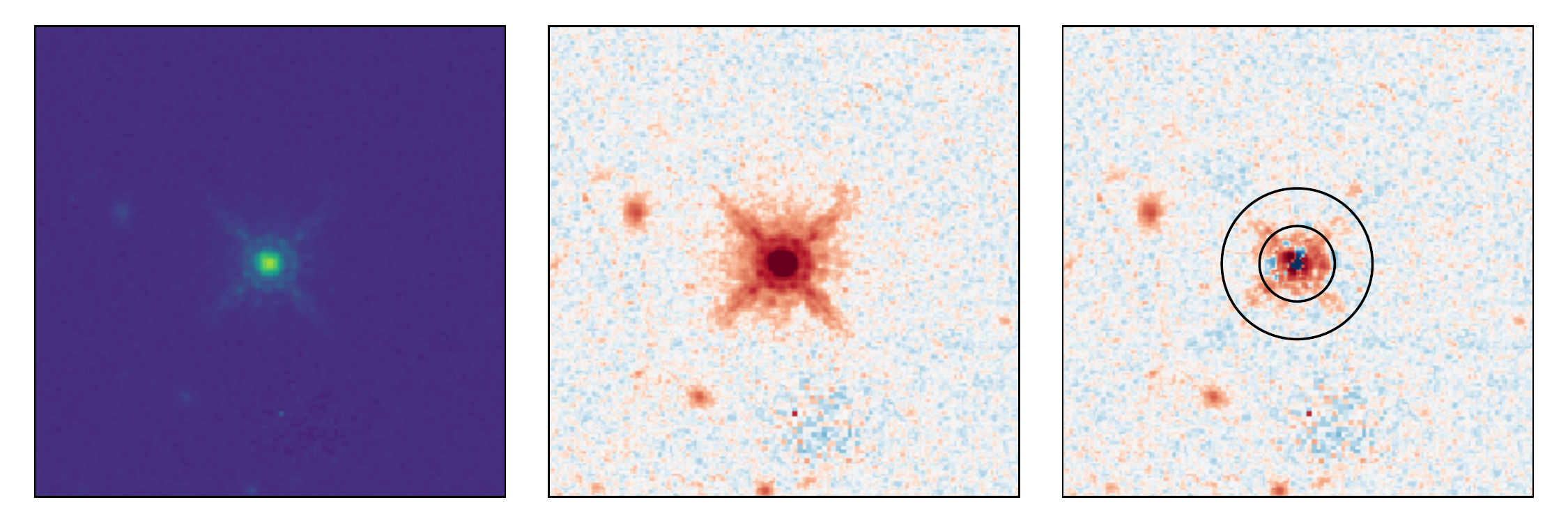}
    \includegraphics[width=3in]{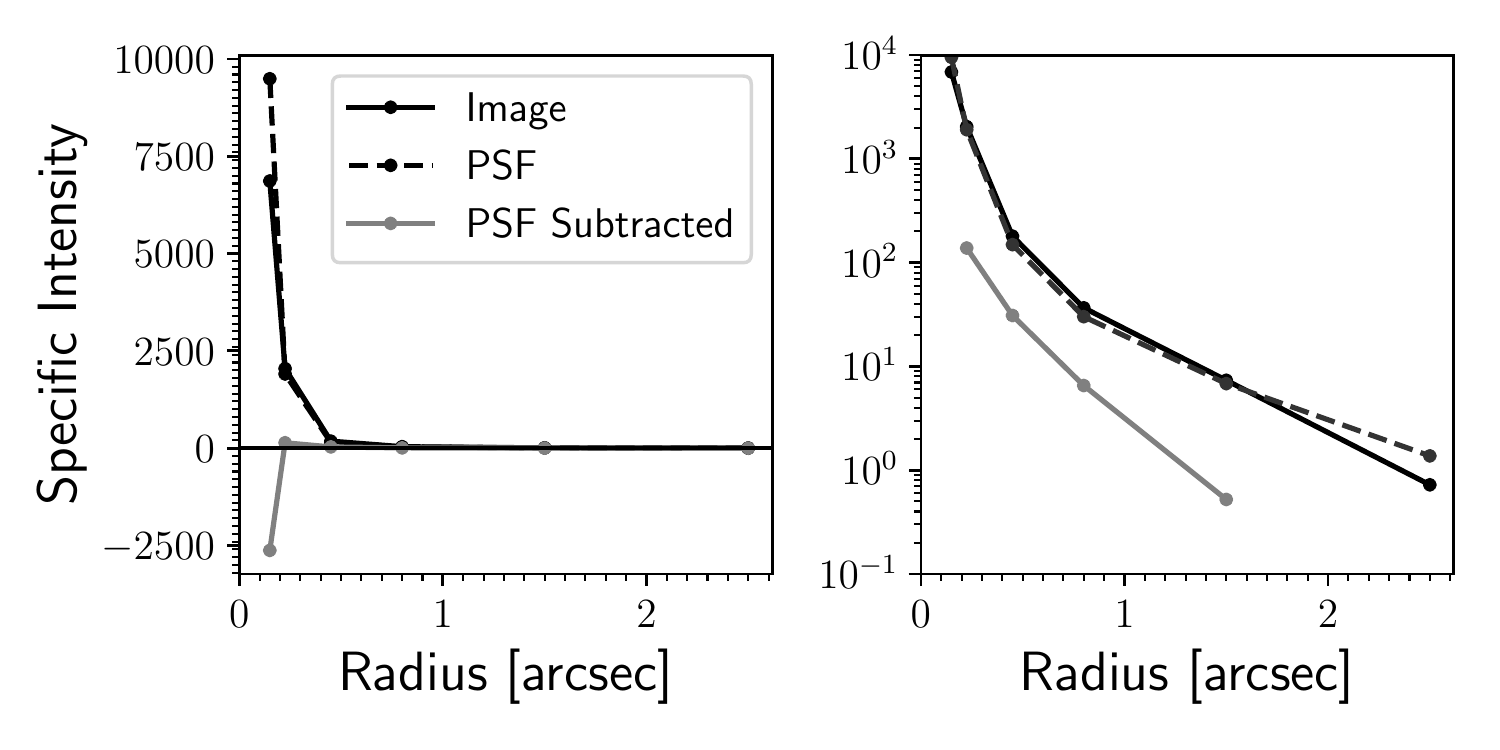}
    }
    \caption{
    PSF subtraction procedure illustrated for a reference star in the field of one of our targets. From left to right: the drizzled WFC3 F140W image, same with a more sensitive brightness stretch, PSF-subtracted residual with circular apertures of radii 1\arcsec\ and 2\arcsec\ superposed. The last two panels show aperture photometry of the original image, PSF model and residuals on linear and logarithmic scales. 
    }
    \label{pic:wfc3:refstar}
\end{figure*}

We use a star in the HST F140W/F160W image of each quasar to evaluate the performance of the PSF subtraction (Figure \ref{pic:wfc3:refstar}). Within a circular aperture of radius 1\arcsec\ centered on the star, the PSF-subtracted residual has negligible bias with a standard deviation of 5\% of the original flux of the star, whereas beyond 1\arcsec, the residuals are $<$10\% of the (already small) extended emission of the PSF. Thus our ability to characterize the extended faint emission around bright quasars is limited by our ability to accurately measure the PSF. 

\section{Host Galaxy Properties}
\label{sec:host}

\begin{figure*}
    \centering
    \vbox{
    \hbox{
    \includegraphics[width=4in]{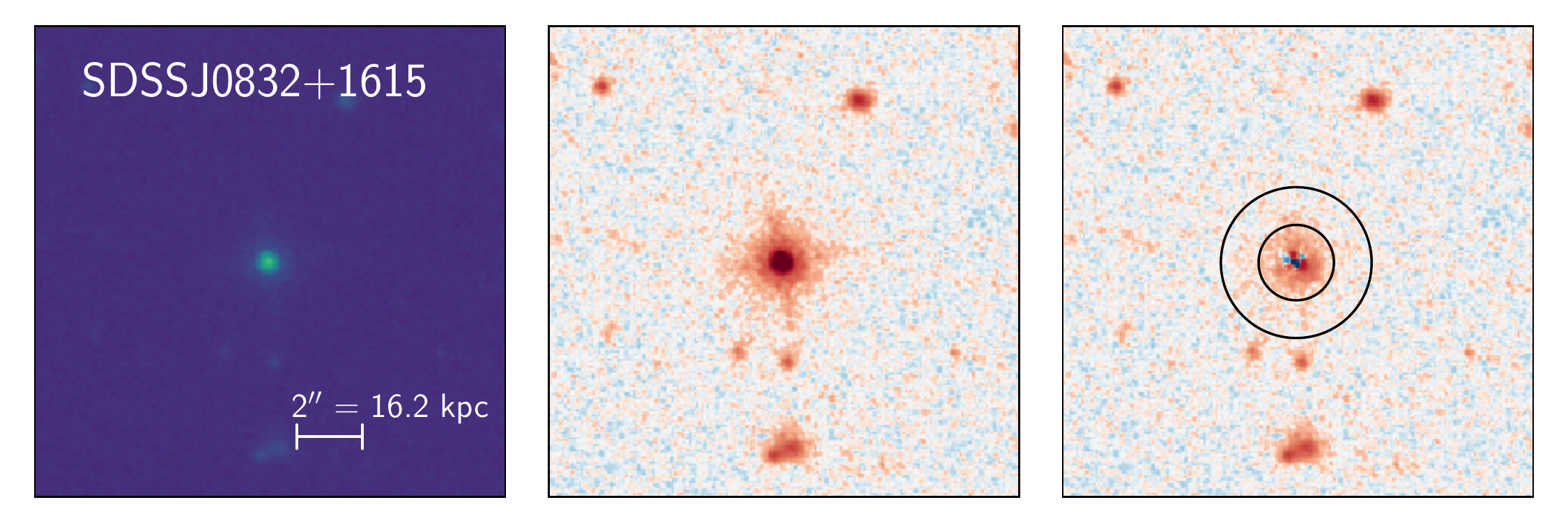}
    \includegraphics[width=3in]{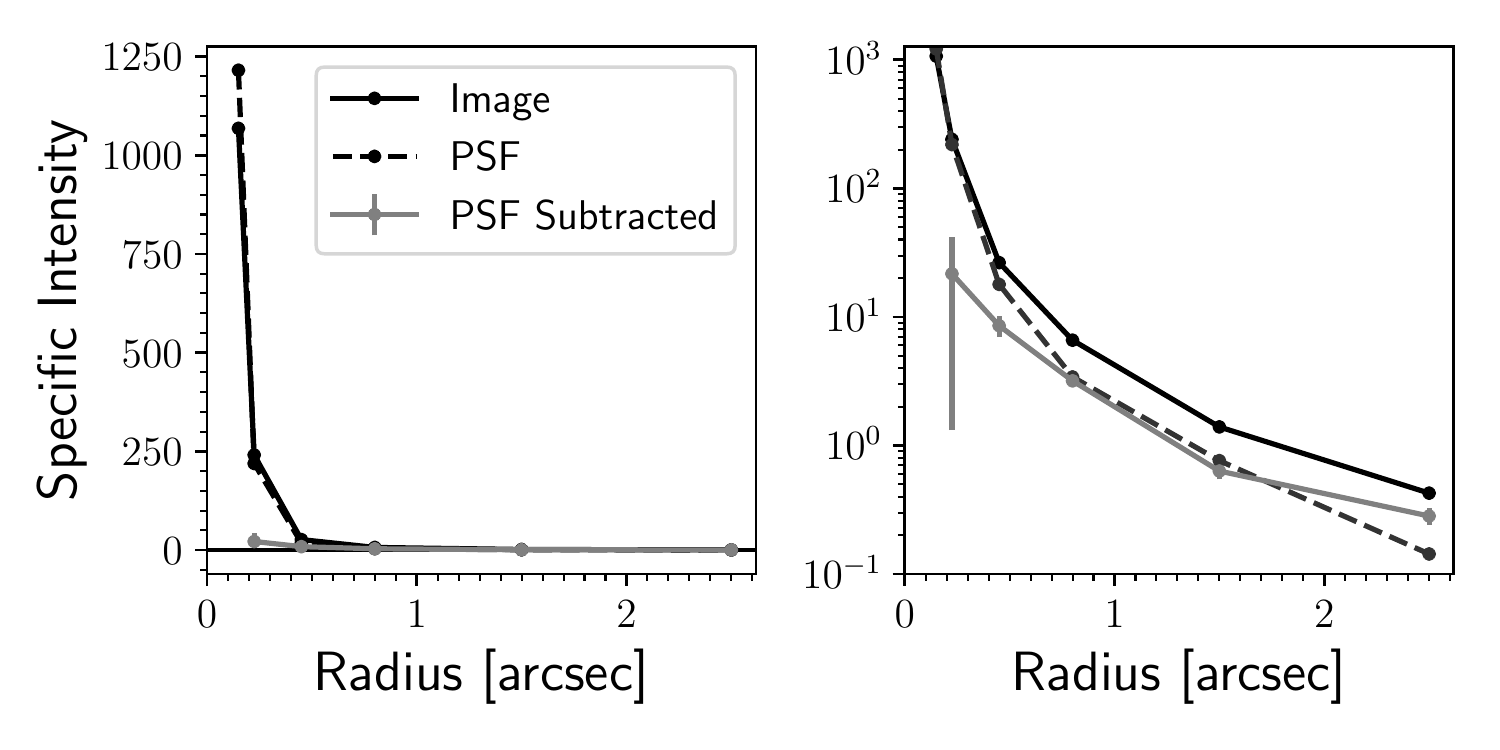}
    }
    \hbox{
    \includegraphics[width=4in]{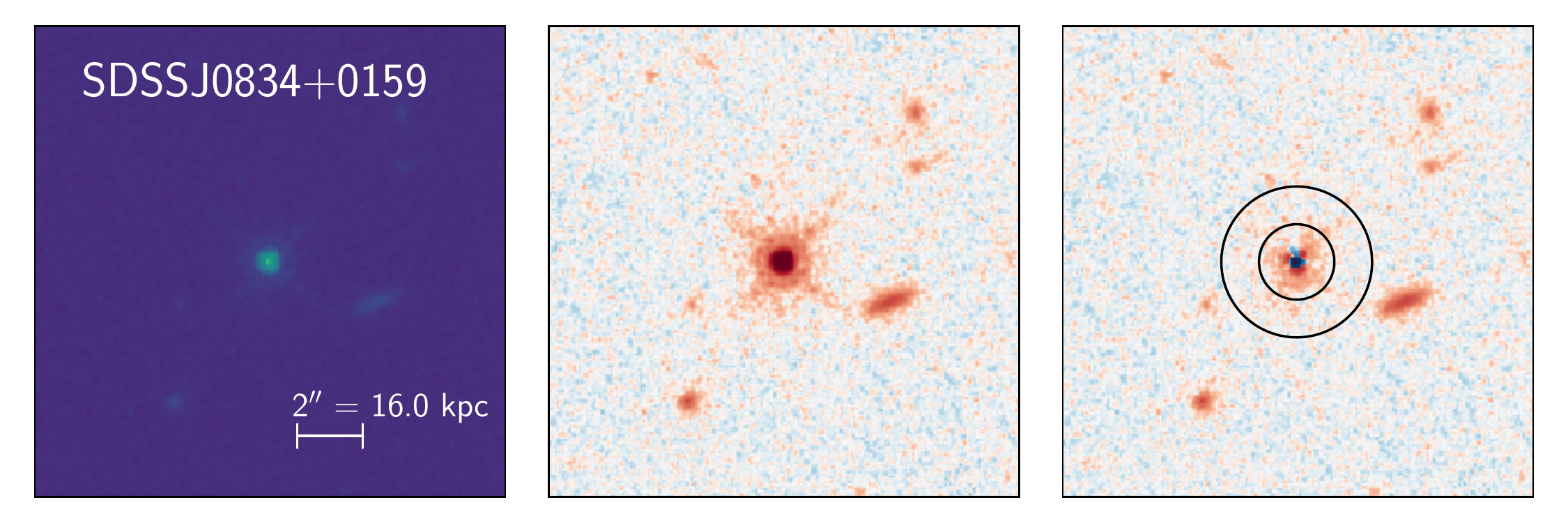}
    \includegraphics[width=3in]{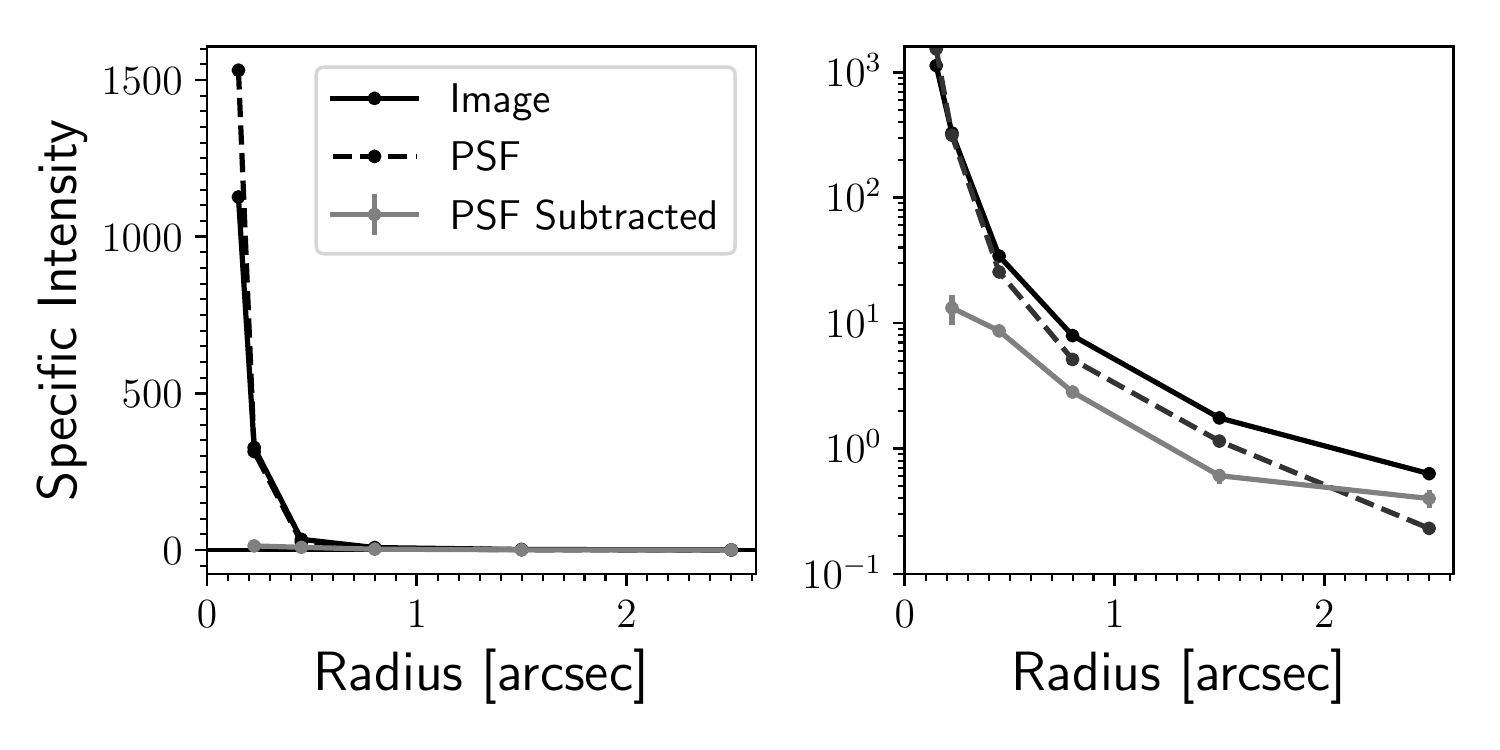}
    }
    \hbox{
    \includegraphics[width=4in]{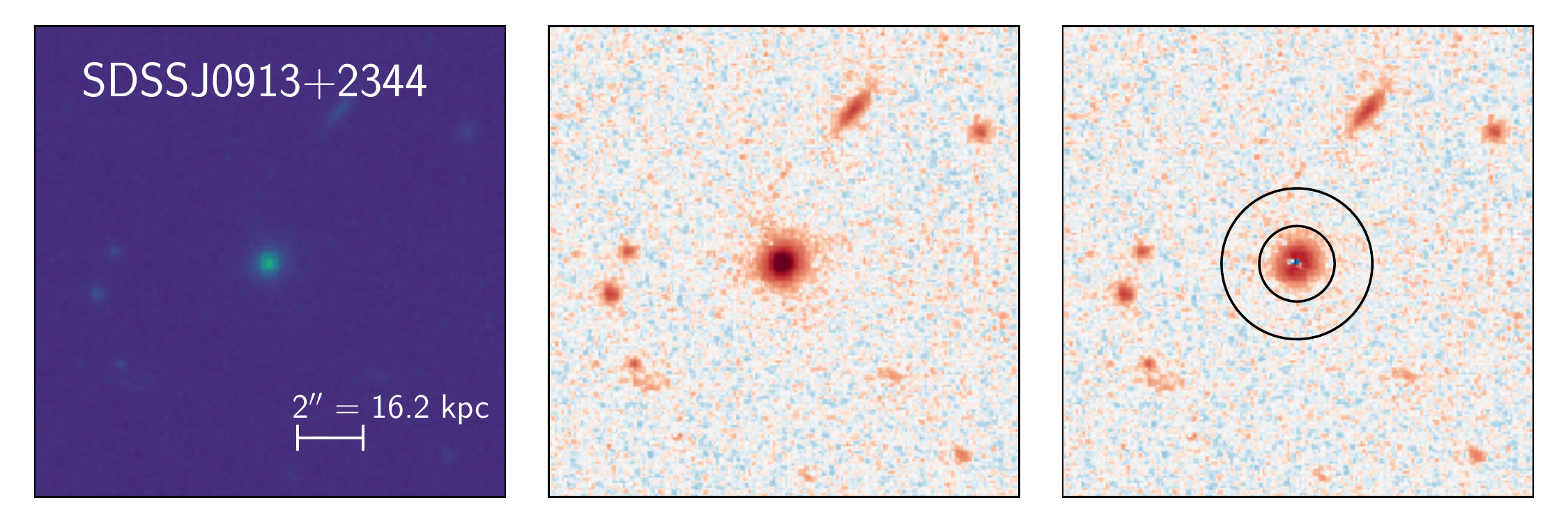}
    \includegraphics[width=3in]{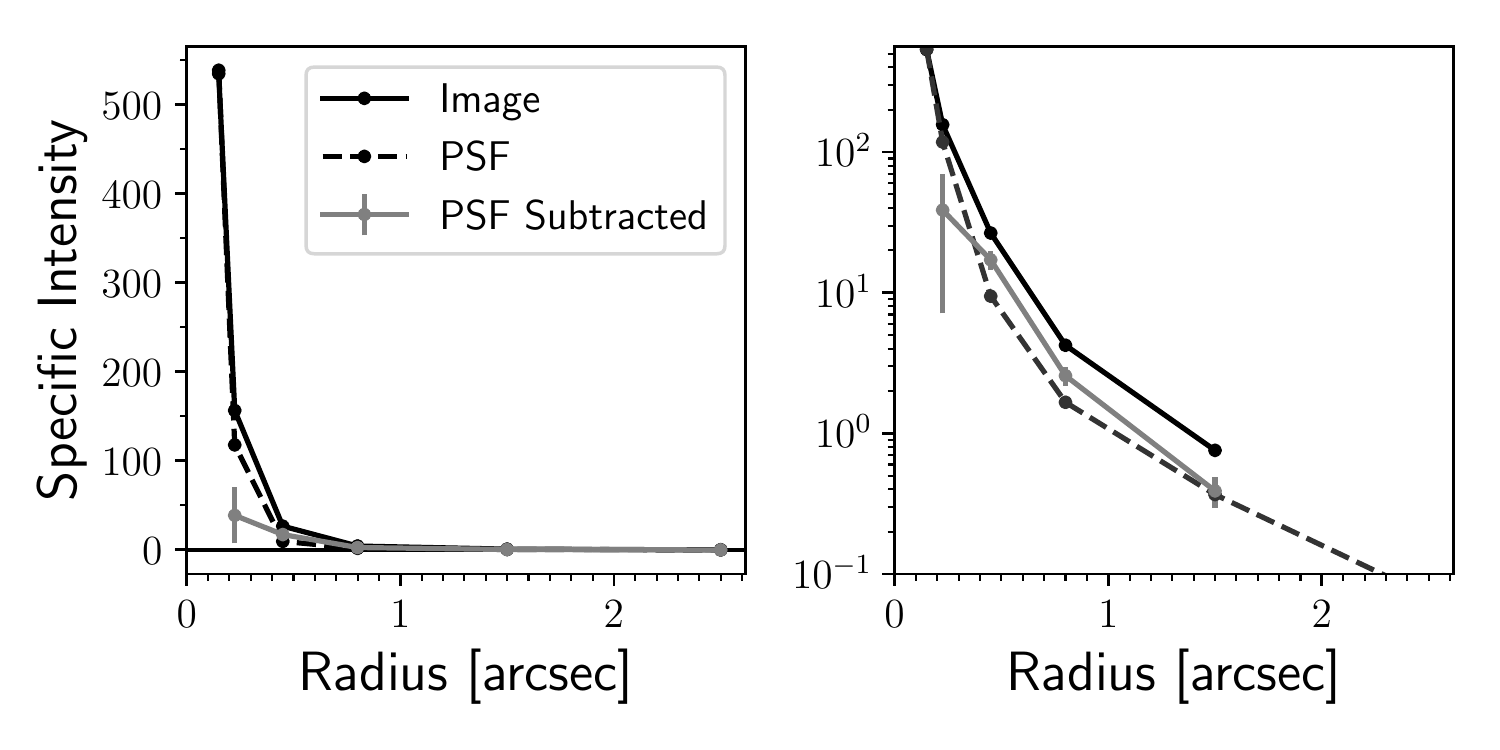}
    }
    \hbox{
    \includegraphics[width=4in]{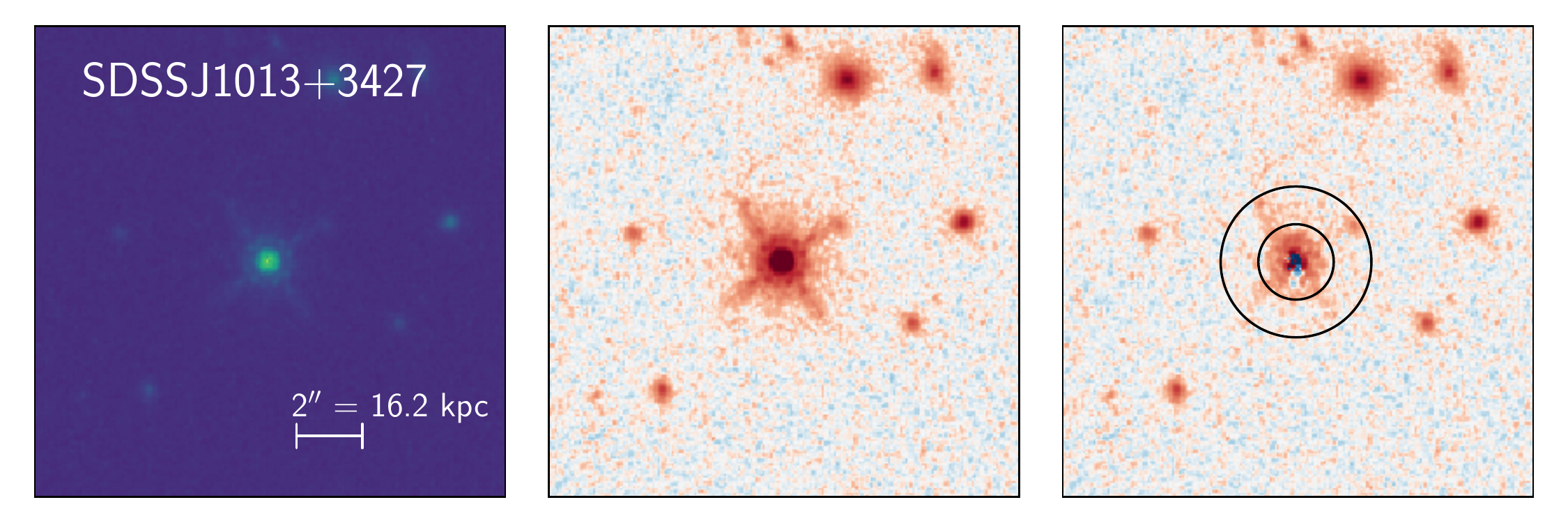}
    \includegraphics[width=3in]{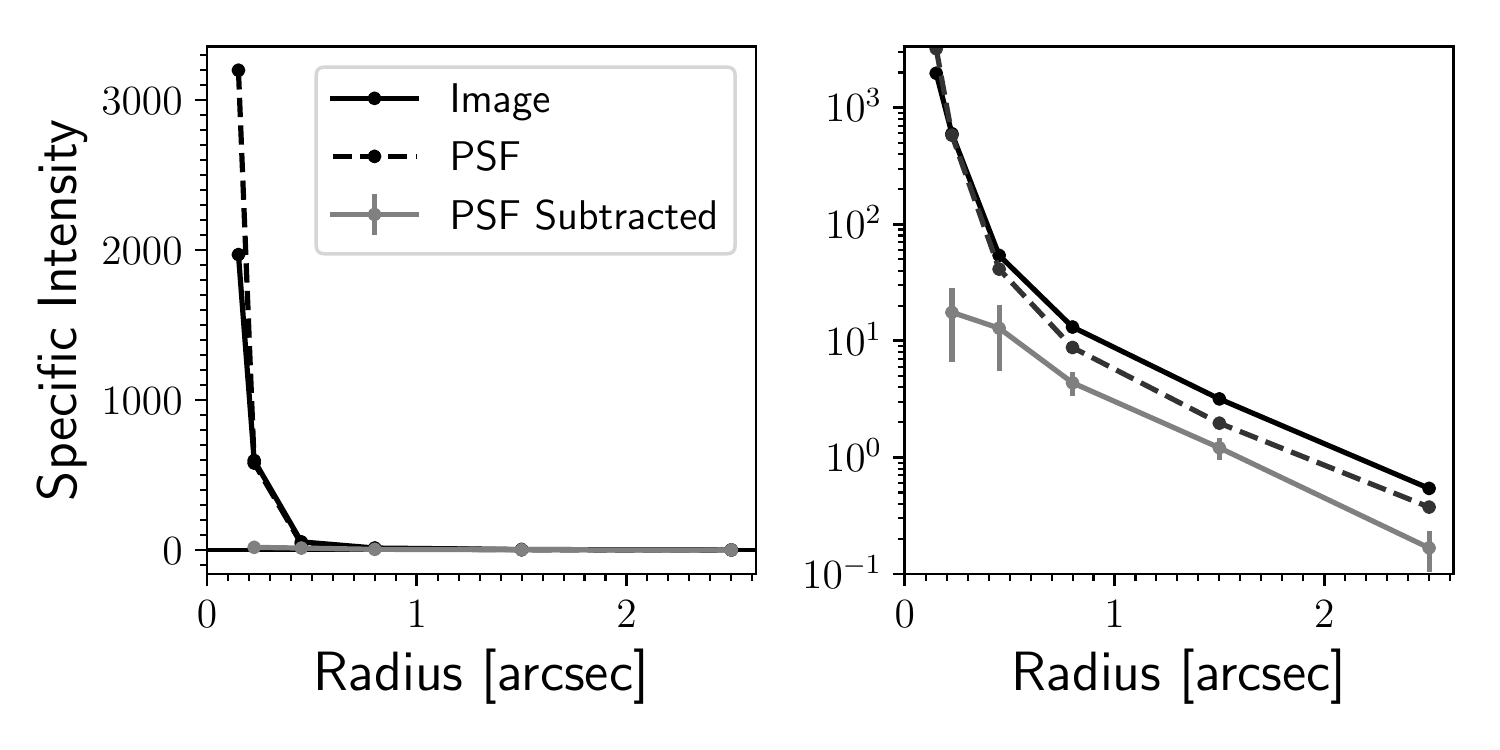}
    }
    \hbox{
    \includegraphics[width=4in]{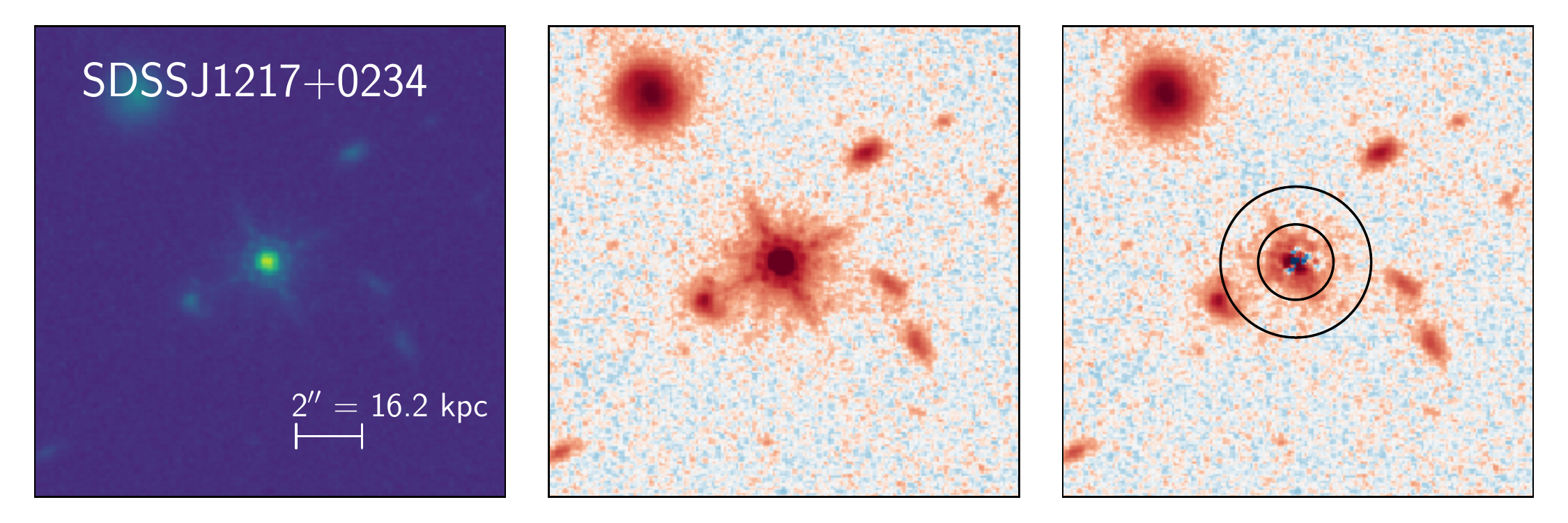}
    \includegraphics[width=3in]{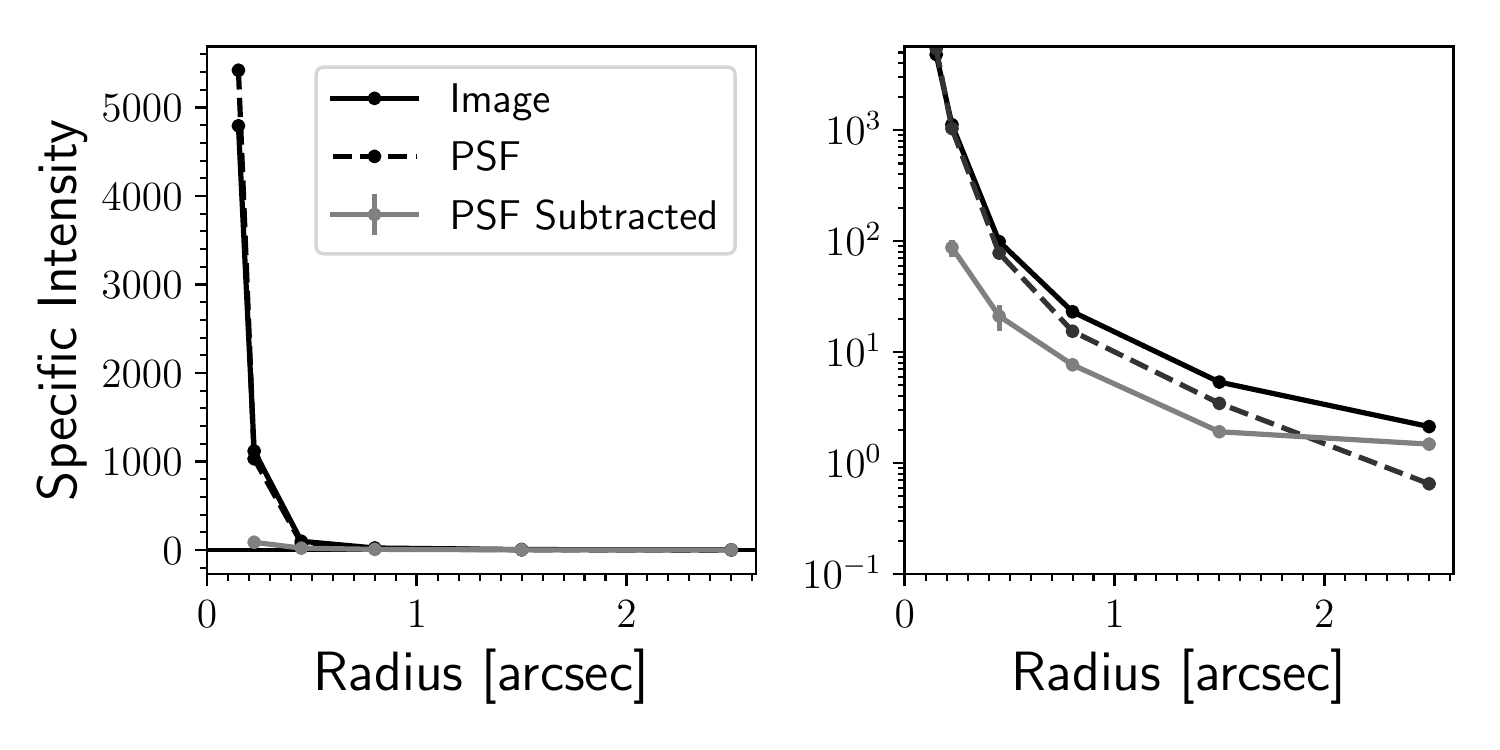}
    }
    }
    \caption{{\it Left:} The WFC3 F140W/F160W images of our sample of extremely red quasars; the filters used for each target are listed in Table \ref{tab:measure}. North is up and East is to the left. The first two panels are the original image and the third is the PSF subtracted image. The second and third panels adopt a colour scheme highlighting faint features showing the negative, zero, and positive values in blue, white, and red. The two circles are apertures of radii 1\arcsec\ and 2\arcsec. 
    {\it Right:} The radial distribution of the specific intensity in units of 
    $1\times10^{-30}$ $\mathrm{erg~s^{-1}~cm^{-2}~Hz^{-1}~arcsec^{-2}}$ of the original image, the PSF model, and the PSF subtracted image. Error bars combine the Poisson noise, the background-subtraction error and the difference between the two methods of PSF subtraction discussed in Sections \ref{sec:host:aperture} and \ref{sec:host:size}. The inner-most radial bin, where the fractional uncertainties of PSF subtraction are highest, is not shown.
    }
    \label{pic:wfc3:erq}
\end{figure*}

In Section \ref{sec:host:expected}, we summarize the relationship between the two subsamples and the expectations for the PSF contribution. We detect extended emission in almost all of our 16 targets, as shown in Figures \ref{pic:wfc3:erq} and \ref{pic:wfc3:t2c}. We conduct two different sets of host galaxy measurements to better understand their uncertainties and the limitations of our observations. In Section \ref{sec:host:aperture}, we discuss our measurements of host fluxes using aperture photometry, in Section \ref{sec:host:size} we discuss model fitting and structural parameters of the host galaxies, and in Section \ref{sec:host:mass} we compare the results from the two methods and estimate host luminosities and masses. 

\addtocounter{figure}{-1}
\renewcommand{\thefigure}{\arabic{figure} (Cont.)}
\begin{figure*}
    \centering
    \vbox{
    \hbox{
    \includegraphics[width=4in]{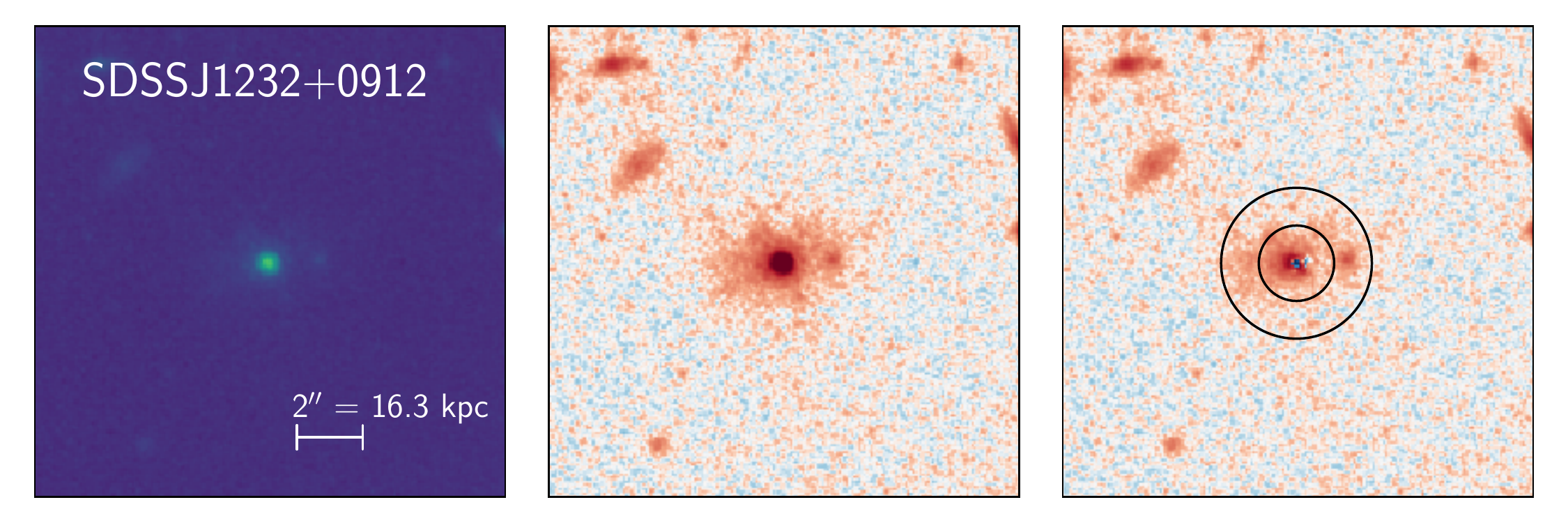}
    \includegraphics[width=3in]{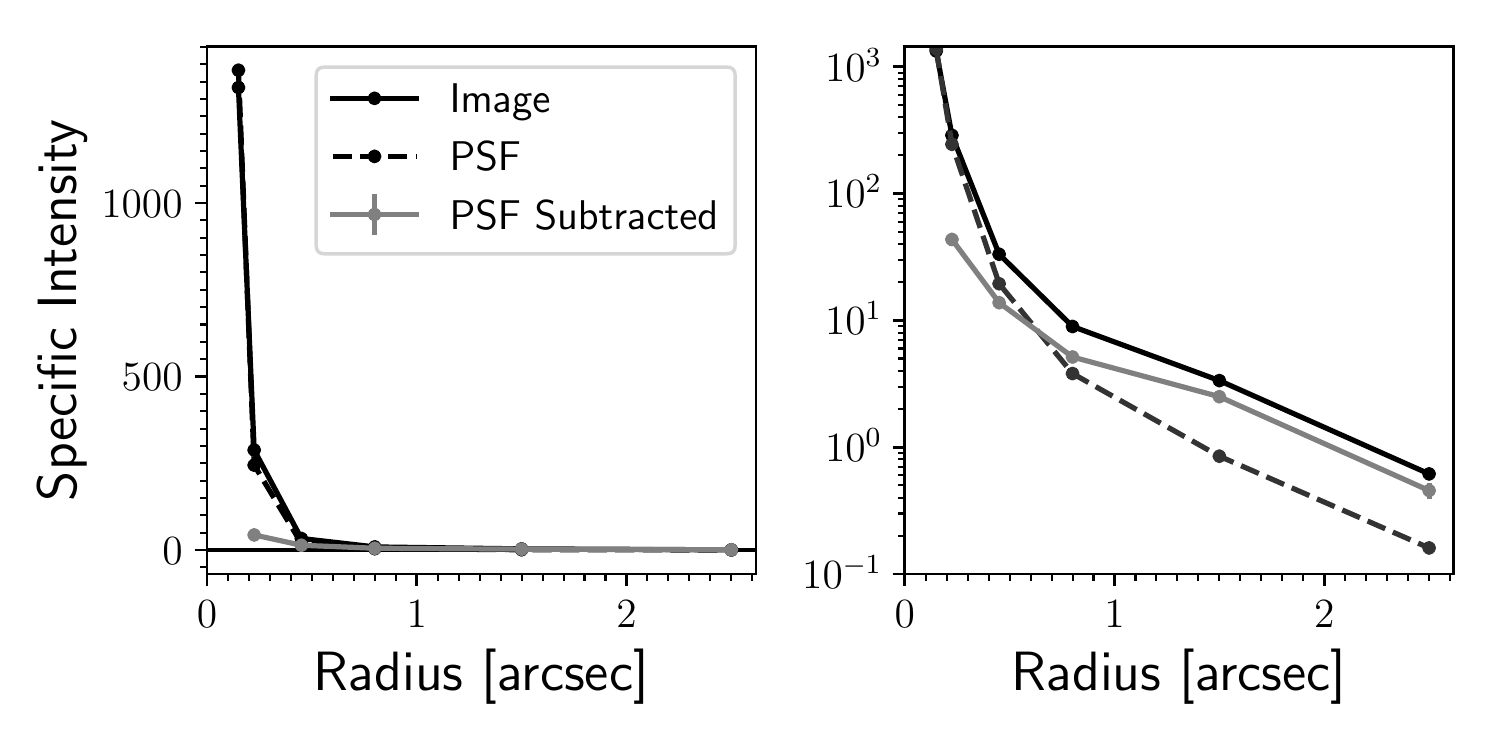}
    }
    \hbox{
    \includegraphics[width=4in]{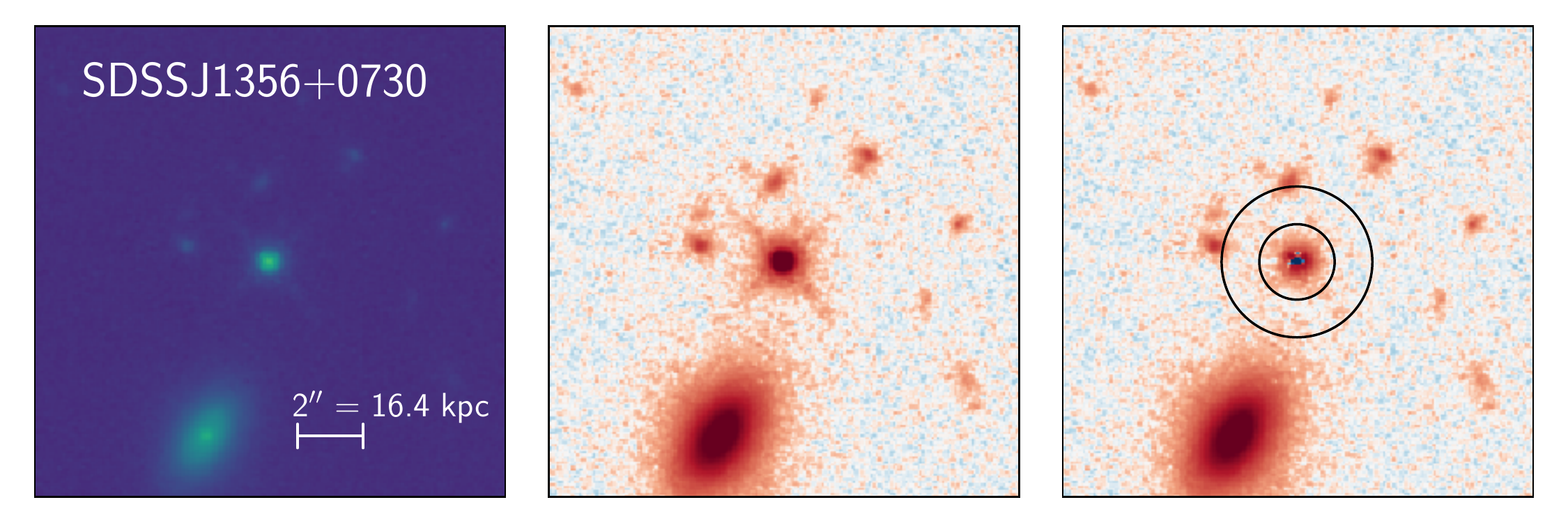}
    \includegraphics[width=3in]{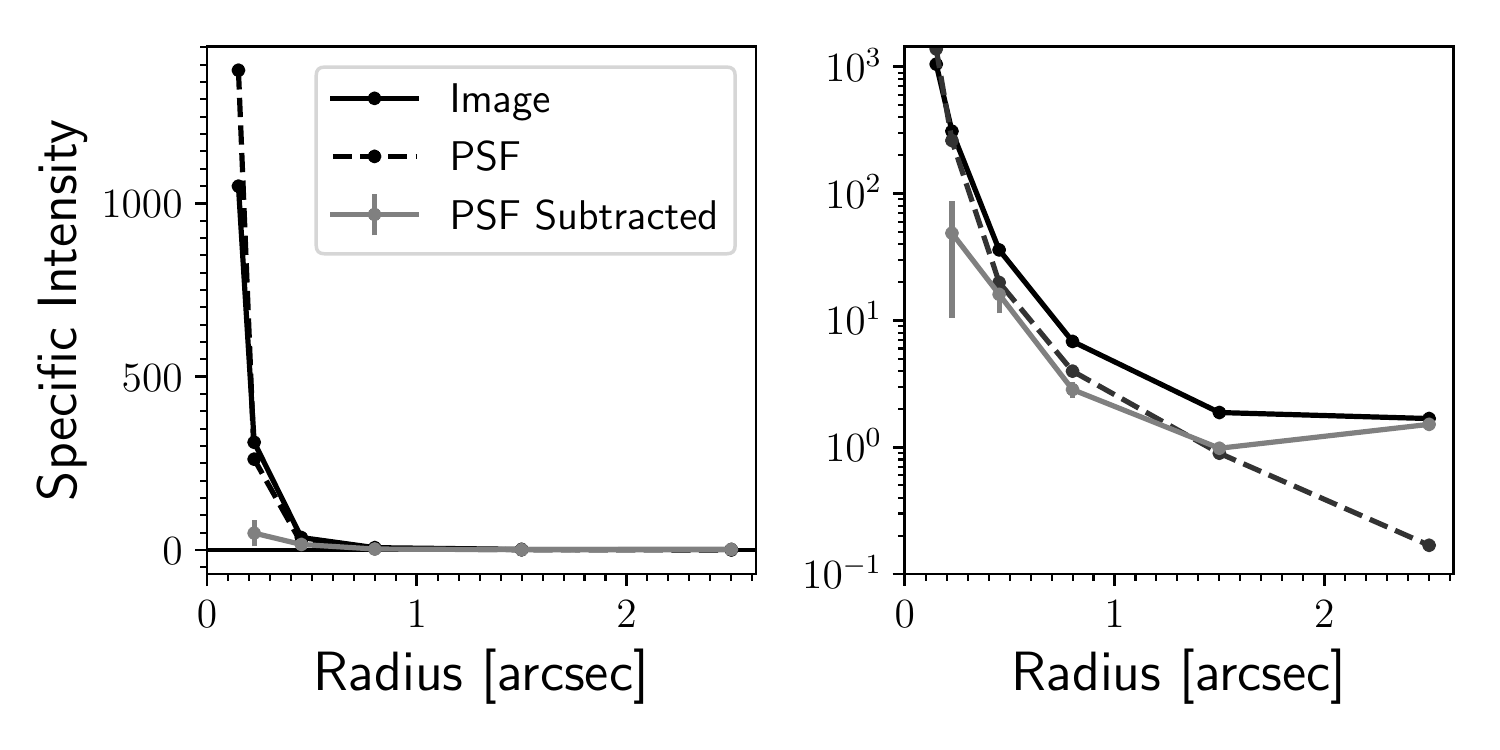}
    }
    \hbox{
    \includegraphics[width=4in]{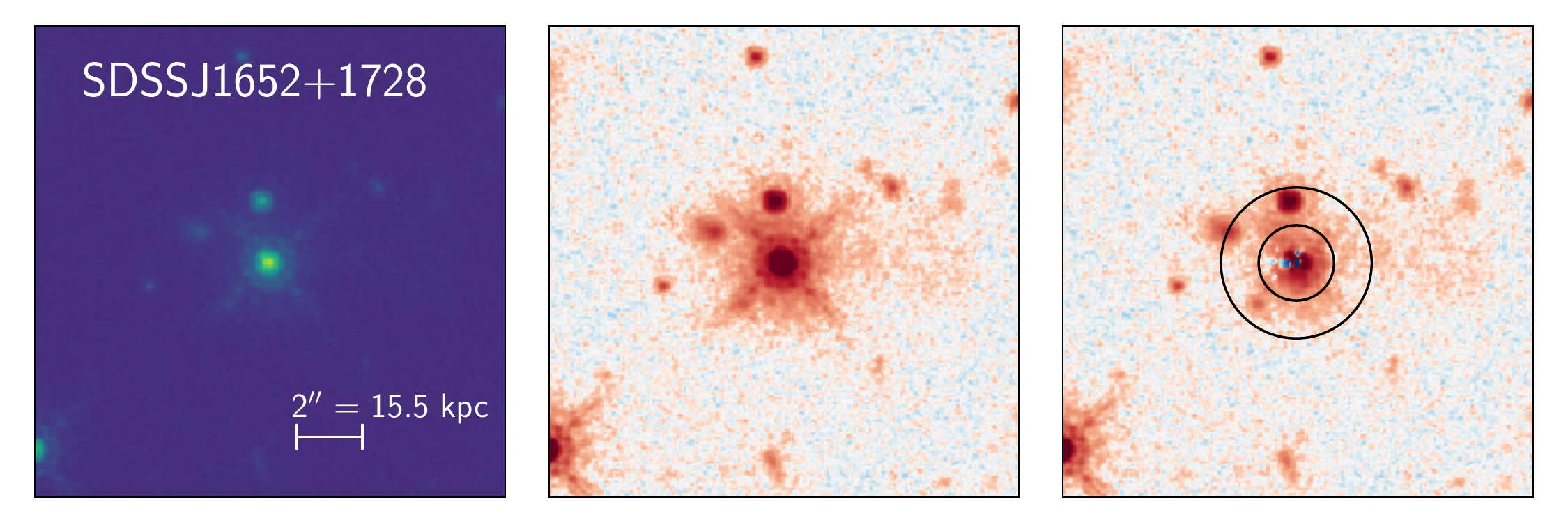}
    \includegraphics[width=3in]{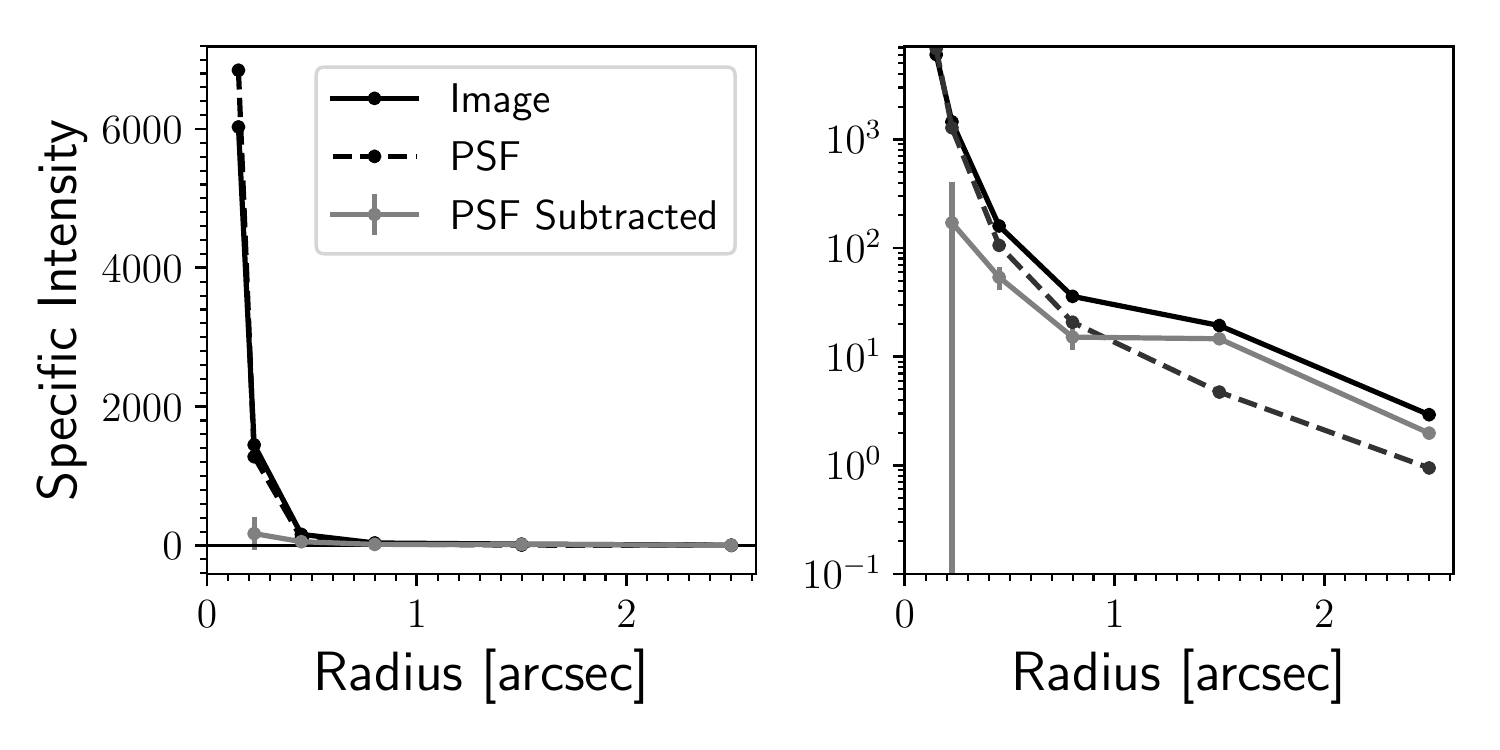}
    }
    \hbox{
    \includegraphics[width=4in]{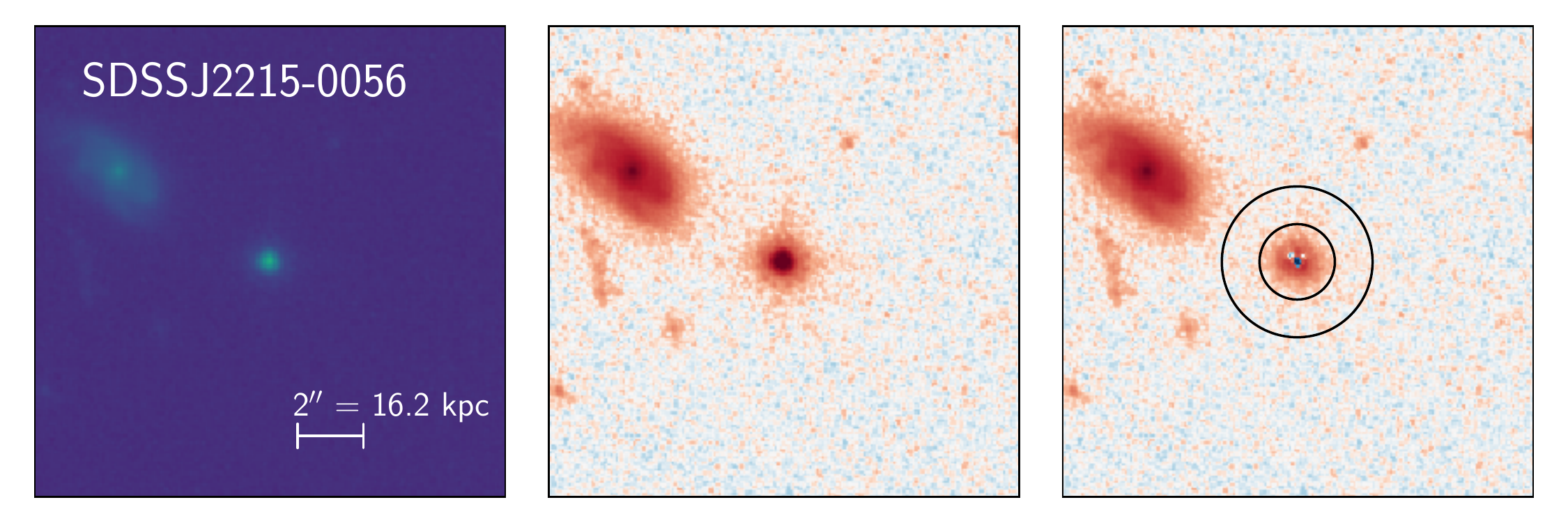}
    \includegraphics[width=3in]{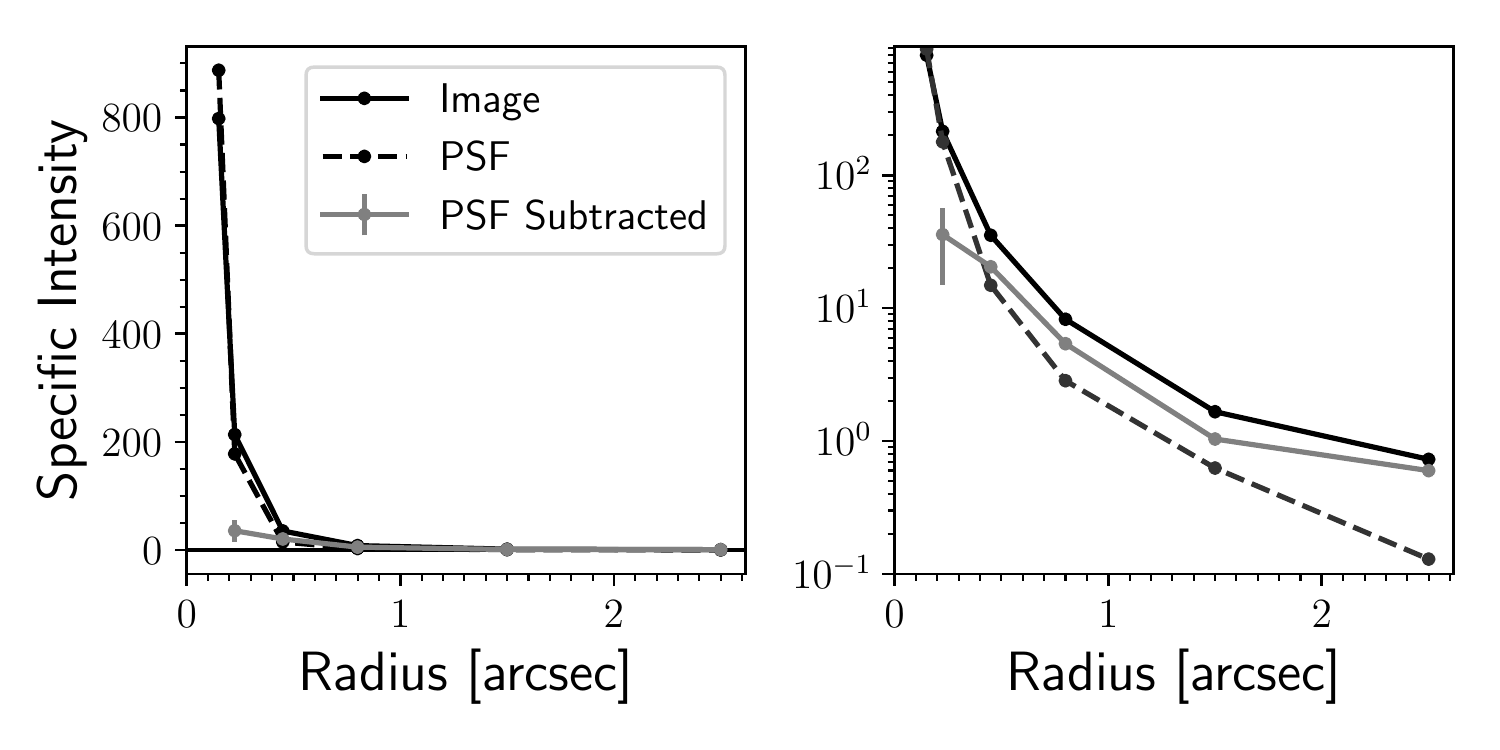}
    }
    \hbox{
    \includegraphics[width=4in]{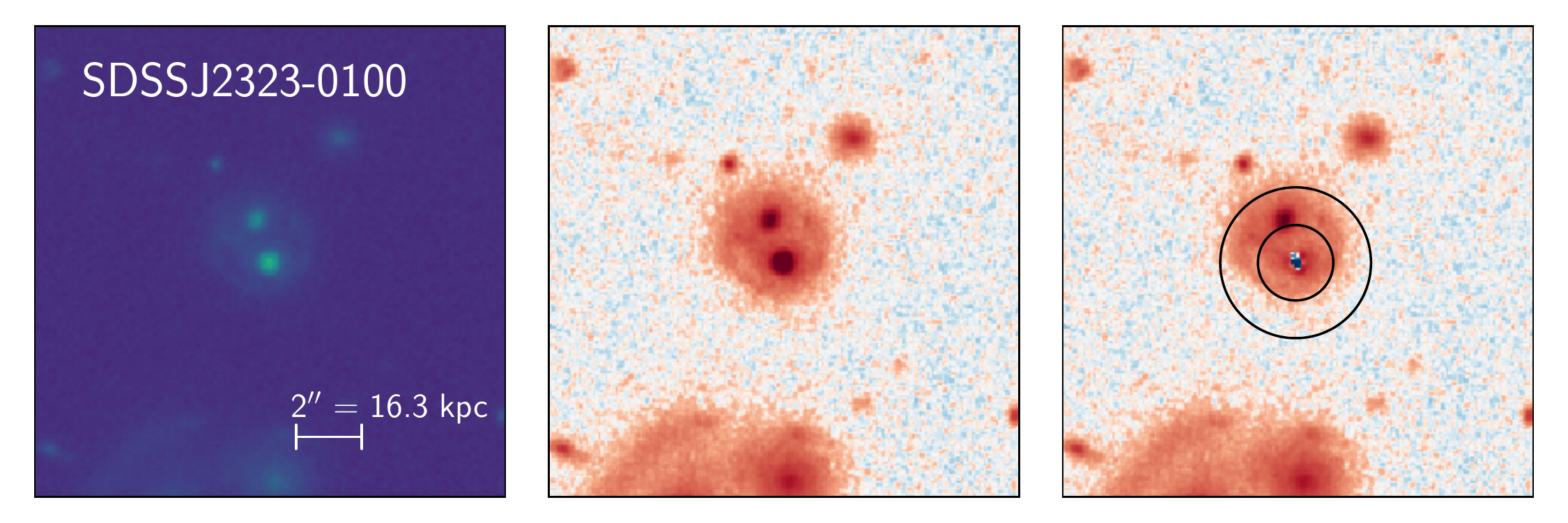}
    \includegraphics[width=3in]{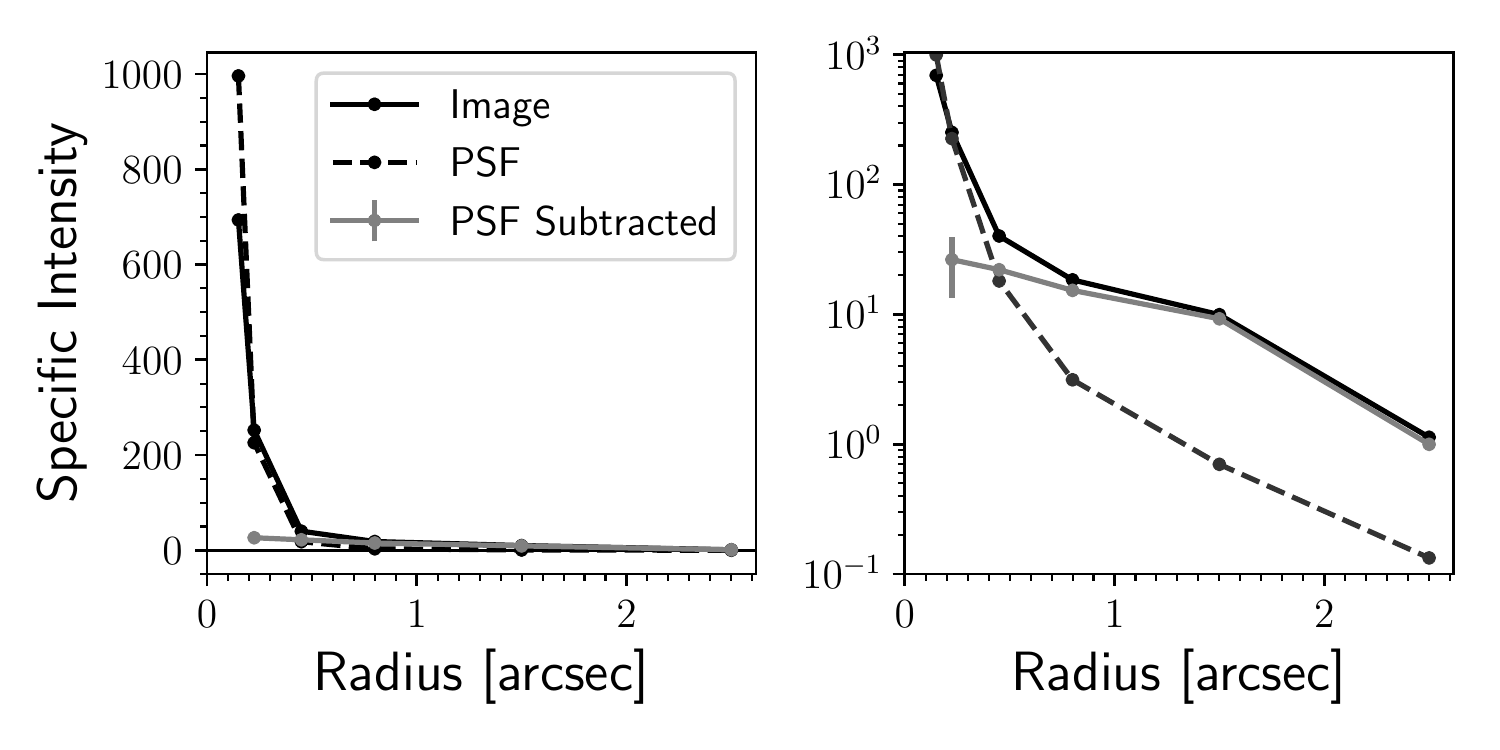}
    }
    }
    \caption{}
\end{figure*}
\renewcommand{\thefigure}{\arabic{figure}}

\begin{figure*}
    \centering
    \vbox{
    \hbox{
    \includegraphics[width=4in]{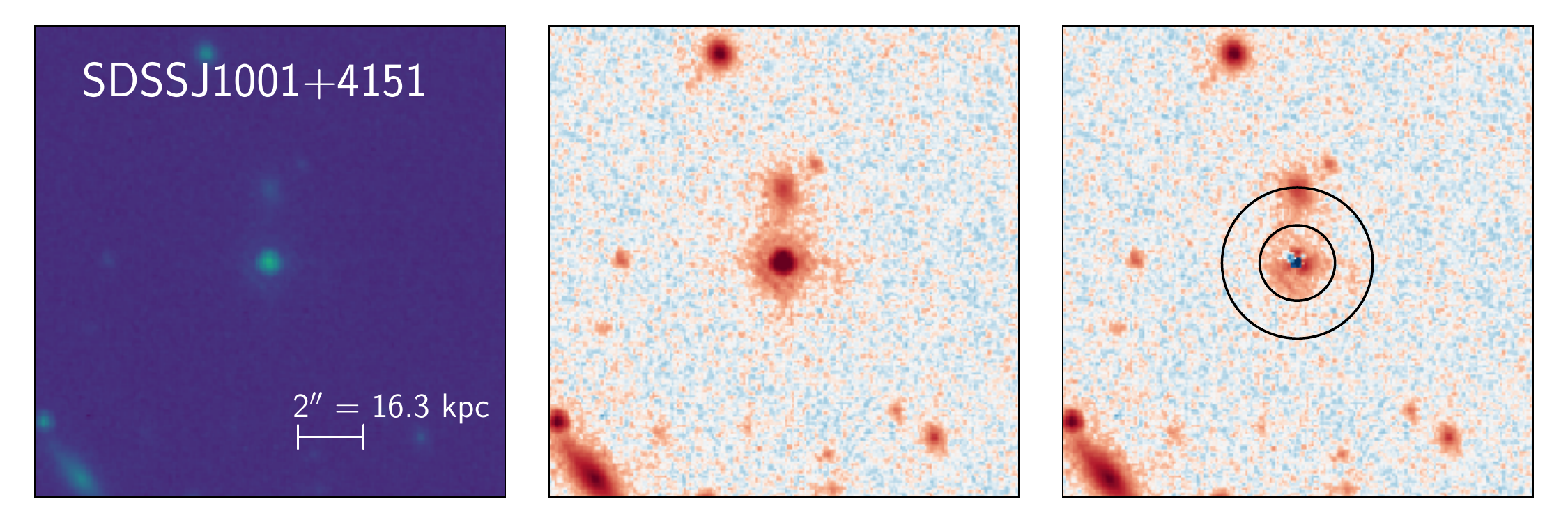}
    \includegraphics[width=3in]{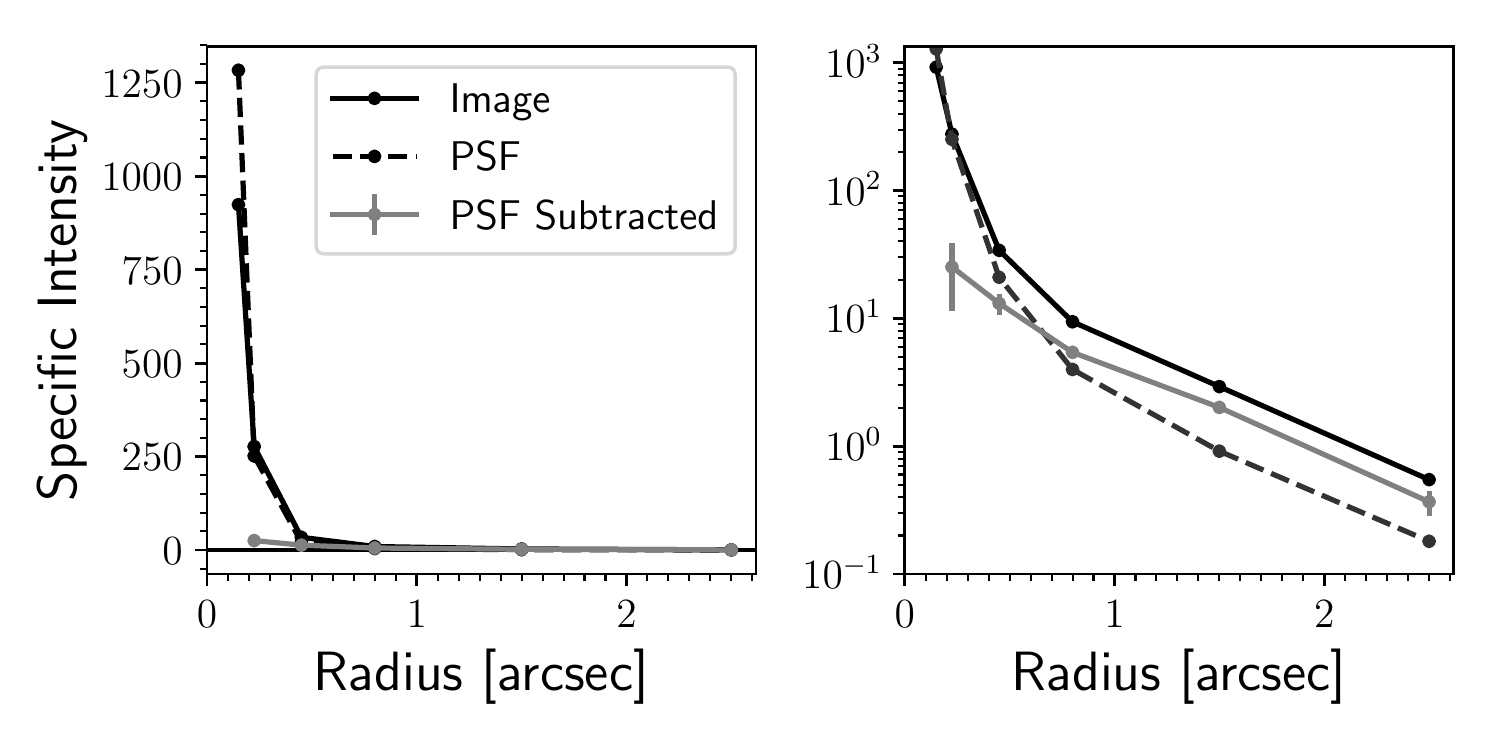}
    }
     \hbox{
    \includegraphics[width=4in]{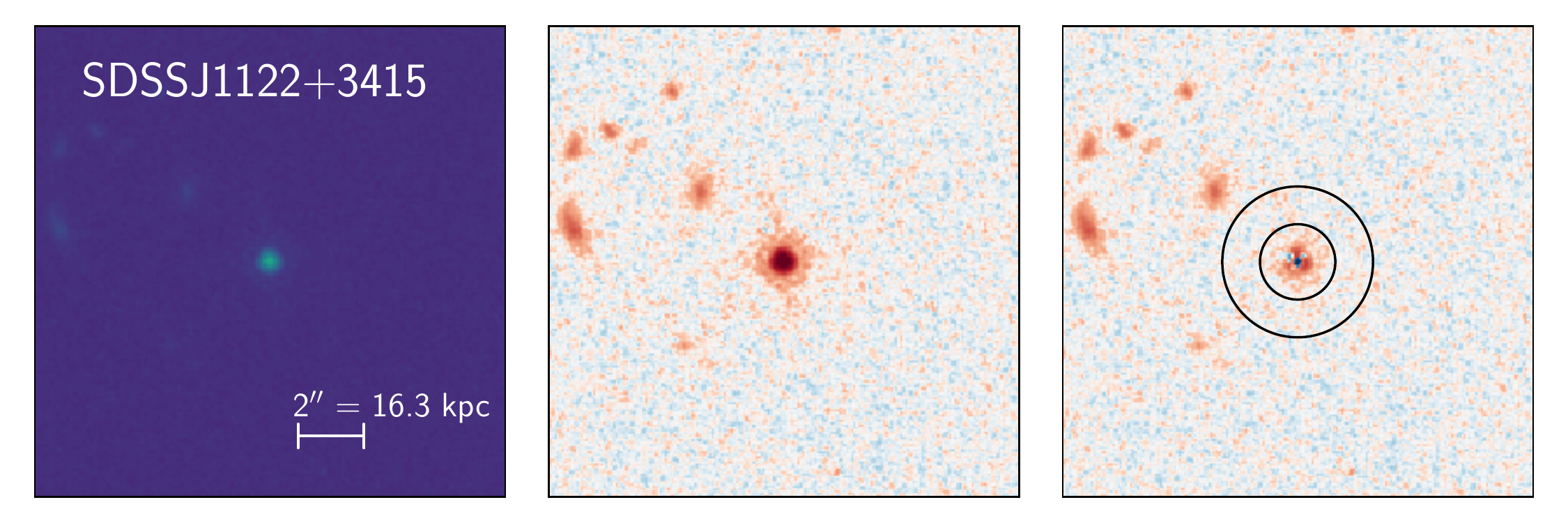}
    \includegraphics[width=3in]{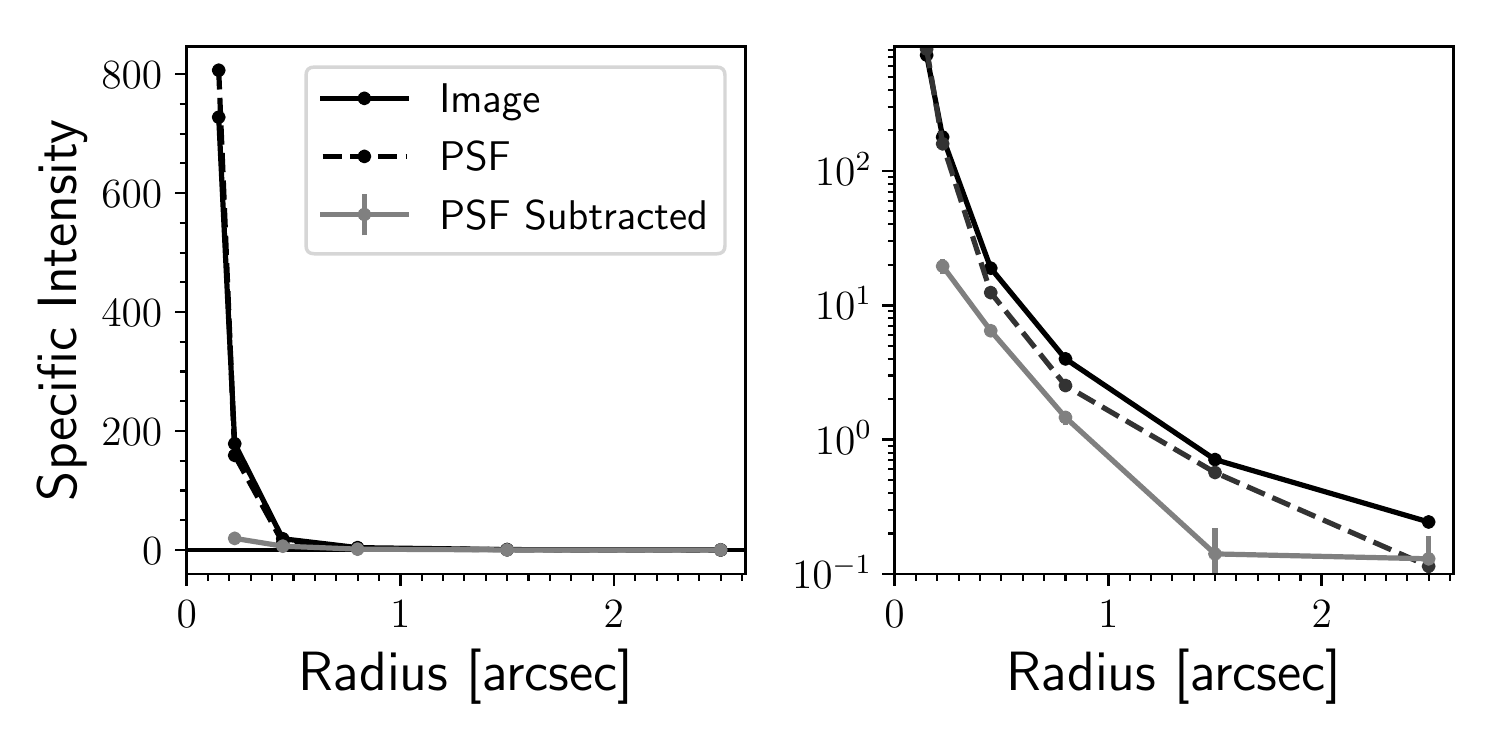}
    }
     \hbox{
    \includegraphics[width=4in]{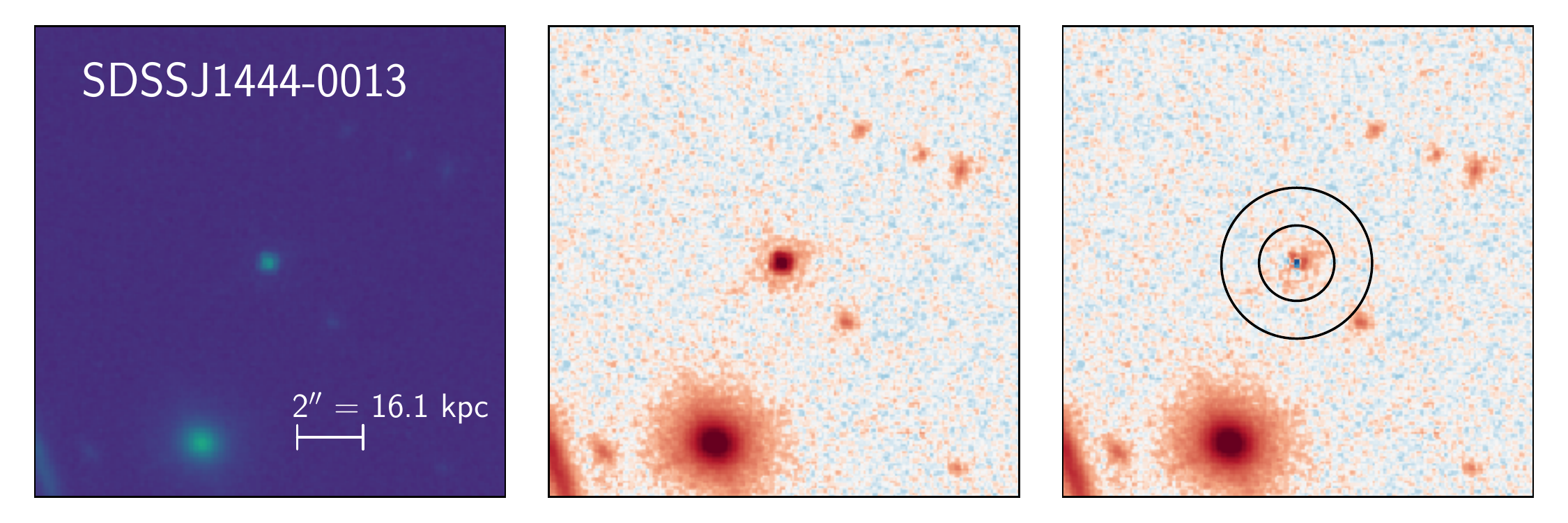}
    \includegraphics[width=3in]{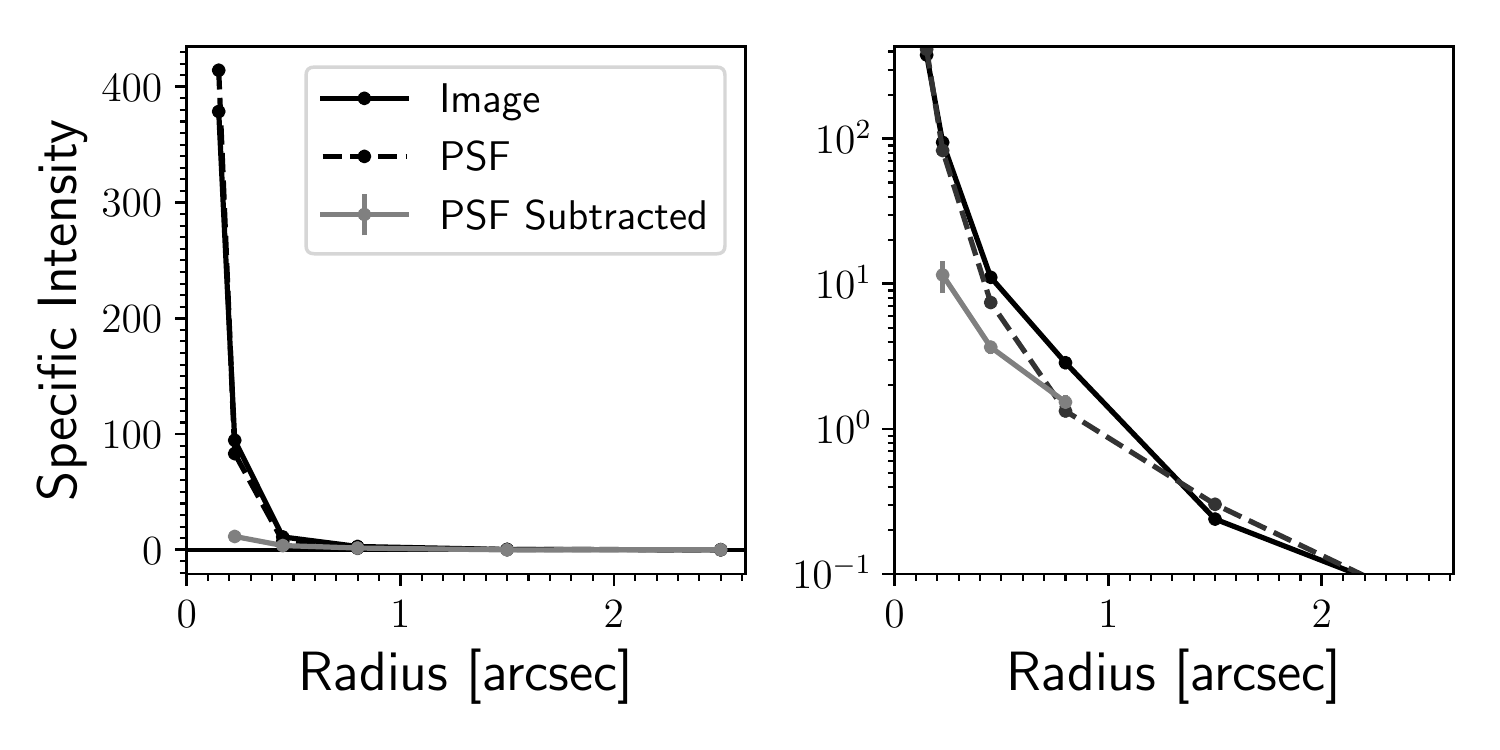}
    }
     \hbox{
    \includegraphics[width=4in]{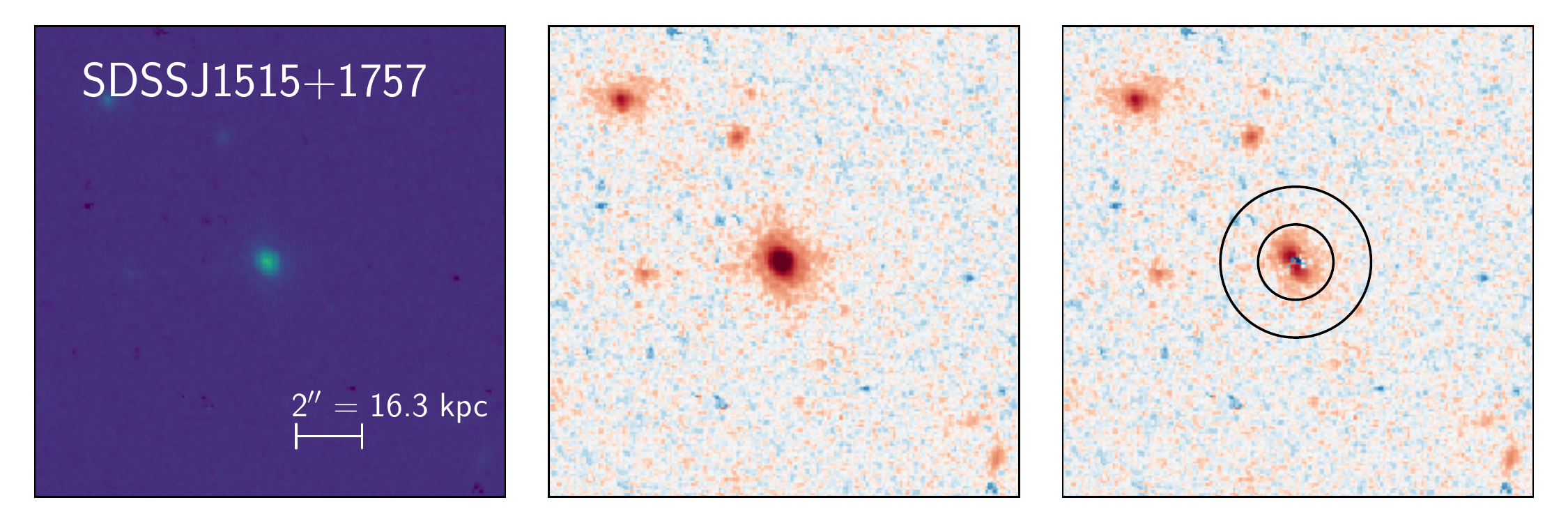}
    \includegraphics[width=3in]{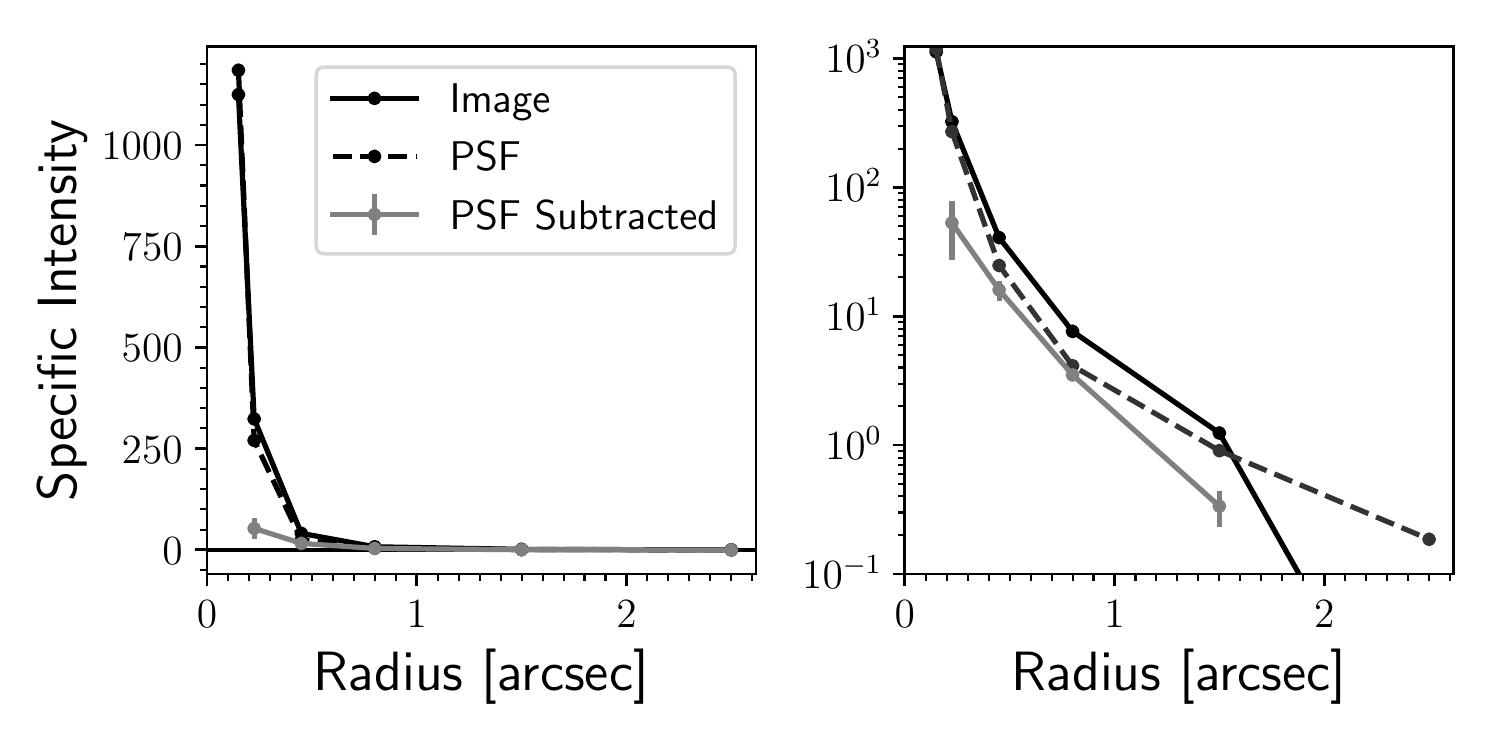}
    }
     \hbox{
    \includegraphics[width=4in]{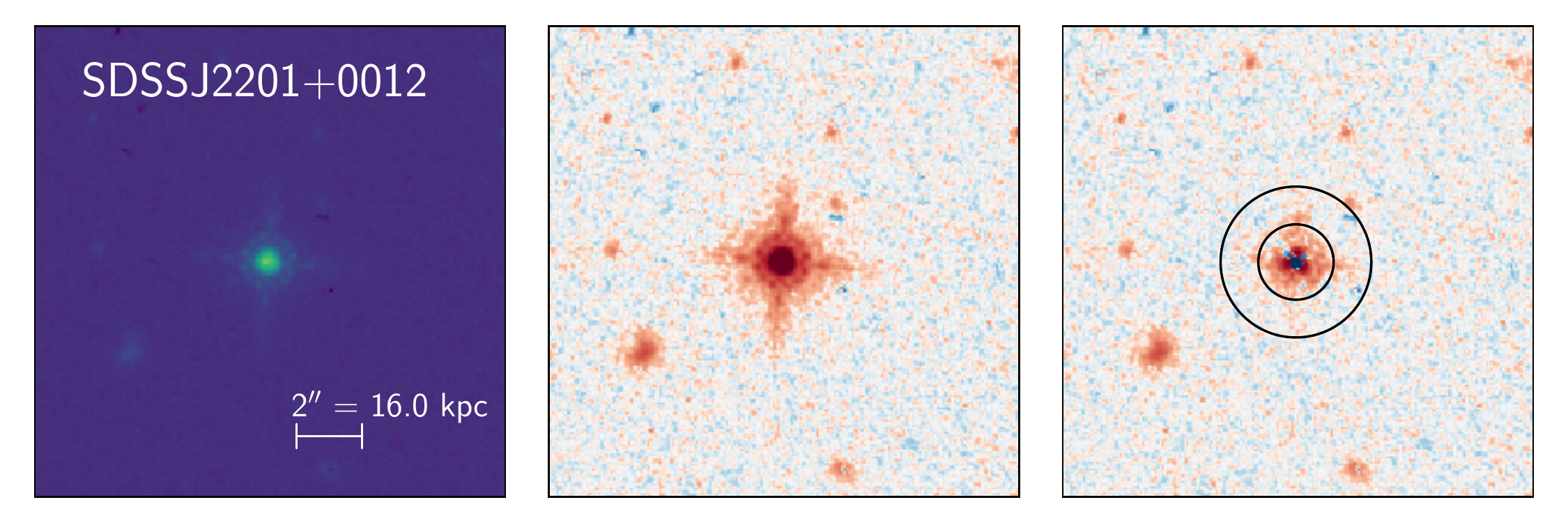}
    \includegraphics[width=3in]{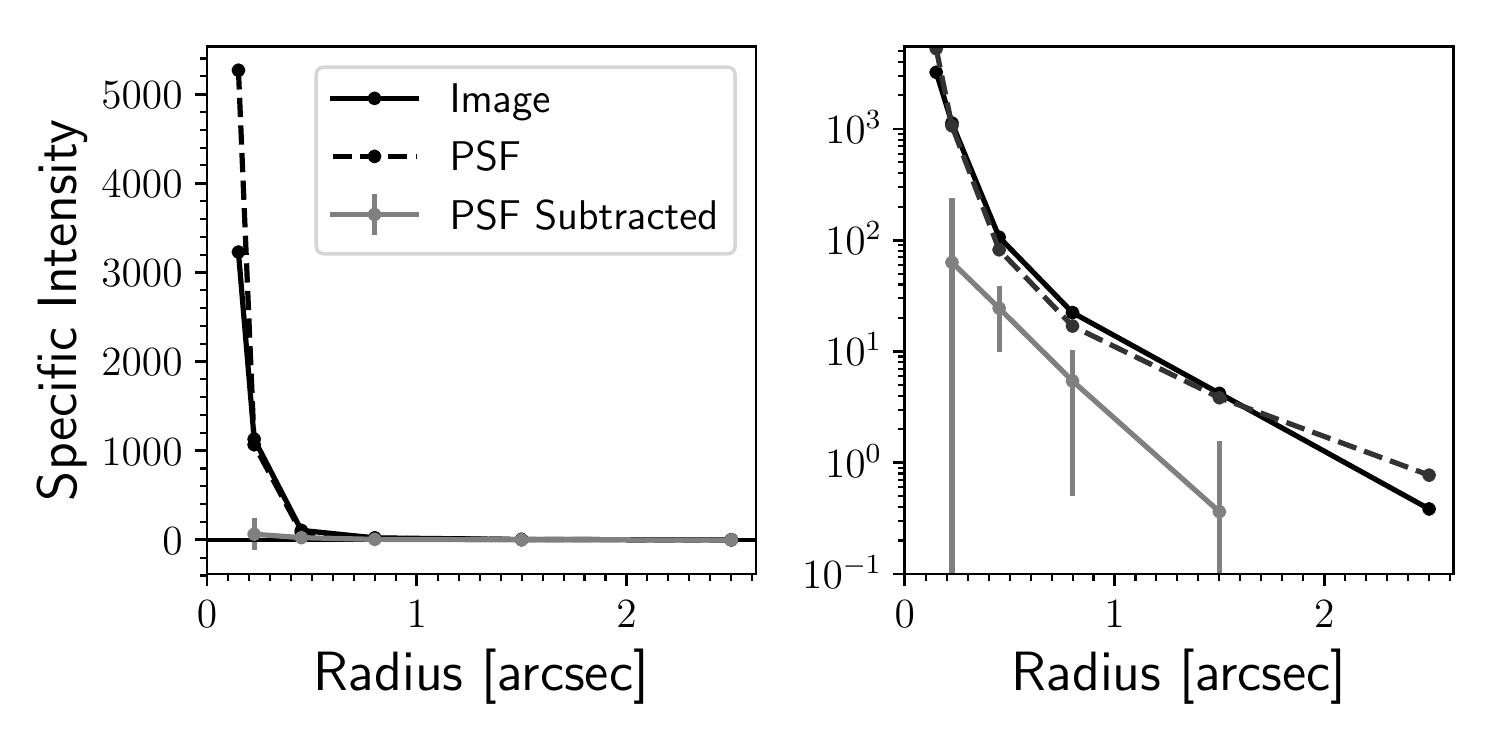}
    }
     \hbox{
    \includegraphics[width=4in]{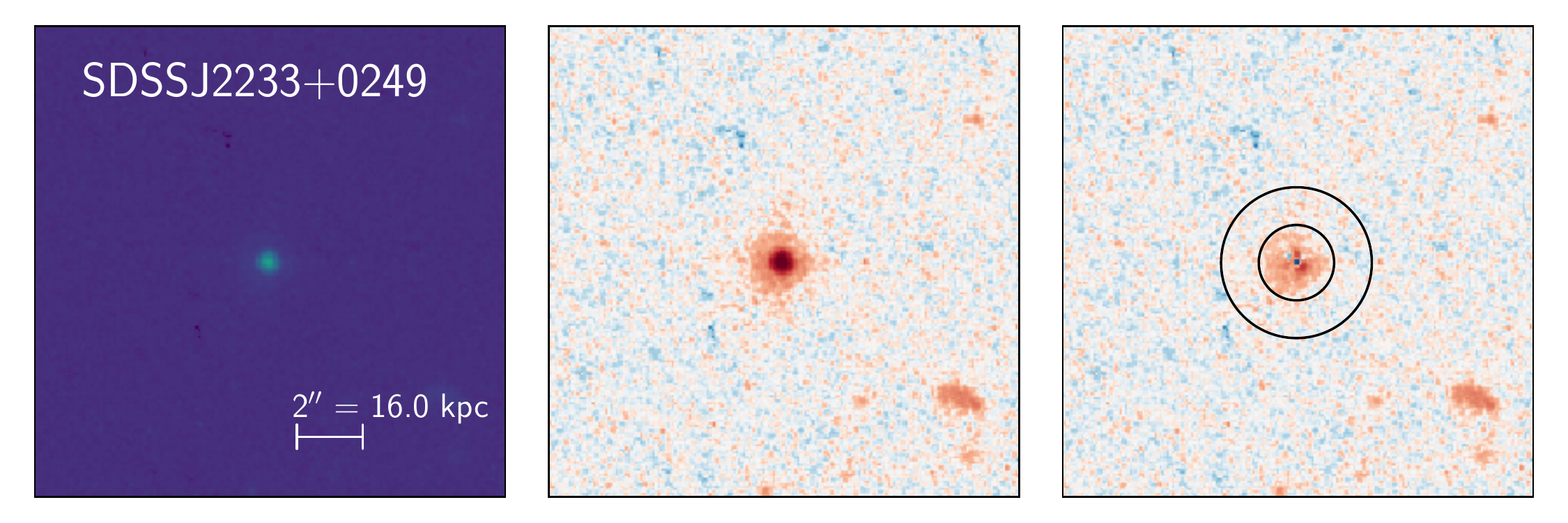}
    \includegraphics[width=3in]{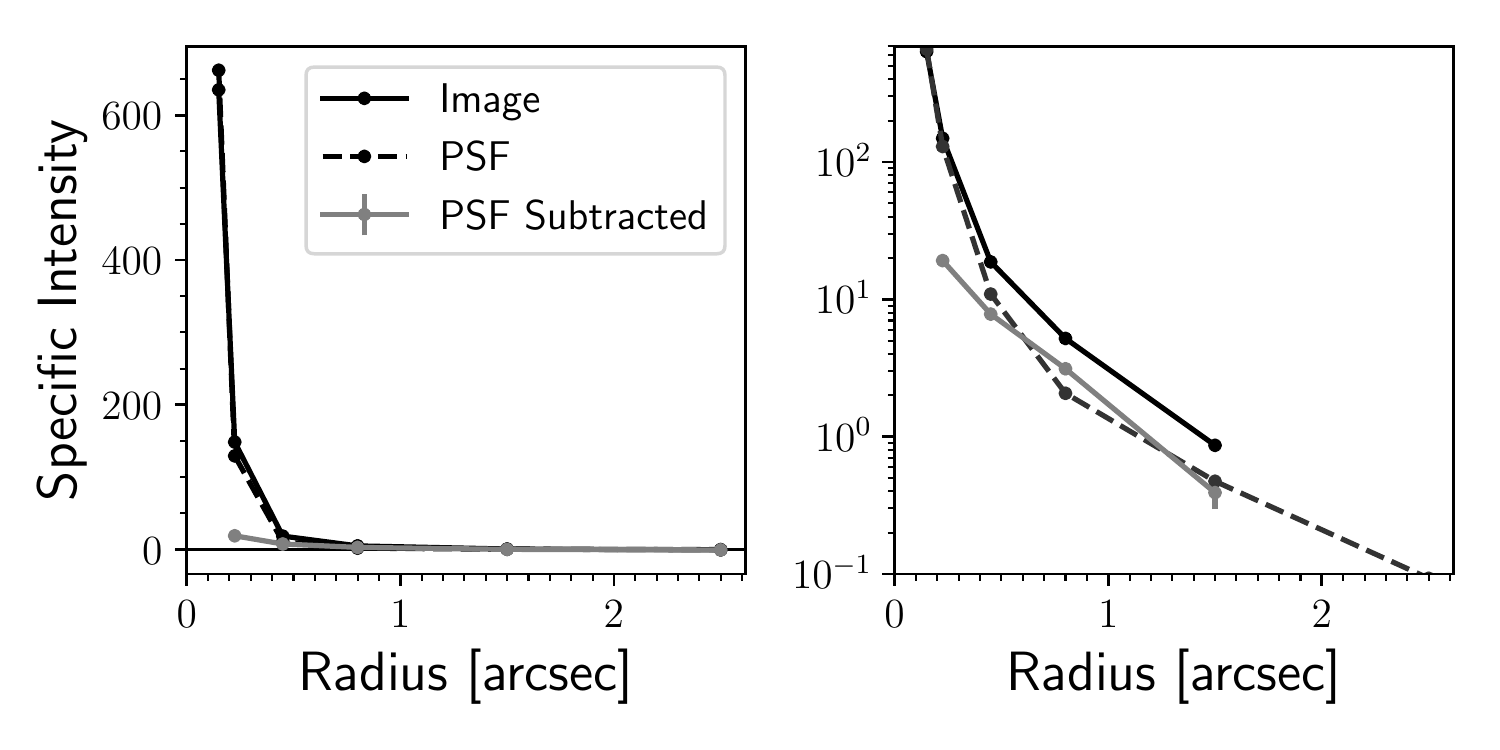}
    }
    }
    \caption{Same as Figure \ref{pic:wfc3:erq}, but for the six type 2 quasar candidates.}
    \label{pic:wfc3:t2c}
\end{figure*}

\subsection{Comparison of type 2 quasar candidates and ERQs}
\label{sec:host:expected}

The two subsamples of objects presented in this paper -- ERQs and type 2 quasar candidates -- were both originally selected from optical and infrared data as candidate luminous obscured quasars. Although the specific selection cuts preclude overlap between the two samples, they show many physical similarities: rest-frame UV continua are likely dominated by scattering \citep{alex18}, intrinsic luminosities are high, and rest-frame optical spectra display signs of obscuration commensurate with the selection criteria. Furthermore, both subsamples were observed using a near-identical HST setup allowing direct comparisons. There are also interesting physical differences between these subsamples: ERQs are more luminous and are ubiquitously associated with powerful [OIII] outflows than type 2 quasar candidates, although the origin of the connection between the selection criteria and the [OIII] kinematics is not yet understood \citep{perr19}. 

Although both subsamples are drawn from an obscured population, the HST images have a strong PSF contribution due to the quasars. While we do not perform SED decomposition for our targets \citep{asse15, farr17}, in Figure \ref{pic:sed} we illustrate plausible components of the SED and normalize host templates to the median host luminosity in our sample calculated in Section \ref{sec:host:mass}. Scattered quasar light is likely the source of optical (rest-frame UV) continuum emission, dominating the contribution from the host at these wavelengths. The origin of the near-infrared (rest-frame optical) continuum is less clear. Because broad H$\alpha$ is sometimes detected in these sources \citep{gree14b, zaka16b}, some direct light from the quasar reaches the observer at these wavelengths, albeit obscured by amounts ranging between $A_V<1$ to several mag. Due to the steep rise of the stellar SED at $>0.4$\micron, the contrast between the quasar and its host galaxy is most favorable at these wavelengths, which is covered by the WFC3 filters, but PSF subtraction is still required to reveal the quasars' underlying host galaxies and companions. 

\subsection{Aperture photometry}
\label{sec:host:aperture}

\begin{figure*}
\includegraphics[scale=0.85,trim=0cm 10cm 10.5cm 0cm,clip=true]{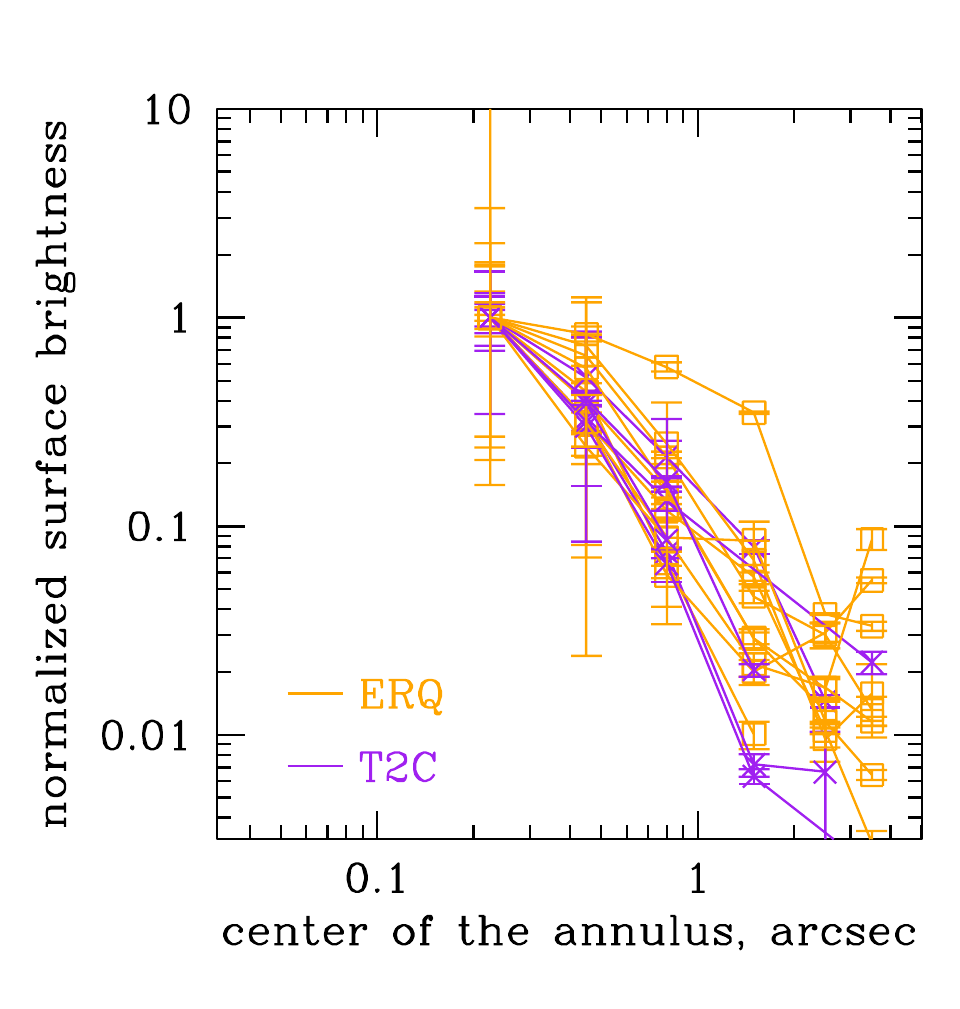}%
\includegraphics[scale=0.85,trim=10cm 10cm 0cm 0cm,clip=true]{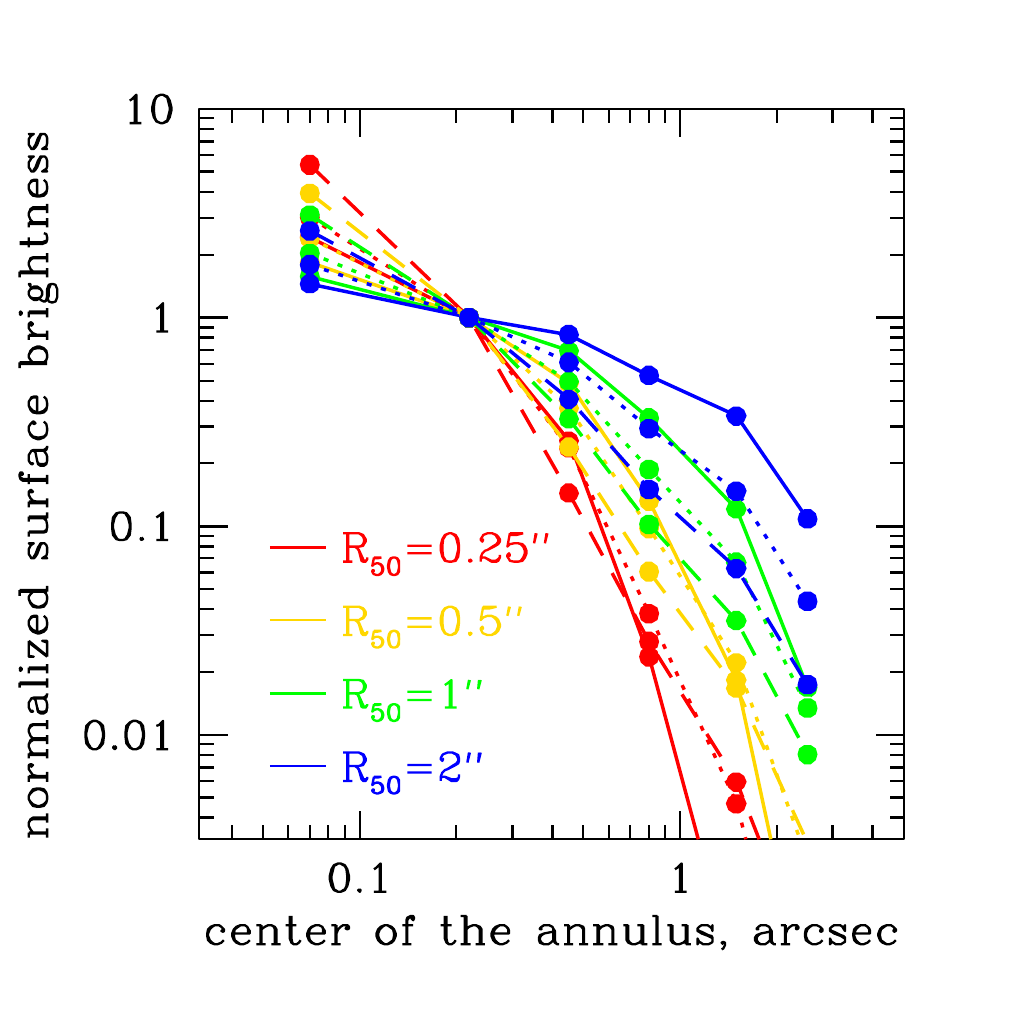}
\caption{{\it Left:} Average surface brightness in annuli for the observed host galaxies after PSF subtraction (orange for ERQs and purple for type 2 quasar candidates), normalized to the second annulus. The large error bars close to the center reflect the PSF subtraction uncertainty. The central aperture is omitted as it often suffers from large PSF residuals and slight over-subtraction. {\it Right:} Model \citet{sers68} profiles for $R_{50}=0.25$\arcsec\ (red), 0.5\arcsec\ (yellow), 1\arcsec\ (green) and 2\arcsec\ (blue) and for $n_s=1$ (solid), $n_s=2$ (dotted) and $n_s=4$ (dashed). The model profiles have been computed the same way as the observed ones -- by convolving with the typical PSF and averaging surface brightness within annuli. Qualitative comparison between the observed and the theoretical profiles shows that we expect typical $R_{50}$ values between 0.25 and 1\arcsec, with most around 0.5\arcsec. }
\label{pic:sb}
\end{figure*}

After PSF subtraction, for every quasar we calculate aperture photometry in annuli bounded by 0.15\arcsec, 0.3\arcsec, 0.6\arcsec, 1\arcsec, 2\arcsec, 3\arcsec\ and 4\arcsec\ from the center of the PSF. To this end, we sum up the flux within annuli, taking into account the fractional contribution of pixels bisected by apertures. For average surface brightness, we then divide the resulting flux by the area of the annulus. The average surface brightnesses in physical units for the 0.3\arcsec-2\arcsec annuli are given in Table \ref{tab:measure}. Surface brightness profiles normalized to the value in the second innermost annulus are shown in Figure \ref{pic:sb}, left. As the PSF-fitting is performed before the aperture photometry, the results of the fits are independent of the aperture choice. 

Next, we use the aperture fluxes of the host galaxies from the PSF-subtracted images to compute the host luminosities. In Figure \ref{pic:sb} it is apparent that the surface brightness profiles, over the range where they are well measured, are close enough to $R^{-2}$ that we need to worry about the convergence of the total flux $\int_0^{\infty} 2 \pi R \langle I(R) \rangle {\rm d}R $ both on the low $R$ and on the high $R$ end. Therefore we cannot extrapolate the observed profiles either close to the nucleus or far away from the nucleus where the profiles are not well measured. 

To obtain an estimate of the host-galaxy flux, we follow the ``minimal flux" approach. We take the PSF-subtracted profile and assume that the surface brightness in the central aperture is the same as in the second aperture (0.15$-$0.3\arcsec). For all realistic galaxy profiles the former should be larger than the latter, so by making this assumption we are underestimating the total galaxy flux. Furthermore, the second aperture surface brightness may in turn be under-estimated because of the over-subtraction of the PSF. Then we use the observed data points and integrate the surface brightness profile out to the largest available aperture of 3\arcsec, but no further, again potentially underestimating the total flux. 

The results are presented in Table \ref{tab:measure} in the form of $\nu L_{\nu}$ at the effective rest-wavelength of our WFC3 observations. Furthermore, we bring all luminosity measurements to the same rest-frame Johnson $B$-band\footnote{We use a generic Johnson $B$-band filter curve from the Filter Profile Service at the Spanish Virtual Observatory \url{http://svo.cab.inta-csic.es/}} by applying k-corrections using the Sb templates from \citet{poll07} shown in Figure \ref{pic:sed} and hereafter report $L_B\equiv \nu L_{\nu}$ at rest-frame 4400\AA. In the wavelength range 0.4$-$0.7\micron, Sb and Ell templates produce similar results, whereas the M82 template, which has more dust obscuration, is appreciably redder in the optical. Because our observations are close to rest-frame $B$-band, k-corrections are in general smaller than other uncertainties -- out of the objects in our sample, two thirds have k-corrections under 0.1 dex and the remaining third under 0.2 dex. The median measured value is $L_B=10^{11.1}L_{\odot}$, and the mean and the sample standard deviation of $\log (L_B/L_{\odot})$ are $11.1\pm 0.4$.

Finally, we can estimate the effective sizes and measures of concentration by comparing the observed profiles with Sersic profiles. A circularly symmetric Sersic profile is a three-parameter function, with surface brightness given by
\begin{equation}
\log I(R)=a_0-(R/b)^{1/n_s},
\label{eq_sersic}
\end{equation}
where $a_0$ is the overall normalization and $n_s$ is the Sersic index: $n_s=1$ for an exponential profile \citep{patt40} and $n_s=4$ for a \citet{deva53} profile. 

The parameter $b$ is proportional to the physical scale of the galaxy, which historically is defined differently for different Sersic indices. One physically meaningful non-parametric value is effective radius $R_{50}$ -- the radius containing half of the total galaxy light. The relation between $R_{50}$ and $b$ can be calculated numerically for every $n_s$, and in the range $n_s=0.5-4$ it is well approximated by a third-order polynomial:
\begin{equation}
\log(R_{50}/b)=-0.2963-0.05122n_s+0.2376n_s^2-0.01896n_s^3.
\end{equation}
For $n_s=4$ (de Vaucouleurs), this equation yields $(R_{50}/b)^{1/4}=3.33$, which corresponds to the classical de Vaucouleurs' parametrization. 

Thus, a circularly symmetric Sersic profile can be completely described by the normalization, index $n_s$ and half-light size $R_{50}$, whereas flux-normalized Sersic profiles form a two-parameter family. In the right panel of Figure \ref{pic:sb} we show model surface brightness distributions obtained from Sersic profiles using the same method as in the observations: i.e., convolving with the PSF, integrating within annuli and then dividing by their areas. We probe a plausible range of Sersic indices ($n_s=1,2,4$) and effective radii $R_{50}$ of 0.25, 0.5, 1.0 and 2.0\arcsec. Qualitatively, most of the observed surface brightness profiles appear to be on scales similar to those of Sersic profiles with $R_{50}=0.25-1\arcsec\simeq 2.1-8.4$ kpc, which is unsurprising considering that that most massive galaxies at $z>2$ have $R_{50}<3$ kpc \citep{vand10}. 

\subsection{Sersic fits and structural parameters}
\label{sec:host:size}

To obtain another measure of the host flux and determine the structural parameters of the host galaxies in our sample, we perform a joint PSF + two-dimensional Sersic fit, which should alleviate the problem of over-subtraction in our previous PSF-only fits. There are many challenges to this analysis. First, as the hosts may be compact, some of their light can be hidden within the central PSF, so we expect some degeneracy between the contributions of the host galaxy and of the quasar. This is a generic difficulty of modeling quasar hosts in imaging data, including at low redshifts \citep{zaka06}. Second, as discussed in Section \ref{sec:data:psf}, the accuracy with which the PSF can be modelled is $\sim 5\%$ in the center and $\sim 10\%$ in the outer parts, limiting the quality of the fits for our sources, as the PSF is larger than the flux of the host galaxy by up to an order of magnitude (Figure \ref{pic:sed}). 

While fitting the \texttt{flt} images, we keep all the parameters of the PSF and Sersic models free. For the PSF model the parameters include its position, amplitude, a Gaussian smoothing width to simulate jitter, and a background level. The two-dimensional Sersic model which represents the host is described by 7 parameters: coordinates of the central position, Sersic index, normalization, half-light radius $R_{50}$, ellipticity and position angle. The equations connecting these parameters to total fluxes and other standard parametrizations of the Sersic profiles are listed in Section \ref{sec:host:aperture} and in \citet{zaka06}. To avoid over-fitting, the Sersic indices are constrained to the range $0-4$ and the ellipticities to the range $0-0.4$. During the fitting, the two-dimensional Sersic model is convolved with a two-dimensional Gaussian best fit to the PSF model. 

\begin{figure*}
    \centering
    \includegraphics[width=7.0in]{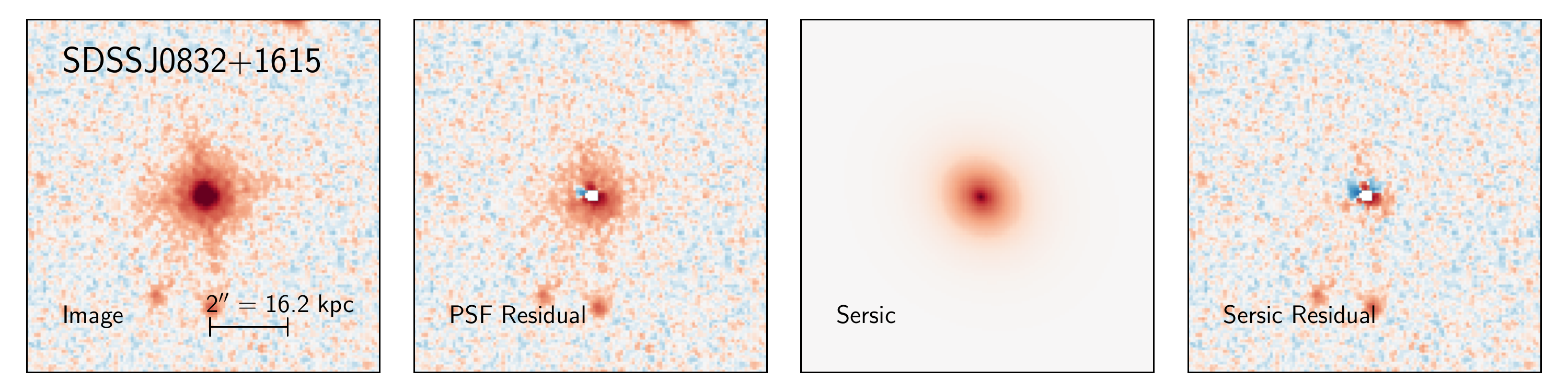}
    \includegraphics[width=7.0in]{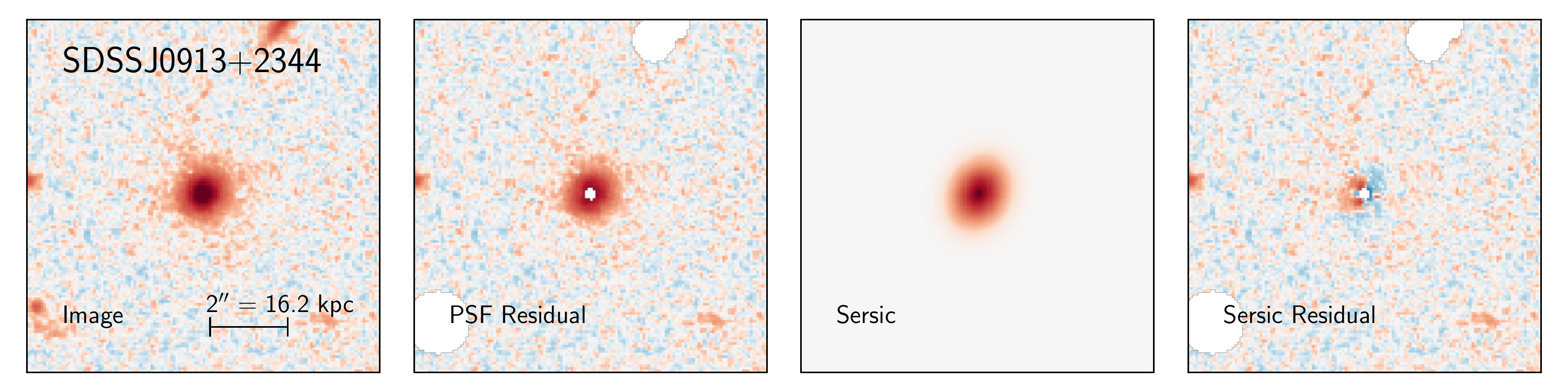}
    \includegraphics[width=7.0in]{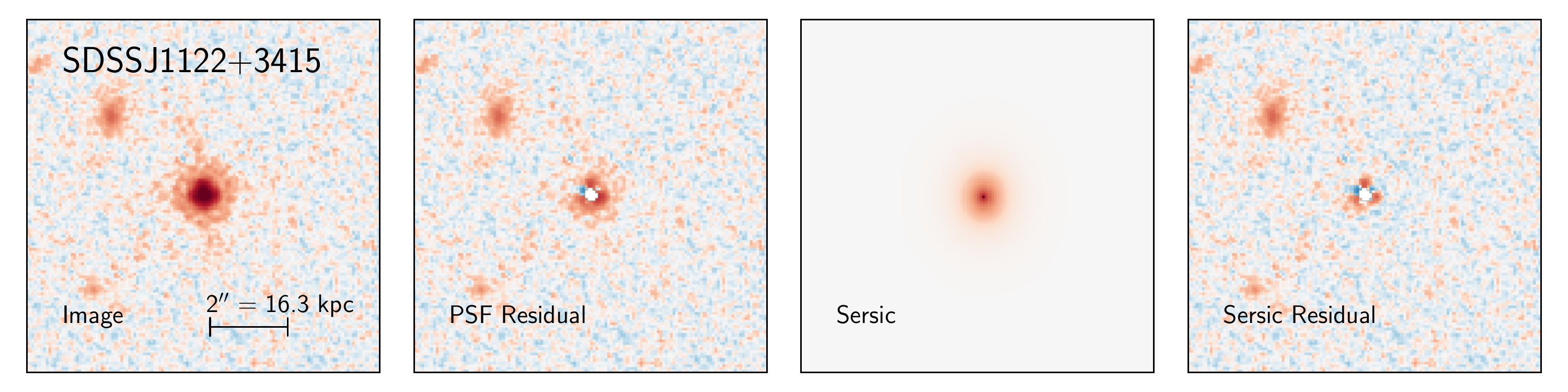}
    \includegraphics[width=7.0in]{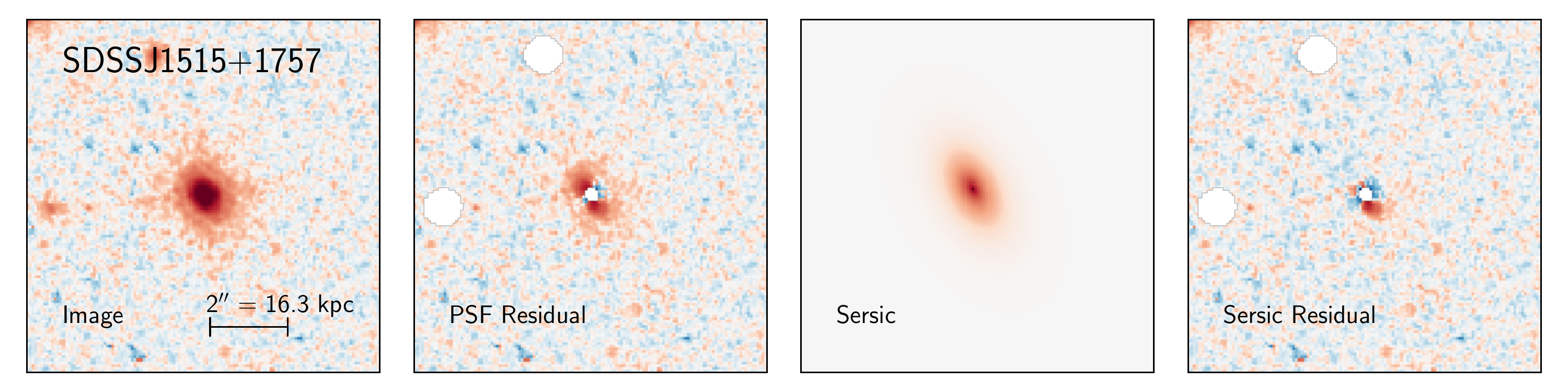}
    \caption{Example joint PSF+Sersic fits of the WFC3 F140W/F160W data: two ERQs on top and two type 2 quasar candidates on the bottom. From left to right: the original image, the image with the best-fit PSF from the joint fit subtracted (pixels masked in fitting are indicated), the best-fit Sersic model, residuals after subtraction of both the PSF and the Sersic component. North is up and East is to the left.}
    \label{pic:sersic}
\end{figure*}

The best-fit PSF models of the \texttt{flt} images are drizzle-combined to form the PSF model of the \texttt{drz} image. This PSF model is then fitted to the \texttt{drz} image with the amplitude as the only free parameter, together with a free Sersic model. The central residuals can be quite large, higher than the galaxy signals, due to both the Poisson noise and the mismatch between the PSF template and the data, so they contribute overwhelmingly to the $\chi^2$. To avoid placing all statistical weight of the fits on the region where the host signal contributes the least, the central pixels that are brighter than 20\% of the peak intensity are masked at this stage. Companions, foreground stars, and galaxies are masked so that they do not affect the orientation and the ellipticity of the Sersic profile. Because of the higher signal-to-noise ratio and the better resolution of the \texttt{drz}-based images, we use \texttt{drz}-based fits with these masks in what follows. Some of these fits are shown in Figure \ref{pic:sersic}, and the fitting parameters are presented in Table \ref{tab:sersic}.

In most cases the joint PSF+Sersic model fits the data reasonably well, leaving only small residuals that may be due to the asymmetry of the host galaxy. There are a few cases where the diffraction spikes of the PSF are not entirely subtracted by the best-fit model, including J0834+0159, J1013+3427, J1217+0234, J1652+1728, and J2201+0012. If we increase the amplitude of the PSF until the spikes are subtracted, the core of the PSF model becomes brighter than the original image. These few targets also have the highest PSF intensity among the sample, suggesting that the core of these images have a nonlinear or saturated response. Residual diffraction spikes and rings would increase the fluxes on the outskirts of the galaxy, thus lowering the measured Sersic index, increasing the radius and leading to additional uncertainties in the flux.

There are multiple sources of systematic uncertainties in our fitting results, including imperfections of the PSF model (e.g., possible saturation issues) and deviations between the real surface brightness profiles and the model assumptions -- for example, that the host galaxy is well represented by a Sersic profile. Such systematic uncertainties dominate the statistical error. We test the sensitivity of the outputs to the assumptions about the PSF in a variety of ways by modifying the fitting procedure or the masking prescriptions. For example, we find that \texttt{flt}-based fluxes are $\sim 0.2$ dex brighter than the \texttt{drz}-based ones, which is likely due to the lower effective resolution of the \texttt{flt}-based images and the resulting higher degeneracy between the nucleus and the host. The standard deviation of the difference between \texttt{flt}-based and \texttt{drz}-based fluxes is 0.25 dex, whereas the variations among the fitting procedures we have tried result in standard deviations $\la 0.2$ dex. We adopt the \texttt{drz}-based fluxes, and use 0.25 dex as a conservative estimate of the Sersic flux error. In Table \ref{tab:sersic} we list both the measured monochromatic luminosities and those k-corrected to the rest-frame $B$-band, finding for the entire sample $\log (L_B/L_{\odot})=11.0,11.0\pm 0.5$. 

Furthermore, we find that the aperture fluxes and the Sersic fluxes show no significant offset relative to one another, with aperture-based fluxes being are on average higher by only 0.1 dex than the Sersic-based fluxes. The standard deviation of the differences between the two sets of fluxes is 0.18 dex, which is within our estimated error on the Sersic fluxes. The two most discrepant fluxes are found in hosts with companions, which are naturally not well modeled by a single Sersic profile. Excluding the five objects with companions named in Sec. \ref{sec:merger:major} reduces the standard deviation of the difference to 0.14 dex. This comparison of fluxes calculated by two completely different methods suggests that despite the challenging PSF subtraction, the extended host flux measurements are quite robust. However, if the host galaxy has a compact circumnuclear starburst, it would be absorbed into our PSF model and not reported as part of the host flux with either the aperture or Sersic method, and we cannot make a realistic estimate of uncertainty associated with this possibility. Emerging integral field unit spectroscopic studies of quasar hosts may be able to circumvent this problem by taking advantage of spectral differences between the quasar and its host \citep{rupk17}.

\begin{figure*}
\includegraphics[scale=0.9,trim=0cm 13.3cm 0cm 0cm,clip=true]{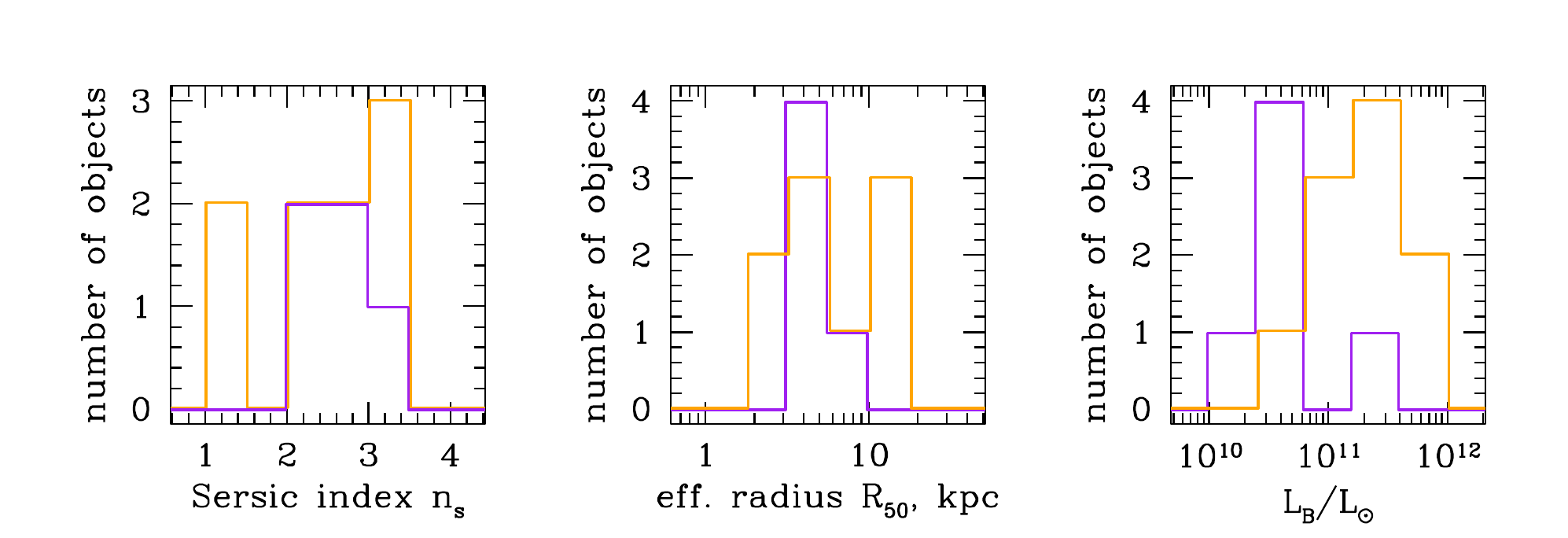}
\caption{Distributions of Sersic indices, effective radii $R_{50}$ and monochromatic luminosities for host galaxies of ERQs (orange) and of type 2 quasar candidates (purple). In the left and in the middle panel, the distributions of structural parameters for the two subsamples are consistent with being drawn from the same population. In the right panel, they are not consistent at $>$95\% confidence level, as measured using the Kolmogorov-Smirnov test.}
\label{pic:structure}
\end{figure*}

We display the distributions of our best-fit Sersic indices $n_s$ and half-light radii $R_{50}$ in Figure \ref{pic:structure} and list them in Table \ref{tab:sersic}. We expect significantly higher relative uncertainties in the structural parameters than in the flux because $n_s$ and $R_{50}$ are somewhat degenerate, with both high $n_s$ and small $R_{50}$ resulting in a compact light distribution. To quantify this uncertainty, we simulate two-dimensional Sersic profiles with a fixed total flux outside of the inner 0.15\arcsec\ (mimicking our PSF contamination). We then find a family of Sersic models with surface brightness at 0.5\arcsec, 1\arcsec\ and 2\arcsec\ within 10\% of the fiducial model with $n_s=2.5$ and $R_{50}=4.5$ kpc. Within this family, whose surface brightness profiles would look essentially indistinguishable to an observer, $n_s$ ranges between 1.8 and 4 -- i.e., it is very poorly constrained. The range of $R_{50}$ is narrower, between 3.5 kpc and 5 kpc (relative error of $\sim 0.1$ dex), due to its stronger sensitivity to the flux constraints outside 0.15\arcsec.

An additional source of error in structural parameters is due to the fitting itself, which we estimate by comparing results from different fitting procedures (e.g., with and without masks), from different \texttt{flt} images, and from \texttt{drz} vs \texttt{flt} images. The resulting fitting uncertainty for each source (estimated as the sample standard deviation from a range of fits) is typically 0.5$-$0.6 for $n_S$ and 0.18 dex for $R_{50}$. As with fluxes, if there is a bright centrally concentrated component (high $n_s$ or small $R_{50}$), the contribution from the central part of the galaxy is degenerate with the quasar PSF, so all information about such component is lost due to the presence of the quasar, and the systematic uncertainties due to this effect cannot be estimated. Given all these considerations, we adopt a nominal error of 1.0 on each $n_s$ measurement and 0.2 dex on each $R_{50}$ measurement except for the 5 most compact objects ($R_{50}\la 4.5$ kpc) where the uncertainty is 0.3 dex. 

A Kolmogorov-Smirnov test shows that the distributions of these structural measurements for ERQ and type 2 quasar candidate hosts are consistent with being drawn from the same population. The fit values of $R_{50}=4.9,4.8\pm 1.3$ kpc or $0.6,0.6\pm 0.15\arcsec$ (median, mean and sample standard deviation) are qualitatively consistent with our expectations from Figure \ref{pic:sb}. When compared to local galaxies, the Sersic indices $n_s=2.7,2.6\pm 0.6$ are intermediate between disk-like and bulge-like, but when compared to distant massive galaxies, these values are typical \citep{vand10}. However, while our estimated structural parameters appear to be similar to those of high-redshift massive galaxies, the uncertainties on these values are too large at this point to warrant a more detailed comparison with the high-redshift populations studied with much deeper data and without the difficulties of the central source contamination. Furthermore, because the individual errors on $n_s$ and $R_{50}$ are comparable to the standard deviation within the population, the measurements are not yet sensitive to any differences between the structural parameters of ERQ and type 2 quasar candidate hosts.

The three objects with measured effective radii in excess of 10 kpc are SDSS~J1217+0234, SDSS~J1356+0730 and SDSS~J2323$-$0100, all of them ERQs. SDSS~J2323$-$0100 is an ongoing merger, where a single-Sersic fit is inadequate. In the other two objects the host galaxy is clearly detected in the PSF-subtracted images, but is not so extended as to warrant such large effective radii. It is likely that either a Sersic profile is an inadequate representation of the galaxy, or the PSF model is not very accurate, and we consider these joint fits to be unsuccessful and omit their structural parameters from Table \ref{tab:sersic}.

\subsection{Host luminosity and stellar mass}
\label{sec:host:mass}

In what follows we use $L_B$ luminosities from Sersic fits, but none of the results are sensitive to which set of luminosities we choose to use. The median luminosity of all host galaxies in our sample is $L_B=10^{11.0}L_{\odot}$. The mass-to-light ratio of stellar populations in the $B$-band is a function of metallicity, age and star formation history. For stellar populations with age $0.3-2$ Gyr (the upper limit being the maximum age at the redshift of our sample) and solar metallicity $Z_{\odot}$ characteristic of our targets \citep{hama17}, the mass-to-light ratio ranges between 0.4 and 4 $M_{\odot}/L_{\odot}$ \citep{mara05}, where we have corrected from the $B$-band luminosity used by \citet{mara05} to the bolometric luminosity $L_{\odot}$ used as a luminosity unit in this paper. Therefore, the median corresponding stellar mass is $10^{10.4-11.4}M_{\odot}$, depending on the age of the galaxies in our sample. A higher mass-to-light ratio is possible if the galaxies are very dusty, and an 0.3 dex lower mass-to-light ratio would be expected for $Z_{\odot}/20$ -- e.g., if the host galaxy has a much lower metallicity than the nucleus \citep{mara05}. 

The measured galaxy fluxes imply that the extended narrow-line region contamination to our imaging (Section \ref{sec:data:hst}) is minor. The total [OIII] luminosity rarely exceeds $10^{10} L_{\odot}$ \citep{gree14b,zaka16b}, which is less than 10\% of the host luminosity falling within the broad filter. Furthermore, only a small fraction of the line luminosity is expected on scales of a few kpc, as its spatially unresolved, centrally-concentrated component is subtracted using our PSF subtraction procedure. Thus the extended line luminosity makes a negligible contribution to our measured extended host fluxes. In order to detect extended line emission, one would need to use integral field spectroscopy \citep{liu13a} or narrow-band imaging.  

Despite limited statistics we find an appreciable luminosity difference between the hosts of ERQs and those of type 2 quasar candidates in our sample (Figure \ref{pic:structure}, right). The median, the mean and the standard deviation of the log host luminosity for the ERQ sample are $\log (L_B/L_{\odot})=11.2, 11.3\pm 0.4$, whereas for the type 2 quasar candidate sample these values are $\log (L_B/L_{\odot})=10.6, 10.7\pm 0.4$. The Kolmogorov-Smirnov test shows that the two distributions are statistically inconsistent; the probability that the two host luminosity distributions are drawn from the same underlying one is 0.4\%. Since each distribution is likely a convolution of the intrinsic distribution with the measurement errors of $\sim 0.25$ dex, accounting for errors would make the apparent difference even more statistically significant. Thus we find evidence that ERQ hosts are either more massive or more strongly star-forming than the hosts of type 2 quasar candidates, resulting in a greater $B$-band luminosity. 

For comparison, the $B$-band luminosity $L^*$ of the break of the galaxy luminosity function at $z=2.5$ is at $m_{AB}=-21.74$ mag or $10^{10.6}L_{\odot}$ \citep{gial05}. Therefore, type 2 quasar candidates' hosts have luminosities right around $L^*$ at their redshift, but ERQ hosts are on average four times more luminous. In Section \ref{sec:disc:scat} we demonstrate that there is no evidence for scattered light contamination of the extended emission in the rest-frame $B$-band. Another possible explanation for the difference in host luminosities is that type 2 hosts might suffer from a greater level of galactic-scale extinction than do ERQs. While we cannot conclusively rule out this possibility, we see no evidence for any differences in structural parameters or morphology between the two subsamples. Thus in what follows we take the difference between ERQ and type 2 host luminosities at face value. 

The host-to-PSF flux ratio is $0.4, 0.3 \pm 0.3$ for the ten ERQs (median, mean, sample standard deviation) and $0.18, 0.17 \pm 0.05$ for the six type 2 candidates (the large standard deviation of the ratio for the ERQs is partly because SDSS~J2323-0100 has a much larger host-to-PSF ratio than the rest of the sample). The consequence is that the nuclei of the type 2 quasar candidates in our sample are nearly an order of magnitude less luminous in the rest-frame optical than the nuclei of ERQs. This is not surprising: the majority of ERQs show evidence for a broad-line region in their optical spectra \citep{zaka16b, perr19} and therefore have a type 1 optical classification. In contrast, type 2 quasar candidates typically have a type 2 or a type 1.9 classification (Section \ref{sec:data:t2}) and therefore a smaller contribution of the nuclear light to the rest-frame optical spectra. 

\subsection{Quantifying signs of merger activity}
\label{sec:merger:major}

Models of co-evolution of black holes and their hosts often appeal to mergers as a leading agent in triggering the active nucleus and the formation of the spheroid \citep{hopk06}. Some models predict that nearly 100\% of quasars with $L_{\rm bol}>10^{47}$ erg/sec reside in galaxies that are merging, interacting or irregular \citep{trei12, hick14}. However, there is no universal agreement on how to quantify such activity in quasar hosts, and identifying major merger activity in quasar hosts at high redshifts remains an especially difficult problem. 

At low redshifts, one of the most identifiable hallmarks of galactic interactions is the presence of tidal tails, but these are hard to see at $z=2.5$ due to cosmological surface brightness dimming. Furthermore, the density of sources in WFC3 images is quite high, and two galaxies that are close together on the plane of the sky can be a chance superposition. Finally, the disturbed morphology of the host may be a sign of clumpy star formation characteristic of high-redshift galaxies rather than a recent merger \citep{ager09}, especially in the rest-frame $B$-band probed by our WFC3 observations. Now that several studies of hosts of extremely luminous quasars are available in the literature, our primary interest is to determine whether the wide variations in the reported merger fractions are related to any quasar properties. The comparison among different studies is complicated by the use of widely different methods -- often just visual classification without any PSF subtraction (e.g., \citealt{fan16}) -- in the literature. 

Keeping this in mind, we classify a target quasar as an ongoing major merger if it has a galaxy within 2.5\arcsec\ with a flux that is greater than 1/4 of the estimated flux of the quasar host galaxy \citep{goul18b, mcal18}. While this definition likely provides a lower limit on major merger activity, it has the advantage of being nearly redshift-independent, enabling us to compare studies in the range $1<z<3$. Indeed, it does not rely on diffuse features, such as tidal tails or shells. For comparison, over 90\% of local ultraluminous infrared galaxies (ULIRGs) with a double-nucleus morphology have a flux ratio above 1/4 \citep{kim02}. Furthermore, the angular diameter distance is within 5\% of the $z=2.5$ value in this range. Finally, by requiring a high flux ratio between the host and the companion and a small separation distance, we reduce the probability of counting chance superpositions and small companions without dynamical influence on the host \citep{mcal18}. The major drawback of this definition is that a late-stage merger without two separate galactic nuclei would not be identified as such. For example, only 42\% of the local ULIRGs have a double-nucleus morphology we would recognize in our observations, with the rest being single late-stage remnants \citep{kim02}.

While there are many faint objects in our images, only three targets out of 16 have a companion within 2.5\arcsec\ with a flux above 1/4 of our Sersic-fitted flux: two are type 2 quasar candidates (SDSS~J1001+4151 and SDSS~J1444$-$0013) and one is an ERQ (SDSS~J2323$-$0100). ERQ SDSS~J1217+0234 has a companion (itself a clumpy, disturbed galaxy) with a flux of 0.24$\times$ the flux of the host galaxy, just under the 1:4 cutoff, and could be a major merger within the uncertainty of our flux measurements. For each image, we calculate the probability of a chance superposition within 2.5\arcsec\ to be 0.11 using WFC3 number counts by \citet{guo13} and the range of fluxes between those of our targets and our cutoff for the companion (1/4). We further use WFC3/ACS flux ratios and redshifted \citet{poll08} templates to confirm that the identified companions are at $z>1$ (the ACS images lack the depth for better redshift constraints), which decreases the probability for a chance superposition by about a factor of 2.

\begin{figure}
\includegraphics[scale=0.85, trim=15cm 0cm 0cm 0cm, clip=true]{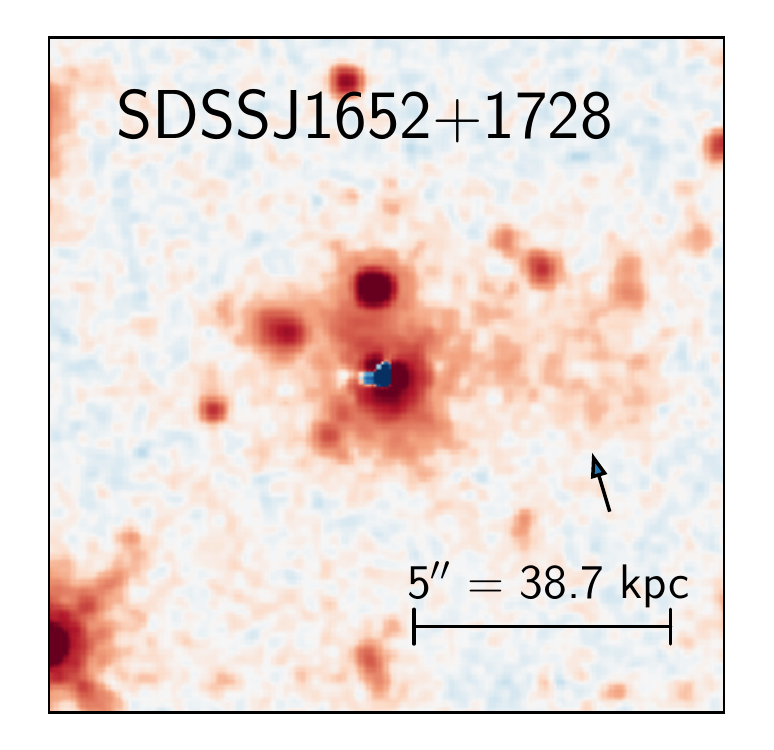}
\caption{PSF-subtracted and smoothed image of SDSS~J1652+1728 with enhanced contrast to highlight the candidate tidal tail marked with an arrow (North is up).} 
\label{pic:tidal}
\end{figure}

In addition to the major merger candidates selected due to the presence of a companion, we find an additional candidate -- ERQ SDSS~J1652+1728. The analysis of this target is complicated by the presence of a bright compact source just North of our object (Figure \ref{pic:tidal}); from the combination of optical colours from the SDSS, our derived HST near-infrared colours and Keck spectroscopy \citep{alex18} which shows a featureless spectrum, we conclude that this object is most likely a foreground star. Excluding this chance superposition from consideration, we find that SDSS~J1652+1728 has multiple faint companions within 2\arcsec\ -- which could well be giant clumps of star formation characteristic of high-redshift star-forming galaxies  \citep{elme07, fors09} -- and a pronounced tidal tail extending over tens of kpc, marked with an arrow in Figure \ref{pic:tidal}. Therefore, we flag this object as a major merger candidate even though it does not meet our formal companion-based definition. Thus our final count of ongoing major mergers among ERQs is $2\pm 1$ out of 10 objects, and it is 2/6 for the type 2 quasar candidates. There are no other strongly disturbed sources that we would be tempted to classify as major mergers, although a late-stage single-nucleus merger remnant can be easily missed by our classification. 

In a late-stage merger, a separate companion would not necessarily be identifiable and tidal features fade, but the host galaxy could still show morphological disturbances. Various non-parametric quantitative measures of morphology exist, such as $M_{20}$, Gini coefficient and asymmetry \citep{lotz04} defined as
\begin{equation}
A=\frac{\Sigma_{ij}|I_{ij}-I_{ij, 180}|}{2\Sigma_{ij} |I_{ij}|}. 
\label{eq:asym}
\end{equation}
Here $I_{ij}$ is the original 2D image, while $I_{ij, 180}$ is the image rotated by 180$^{\rm o}$ about the central point -- in our case, the quasar. The key complication in using these parameters for our sample is the PSF subtraction. In particular, both $M_{20}$ and Gini coefficients rely upon the relative distributions of bright and faint galaxy light. Since the bright center of the host galaxy is blocked by the PSF, these measures are not suitable for our sample where the PSF subtraction uncertainties dominate these regions. The asymmetry is more promising, because if the PSF is symmetric, its residuals would not contribute to the numerator of equation (\ref{eq:asym}). 

However, we find that the asymmetry measurements vary by over a factor of 2 depending on the exact method of PSF subtraction and to the range or radii over which they are calculated. The same difficulty was also pointed out by \citet{glik15}. We conclude that non-parametric measurements of morphology are not yet well tailored to galaxies with a contaminating central source. We use asymmetries with extreme caution and only for comparison with \citet{glik15} when we follow the same measurement procedure as they do. This involves computing asymmetry within 1.5$\times$ the \citet{petr76} radius calculated from our Sersic fits, while masking a 1\arcsec$\times$1\arcsec region around the nucleus and not masking the companions, resulting in median, mean and sample standard deviation of asymmetries of 0.58, 0.53$\pm$0.15. While ERQs and type 2 quasar candidates are statistically indistinguishable in this parameter as per the Kolmogorov-Smirnov test, this may not be a meaningful comparison given the restrictions imposed on our measurement procedure and large individual errors. We discuss the implications of our morphological measurements in the next section. 

\section{Discussion} 
\label{sec:discussion}

With our host galaxy measurements in hand, we discuss the implications of our results and compare the hosts of our targets with several other samples. In Section \ref{sec:disc:scat} we determine whether light scattered off the interstellar medium of the host galaxy can affect our measurements. In Section \ref{sec:disc:bh} we discuss the black hole vs host galaxy relationship of our sample and in Section \ref{sec:disc:sfr} we discuss the constraints on the star formation rates. In Section \ref{sec:disc:comp} we summarize several comparison samples of powerful active nuclei at similar redshifts with host galaxy measurements. We compare the hosts of these objects with ours in Sections \ref{sec:disc:hosts} and \ref{sec:disc:rl}.

\subsection{Effects of scattered light}
\label{sec:disc:scat}

Even when the direct view to the quasar is obscured -- either by the host galaxy or by circumnuclear dust -- the emission from the quasar can escape along other directions, scatter off the surrounding interstellar matter and reach the observer \citep{anto85,anto93}. \citet{alex18} present Keck spectropolarimetric observations of ERQs and type 2 candidates, four of which overlap with our HST observations. Three of these four -- SDSS~J1232+0912, SDSS~J1515+1757 and SDSS~J1652+1728 -- show high continuum polarization in the optical (rest-frame UV), between 10 and 15\%, while the remaining source SDSS~J2215$-$0056 is too weak in the continuum to reliably measure polarization. While these values are not at the theoretical maximum for polarization of scattered light, they are at the high end of polarization for the obscured quasar population \citep{hine93, smit02a, zaka05, gree12}. Therefore, if these three sources are representative of the entire HST sample, then in principle all of the optical (rest-frame UV) emission can be due to scattered light.

As the wavelength increases, the scattered light contribution decreases while the host galaxy becomes more important, so it is not clear what fraction of the flux and on what scales the scattered light contributes to our WFC3 images. If the extended emission outside of the PSF was dominated by scattered light, we would expect a 90 degree offset between the position angles of the optical emission and the position angle of the E-vector of polarization. Instead, for the first three objects we find differences of $\sim 20$, $\sim 50$, and $\sim 30$ degrees. In the last target neither position angle is well determined, but our best attempt yields $\sim 40$ degrees. The clearest measurement is in SDSS~J1515+1757, with Figure \ref{pic:sersic} demonstrating that the position angle of the major axis of the fitted galaxy is 30 degrees E of N, whereas Figure A2 of \citet{alex18} shows that the reddest polarization position angle measurement is 80 degrees E of N, leading to the offset between the two of 50$\pm$10 degrees.

Thus none of the objects for which we have both HST and polarimetric data show the 90 degree offset expected if the extended emission was dominated by scattered light. This is somewhat surprising in light of the results for lower-redshift obscured quasars \citep{zaka06, obie16}, in which the polarization position angle and the extended blue emission are often accurately orthogonal to each other. Therefore in contrast to the low-redshift obscured quasars, the scattering must predominantly occur in a high-density medium on scales smaller than can be probed with our WFC3 observations ($\la 1$ kpc). In this case the scattered light would be part of the PSF component of our analysis. This conclusion is consistent with the estimates of the spatial scales of scattering presented by \citet{alex18}, who found that the rest-frame UV emission lines must originate on scales that are similar to the scales of scattering, $\la$ a few tens of pc. Thus, even though the scattered light constitutes a major component of the observed SEDs of our sources, it does not appear that it has much impact on the host galaxy properties derived from the WFC3 data. Extended scattered light can be further probed by a careful analysis of the ACS images where the contrast between the scattered light and the host galaxy is more favorable.

\subsection{Black hole masses and the nature of our targets}
\label{sec:disc:bh}

What kinds of black holes power the extremely luminous quasars at these redshifts? While there exist established scaling relationships between the host stellar mass and black hole mass in nearby galaxies \citep{korm95}, comparing them with our objects presents major difficulties. First, there is a potential evolution in the black hole vs galaxy relationship between the epoch corresponding to our targets and the present day. Second, we are observing in the rest-frame $B$-band, which is strongly affected by on-going and recent star formation and may be a poor probe of stellar mass. Third, given the uncertainties in our measured Sersic indices and the fact that their values are intermediate between disc-like and bulge-like, it is not clear whether we are detecting the progenitors of massive bulges. 

With these caveats, we concentrate on our ERQ subsample, where the mid-infrared luminosities are well measured, giving us a good handle on the bolometric luminosity. To avoid systematic errors from conversions between different bands and stellar masses, we use local galaxies with $B-$band luminosities from \citet{savo13}, but they have no objects in their sample with $L_B>10^{11.2}L_{\odot}$. Therefore, the best available comparison subsample of local galaxies from \citet{savo13} has $L_B=10^{10.8-11.2}L_{\odot}$, with a median luminosity 0.3 dex less luminous than ours. The median, mean and standard deviation of the dynamically measured log black hole masses within the comparison subsample are $\log(M/M_{\odot})=8.9,8.7\pm 0.9$.

With these crude estimates of black hole mass in local galaxies with $L_B$ luminosities comparable to those of ERQs, we confront three different possibilities. The first is that the black hole masses of ERQs are also $\sim 10^9M_{\odot}$, in agreement with the local black hole vs host relationships. The rest-frame 5\micron\ luminosities of ERQs of $10^{46}-10^{47}$ erg/sec (Figure \ref{pic:sed}) correspond to bolometric luminosities of $10^{47}-10^{48}$ erg/sec \citep{rich06}. At a $10^9M_{\odot}$ black hole mass, ERQs would have to be up to 10 times brighter than the Eddington luminosity:
\begin{equation}
L_{\rm Edd}=1.3\times 10^{38}\frac{M_{\rm BH}}{M_{\odot}}{\rm erg/sec}=1.3\times 10^{47}{\rm erg/sec}.
\end{equation}
One possible argument against such high inferred Eddington ratios is that we might be over-estimating bolometric luminosities by using the bolometric corrections characteristic of type 1 quasars from \citet{rich06}. However, this is unlikely because in the X-ray vs infrared luminosity space ERQs are close to type 1 quasars \citep{goul18a}. In fact, for an obscured population it is more plausible that the bolometric corrections are under-estimated rather than over-estimated, and somewhat higher bolometric corrections would still be consistent with the X-ray data \citep{goul18a}.

The structure of the outflows in the near-Eddington and super-Eddington regime is sensitive to the complex radiative transfer in the optically thick regime. Strong radiation pressure may lead to powerful outflows \citep{shak73}, or photons can be trapped in the matter and advected into the black hole \citep{bege82}. This regime has only recently been probed by numerical simulations \citep{sado16, dai18}, which demonstrate that at high accretion rates, wide-angle outflows ubiquitously form and carry an appreciable fraction of the accretion power. The extreme wind activity seen in our targets on multiple scales \citep{zaka16b, alex18, perr19} is qualitatively consistent with super-Eddington accretion, so our multi-wavelength observations of ERQs support this scenario, and other populations of candidate super-Eddington accretors also show powerful outflow activity \citep{done07,ni18}. Furthermore, simulations confirm earlier suggestions that due to photon advection, near- and super-Eddington accretion tends to be radiatively inefficient \citep{abra88}. Therefore, to produce observed ERQ luminosities of a few times the Eddington limit, an even higher ratio of the accretion rate to $\dot{M}_{\rm Edd}$ may be required. 

Another possibility is that the black holes in ERQs are appreciably more massive, with $M_{\rm BH}\sim 10^{10}M_{\odot}$, bringing their luminosities just below the Eddington limit. This might occur if the local black hole vs host relationship still holds for ERQs, but the $B$-band luminosities of our sample are under-estimated by a factor of ten (e.g., due to strong host galaxy obscuration). Alternatively, the $B$-band host luminosities are correct, but the black holes are ten times more massive than suggested by the local black hole vs host relationship. Both these options require extremely high black hole masses, at the upper limit of those observed in the local universe \citep{mcco11, thom16}. This would imply that in our target ERQs, the observed quasar activity is the last episode of appreciable black hole growth. With a better understanding of the duty cycle of ERQs, the different hypotheses about the masses of their supermassive black holes can be tested by measuring the volume density of ERQs and comparing it to the local volume density of black holes of different masses. 

\subsection{HST constraints on star formation}
\label{sec:disc:sfr}

Here we again concentrate on ERQs whose SEDs are better measured than those of type 2 quasar candidates. The star formation rates of ERQ host galaxies are still not well-constrained with the existing data. If their radio luminosity were due to star formation in the host galaxies, then the observed median radio luminosity implies an extremely high star formation rate of 2700$M_{\odot}$/year \citep{ross15,alex16,hama17}. Another potential argument in favor of extremely high star formation rates is supplied by the model of co-evolution of active galaxies and their hosts by \citet{hick14}. These authors find that for $L_{\rm bol}=10^{47-48}$ erg/sec quasars at $z=2$ the expected infrared luminosity of star formation is in the range $10^{46.2-47}$ erg/sec, nominally corresponding to star formation rates of $700-4500$ $M_{\odot}$/year \citep{bell03}. 

Such extremely high star formation rates are well above those seen in sub-millimeter galaxies \citep{blai02, chap05}, but have been seen in the most extreme high-redshift starbursts \citep{rowa18} and may not necessarily be out of place in the hosts of ERQs, which are after all some of the most rapidly accreting active nuclei in the universe. Nonetheless, we previously argued that such high star formation rates are unlikely because the radio luminosities of ERQs are dominated by the active nuclei, not by the host galaxies \citep{hwan18}. Furthermore, the co-evolution models are not yet well probed at $L_{\rm bol}>10^{47}$ erg/sec, and the existing observations do not support the predicted extremely high star formation rates at the high luminosity end \citep{hick14}. 

Thus the star formation rates in ERQ hosts remain controversial. We now consider whether our HST data provide any additional constraints on the star formation rates. For the nominal star formation rate of 2700$M_{\odot}$/year necessary to account for the radio emission, in the rest-frame $U$-band we expect a luminosity of $L_U\equiv \nu L_{\nu}[U]=6\times 10^{12} L_{\odot}$ due to hot young stars \citep{mous06}. For an O5 stellar template \citep{pick98}, we find $L_B/L_U=0.6$, and this ratio is higher for all later stellar types. Therefore, the median $B$-band luminosity at the star formation rates necessary to explain the observed radio emission and to bring our objects in agreement with models by \citet{hick14} is 1.4 dex higher than observed in our HST data. This is yet another argument against extremely high star formation rates in ERQs. However, much of star formation could proceed in an obscured fashion, as seen in other obscured quasars at the same redshifts \citep{weth18}; therefore, we still need alternative measures of star formation in our sources, e.g., from the far-infrared or submm continuum, to settle this issue. 

\subsection{Introducing comparison samples}
\label{sec:disc:comp}

Despite the strong contribution of the quasar PSF to the observed emission, we successfully detect almost all host galaxies of our target quasars. The ability to routinely detect host galaxies of powerful quasars at $z>2$ is a major success of WFC3 sensitivity, image quality and improved analysis tools, and there are not yet many samples available for comparison with ours. Here we focus on the samples that are most directly comparable in terms of quasar luminosity, redshift and observing mode, but they span a range of obscuration and other properties. 

\citet{mech16} analyze host galaxies of 19 radio-quiet type 1 (unobscured) quasars at $z=1.9-2.1$ selected based on their luminosity and black-hole mass. The median optical luminosity of the quasars in this sample is $L_{\rm 3000\AA}=10^{46.4}$ erg/sec or $L_{\rm bol}\simeq 10^{47.1}$ erg/sec, using bolometric corrections by \citet{rich06}. These objects are observed for one orbit with the F160W filter of WFC3, and among other analyses \citet{mech16} perform PSF+Sersic decomposition and merger classification of their samples. 

\citet{fan16} and \citet{farr17} analyze one-orbit F160W or F110W images of hot dust-obscured galaxies (HotDOGs) at $z=1.8-3.9$ which are selected based on their red WISE colours \citep{eise12, tsai15, asse15}. X-ray observations indicate that these objects host powerful obscured quasars \citep{asse16, ricc17, vito18} which likely dominate the bolometric output \citep{farr17}. HotDOGs have similar luminosities and shapes of spectral energy distributions (SEDs) to ERQs \citep{goul18a}, but HotDOGs likely have higher obscuration levels as their SED rise is at longer wavelengths than in ERQs \citep{hama17}. Furthermore, in the admittedly limited sample of five objects with follow-up near-infrared spectroscopy \citep{wu18}, roughly half of HotDOGs are consistent with a type 2 optical classification -- i.e., with permitted lines of the same shape as the forbidden lines and no evidence for a separate broad-line region \citep{reye08}. In contrast, the majority of ERQs show evidence for a broad-line region in their optical spectra \citep{zaka16b, perr19}. As a result, \citet{fan16} and \citet{farr17} assume that the nucleus does not contribute at rest-frame optical wavelengths and do not perform PSF subtraction. 

\citet{glik15} analyze HST images of radio+infrared selected red quasars with bolometric luminosities $L_{\rm bol}=10^{47.3-48.3}$ erg/sec at $z=1.7-2.3$. Their observations are obtained with F160W and a shorter wavelength filter (F105W or F125W) over one or two orbits. They conduct PSF+Sersic decomposition, as well as merger classification and asymmetry measurements, and our observation and analysis methods are most directly comparable to this sample. 

Finally, \citet{hilb16} present the initial results of their study of hosts of radio-loud type 2 and type 1 active nuclei with F140W at $z\ga 1$. The redshifts of their objects are lower than ours, but they have three radio galaxies at $z=1.8-2.5$ which we include in our qualitative comparisons: 3C257, 3C326.1 and 3C454.1. In these three cases we assume that the nucleus does not contribute to the integrated photometry due to the type 2 spectrum in the rest-frame optical (e.g., \citealt{nesv17} for 3C257). 

\subsection{Comparison of our sample with others}
\label{sec:disc:hosts}

Despite the uncertainties in the fitting parameters, our $B$-band luminosities seem robust because the results of the parametric (Sersic) fits agree well with the results of the non-parametric (aperture photometry) estimates. Given that the stellar mass-to-light ratios are not known either for our sample or for any of the comparison ones, we restrict ourselves to direct comparison of host luminosities, not stellar mass. Furthermore, we calculate k-corrections as described in Section \ref{sec:host:mass} for all comparison datasets, so that we can compare $L_B$ among the different samples. 

Three of the comparison samples have $B$-band host galaxy luminosity distributions that are statistically consistent with those of ERQs in the sense of the Kolmogorov-Smirnov test. Aperture fluxes of HotDOGs measured by \citet{farr17} correspond to $\log(L_B/L_{\odot})=11.1,11.2\pm0.5$ (median, mean and sample standard deviation). These luminosities may be upper limits, since in the absence of near-infrared spectroscopy it is not yet known whether the quasars make an appreciable contribution to this luminosity and it is not clear whether the apertures exclude close companions. For their sample of type 1 quasars, \citet{mech16} present $V$-band luminosities which we k-correct to $B$-band by $-0.19$ dex. For their sample we find host luminosities $\log(L_B/L_{\odot})=11.1,11.1\pm 0.3$, again quite consistent with those of ERQs. The three radio galaxies in \citet{hilb16} that have redshifts from $z=2.5$ to $z=1.8$ have luminosities $\log(L_B/L_{\odot})=11.6,11.3\pm 0.47$, somewhat larger than, but statistically consistent with, those of ERQs given the low number counts. This latter comparison comes with a caveat that 1 Gyr elapsed between these two epochs, so the two populations may be quite distinct, especially if the hosts of our targets average star formation rates of 100 $M_{\odot}$/year or more.  

We estimate that the median stellar mass of the host galaxies of our targets ranges between $10^{10.4}-10^{11.4}M_{\odot}$ by assuming a range of plausible mass-to-light ratios \citep{mara05}. The stellar masses could be higher if the host galaxies are extremely obscured on several kpc scales. Since all the comparison studies listed above derive their estimated host stellar masses from rest-frame $B$ or $V$-band data similar to ours, their estimates are subject to similar types of systematic uncertainties. Interestingly, our range of stellar masses is consistent with the stellar masses of high-redshift radio galaxies derived using completely different methods \citep{seym07}.

The hosts of radio-selected red quasars from \citet{glik15} are marginally brighter than those of ERQs, with $\log(L_B/L_{\odot})=11.6,11.5\pm0.6$. However, out of the 10 radio-selected red quasars in the \citet{glik15} sample, only five hosts are well-measured using PSF Sersic fits, and therefore the average $L_B$ quoted above is likely overestimated. The surprising property of this sample is that in about half of the sources, the quasar host galaxies are fainter than the clearly detected companions, whereas in our sample there is not a single object where the companion is brighter than the host. 

We do not find strong evidence that the host galaxies of ERQs or type 2 quasar candidates are associated with ongoing major mergers. The fraction of objects with bright nearby companions in either one of our two subsamples is only 20$-$30\%, short of theoretical expectations from numerical simulations \citep{stei18, mcal18}, especially for the most luminous quasars at this redshift \citep{hick14}. In observations, some recent works \citep{glik15, fan16, hilb16} find a 80$-$100\% merger incidence in luminous obscured populations. For HotDOGs, \citet{farr17} argue that the high merger incidence derived by \citet{fan16} may be over-estimated. Visually, about 40$-$50\% of \citet{farr17} sources would likely be classified by us as ongoing major mergers, which (given the small number statistics) may be consistent with our sample. Their asymmetries are even lower than ours, though they are not directly comparable due to PSF treatment differences. In contrast, high-redshift radio-loud samples may show genuinely higher merger fraction than our sources, as we discuss in Section \ref{sec:disc:rl}.

Submm observations of a HotDOG by \citet{diaz18} and of a luminous type 1 quasar by \citet{bisc18} reveal a multitude of gas companions and tidal tails in host galaxies which are not detectable in the rest-frame optical data (similar techniques are also used in uncovering the environments of extremely high-redshift quasars, \citealt{deca17}). Perhaps in ERQs merger activity can similarly be hidden in rest-frame optical observations. Alternatively our targets could be associated with a late stage of the merger. They are less obscured than HotDOGs, which may or may not be an orientation effect, so the marginally lower merger incidence rate in our sample compared to HotDOGs may indicate that ERQs are more likely to occur in a later phase \citep{hama17}. Finally, perhaps extremely luminous quasar activity does not have to be associated with mergers at all and might result from processes within the galaxy \citep{shlo89}. Whether relevant amounts of gas can be supplied by such processes to power the likely super-Eddington ERQs should be explored with numerical simulations. Type 2 quasar candidates in our sample do not show luminosities that are high enough to suggest super-Eddington accretion, and they may be suitable candidates for such galactic processes. 

\subsection{Comparison with other samples: red radio quasars and 3C sources}
\label{sec:disc:rl}

The merger incidence of ERQs and type 2 quasar candidates is not high and is not statistically distinguishable from that of \citet{farr17} or \citet{mech16} with the current data, and \citet{mech16} in turn suggest that the incidence of mergers in their quasar sample is indistinguishable from that of high-redshift field galaxies of the same mass. Either the types of quasar activity we target in this paper are not predominantly associated with ongoing major mergers, or they may be associated with a post-merger phase we cannot identify using our data. The sample by \citet{glik15} shows qualitatively different behavior. These objects are frequently found in ongoing mergers of highly disturbed hosts and are often hosted by the less massive companion in a pair, and their asymmetries (measured exactly using the same method as ours) are in the range $0.5-0.8$, marginally higher than our values, with host galaxy images visually more disturbed than ours. However, just like \citet{glik15}, we do not regard our asymmetries as reliably measured and are including them in our discussion mostly to emphasize the difficulties of quantitative morphology measurements in the presence of a bright central source.

These targets are similar to our ERQs in their bolometric luminosities and red colours, but have somewhat lower obscuration \citep{glik17}. So why are we reaching different conclusions regarding the morphologies of our ERQ sample? An interesting possibility is that the host morphology may be tied to the radio properties. The \citet{glik15} sources are radio-selected, and their median radio luminosity measured from FIRST \citep{beck95} survey fluxes and k-corrected to the rest-frame 1.4 GHz using $\alpha=-0.7$ is $\nu L_{\nu}=10^{41.9}$ erg/sec. In contrast, neither ERQs nor type 2 quasar candidates are radio-selected. The median luminosity of the ten ERQs presented here is $\nu L_{\nu}$[1.4GHz]$=10^{40.9}$ erg/sec \citep{hwan18}, an order of magnitude lower, and type 2 quasar candidates are typically even less radio luminous \citep{alex16}.  

There are other indications that radio loudness may be related to host properties in high-redshift populations, with very high merger fractions reported for radio-loud quasars by \citet{ivis12} and \citet{noir18}. \citet{chia15} find that at $z\simeq 1$ obscured (type 2) 3C radio galaxies are predominantly associated with mergers, whereas radio-quiet type 2 quasars are not any more likely to be in mergers than are non-active galaxies. \citet{hilb16} present a larger sample of obscured and unobscured radio-loud quasars at $1<z<2.5$. They find that all obscured radio-loud hosts, where PSF subtraction is not necessary for evaluating the morphology, are highly disturbed, have multiple companions and appear to be associated with mergers. Of the three objects at the same redshifts as our targets, two would be classified by us as ongoing major mergers and the third has several faint companions. The analysis of the unobscured radio-loud quasars from this sample \citep{chia18} which uses PSF subtraction finds a strong association with mergers among these objects as well. More indirectly, radio-loud sources, including those at redshifts comparable to those of our sample, are found in more dense large-scale environments \citep{wyle13} than comparable radio-quiet objects, which may naturally result in a higher merger fraction.

We conclude that radio luminosity is an important parameter to control for in future studies of quasar hosts at high redshift. As argued by \citet{hwan18}, the radio luminosity of ERQs may be on the upper end of what may be considered a ``radio-quiet'' population not associated with powerful jets. Therefore, even though the radio luminosity difference between ERQs and \citet{glik15} is only an order of magnitude, we speculate that the two samples may have dramatically different mechanisms of producing radio emission, with jets -- relativistic, collimated outflows -- in the \citet{glik15} sources and winds -- non-relativistic, non-collimated outflows -- in ERQs. The 3C objects from \citet{hilb16} and \citep{chia18} are on the extreme bright end of the radio-loud distribution and are clearly jet-dominated. Correspondingly, it is possible that either the parameters of the supermassive black holes (e.g., spins) or the accretion parameters might be quite different in the populations with and without jets, despite having similar bolometric luminosities. Thus it may not be surprising that the host properties and the larger scale environments may be different as well. 

Another possibility for the difference between our sample and those of \citet{glik15} and \citet{hilb16} is that the merger fraction may be higher in populations selected in the infrared and/or in the radio without regard to optical (rest-frame UV) fluxes. Such selections may naturally result in a bias toward populations with higher galaxy extinction and therefore toward a higher merger fraction. In support of this possibility is the conclusion reached by \citet{glik17} that the obscuration in their sample arises on the scales of the host galaxy, whereas in ERQs it is likely circumnuclear \citep{goul18a}. However, this hypothesis fails to account for the modest merger fraction of HotDOGs (selected based on the mid-infrared alone) which is comparable to that seen in our sample \citep{farr17}. Therefore, the connection between radio loudness and host merger activity at high redshift is well worth exploring with future observations with HST and JWST. 

\section{Conclusions}
\label{sec:conclusions}

In this paper we present HST observations of ten ERQs and six type 2 quasar candidates at $2<z<3$ -- the peak epoch of quasar activity and galaxy formation. ERQs are extremely luminous (up to $L_{\rm bol}\la 10^{48}$ erg/sec) near- or super-Eddington accretors, with unprecedented ionized gas outflow activity seen both on small and large scales. No analogs to these objects have been found in the low-redshift universe. ERQs are ideal candidates for being observed during the early short-lived feedback phase of quasars when they are still dust enshrouded but driving extremely powerful outflows that expel the gas and dust, disrupt star formation, and halt mass assembly in the host galaxies. While this evolution is expected to be triggered by galaxy mergers and interactions \citep{hopk06}, there is not yet a definitive observation of this connection during the epoch of peak galaxy formation and quasar activity. Type 2 quasar candidates are less powerful, show moderate amount of ionized wind activity in their [OIII] emission lines, and resemble low-redshift type 2 quasars, except perhaps at lower levels of obscuration due to their original selection at rest-frame ultraviolet wavelengths. 

In our HST programmes, we image both samples in the ACS F814W and WFC3 F160W (or F140W) bands. We use a careful PSF subtraction procedure on the WFC3 data to reveal the host galaxy emission in almost every object. We also perform joint PSF+Sersic fits to determine the structural parameters of the host galaxies.

The median rest-frame $B$-band luminosities of the hosts are $10^{11.2}L_{\odot}$ and $10^{10.6}L_{\odot}$ for ERQs and for type 2 quasar candidates, respectively, with the latter value similar to the $L^*$ luminosity at this redshift. Despite uncertainties in PSF subtraction and difficulty of joint PSF+Sersic fits, there is good agreement between several different measurements of host luminosities, and we estimate the uncertainties in luminosity of each object at $\la 0.25$ dex. These luminosities are similar to the $B$-band luminosities found for other luminous quasar hosts using similar data. The structural parameters, such as effective radii and Sersic indices, are more uncertain and more sensitive to the exact fitting procedure. Most fits indicate that the galaxies in our sample are early-type, with Sersic indices between 2 and 3 and a median effective radius just under 5 kpc; however, Sersic indices have especially large estimated uncertainties ($\sigma(n_s)\simeq 1$) and are not yet well constrained. These properties are similar to those of massive high-redshift galaxies in \citet{vand10}, except our hosts are somewhat less compact -- with $R_{50}$ of 5 kpc instead of $<$3 kpc -- but this may well be within the large uncertainty on $R_{50}$. 

Only $20-30$\% of ERQs and type 2 quasar candidates are in ongoing major mergers with a flux ratio greater than 1:4 and with components separated by $<$2.5\arcsec. This is in contrast to the $80-100$\% merger fraction expected for quasars at this luminosity and redshift and found among the high-redshift radio-loud population. We have discussed three possibilities for this unexpected result. (i) Merger signatures can be hidden at rest-frame optical wavelengths in high-redshift dusty galaxies and may be revealed by future submm observations. (ii) Objects in our sample might not be associated with mergers at all, and may instead be powered by gas instabilities within galactic discs. This possibility is more likely to apply to the type 2 quasar candidates which are less luminous than ERQs and are thus less likely to be super-Eddington accretors. Numerical simulations are necessary to determine whether super-Eddington accretion of ERQs can result from galactic processes in the absence of mergers. (iii) Objects in our sample might be associated with late-stage mergers that cannot be identified using our HST data due to the faintness of the tidal features. Higher sensitivity JWST observations can be used to probe this scenario. 

\section*{Acknowledgements}

Based on observations made with the NASA/ESA HST, obtained at the Space Telescope Science Institute, which is operated by the Association of Universities for Research in Astronomy, Inc., under NASA contract NAS 5-26555. These observations are associated with programmes GO-14608 and GO-13014. Support for these programmes was provided by NASA through grants HST-GO-14608 and HST-GO-13014 from the STScI. 

NLZ acknowledges support by the Catalyst Award at Johns Hopkins University and by the Deborah Lunder and Alan Ezekowitz Founders' Circle Membership at the Institute for Advanced Study where part of this work was conducted. WNB acknowledges support by the NASA ADP program and NSF grant AST-1516784. GL is supported by the National Thousand Young Talents Program of China and acknowledges the grant from the National Natural Science Foundation of China (No. 11673020 and No. 11421303) and the Ministry of Science and Technology of China (National Key Program for Science and Technology Research and Development, No. 2016YFA0400700). NPR acknowledges support from the STFC and the Ernest Rutherford Fellowship scheme.




\bibliographystyle{mnras}

\begin{thebibliography}{}
\makeatletter
\relax
\def\mn@urlcharsother{\let\do\@makeother \do\$\do\&\do\#\do\^\do\_\do\%\do\~}
\def\mn@doi{\begingroup\mn@urlcharsother \@ifnextchar [ {\mn@doi@}
  {\mn@doi@[]}}
\def\mn@doi@[#1]#2{\def\@tempa{#1}\ifx\@tempa\@empty \href
  {http://dx.doi.org/#2} {doi:#2}\else \href {http://dx.doi.org/#2} {#1}\fi
  \endgroup}
\def\mn@eprint#1#2{\mn@eprint@#1:#2::\@nil}
\def\mn@eprint@arXiv#1{\href {http://arxiv.org/abs/#1} {{\tt arXiv:#1}}}
\def\mn@eprint@dblp#1{\href {http://dblp.uni-trier.de/rec/bibtex/#1.xml}
  {dblp:#1}}
\def\mn@eprint@#1:#2:#3:#4\@nil{\def\@tempa {#1}\def\@tempb {#2}\def\@tempc
  {#3}\ifx \@tempc \@empty \let \@tempc \@tempb \let \@tempb \@tempa \fi \ifx
  \@tempb \@empty \def\@tempb {arXiv}\fi \@ifundefined
  {mn@eprint@\@tempb}{\@tempb:\@tempc}{\expandafter \expandafter \csname
  mn@eprint@\@tempb\endcsname \expandafter{\@tempc}}}

\bibitem[\protect\citeauthoryear{{Abramowicz}, {Czerny}, {Lasota}  \&
  {Szuszkiewicz}}{{Abramowicz} et~al.}{1988}]{abra88}
{Abramowicz} M.~A.,  {Czerny} B.,  {Lasota} J.~P.,   {Szuszkiewicz} E.,  1988,
  \mn@doi [\apj] {10.1086/166683}, \href
  {https://ui.adsabs.harvard.edu/abs/1988ApJ...332..646A} {332, 646}

\bibitem[\protect\citeauthoryear{{Agertz}, {Teyssier}  \& {Moore}}{{Agertz}
  et~al.}{2009}]{ager09}
{Agertz} O.,  {Teyssier} R.,   {Moore} B.,  2009, \mn@doi [\mnras]
  {10.1111/j.1745-3933.2009.00685.x}, \href
  {https://ui.adsabs.harvard.edu/#abs/2009MNRAS.397L..64A} {397, L64}

\bibitem[\protect\citeauthoryear{{Alexandroff} et~al.,}{{Alexandroff}
  et~al.}{2013}]{alex13}
{Alexandroff} R.,  et~al., 2013, \mn@doi [\mnras] {10.1093/mnras/stt1500},
  \href {http://adsabs.harvard.edu/abs/2013MNRAS.435.3306A} {435, 3306}

\bibitem[\protect\citeauthoryear{{Alexandroff}, {Zakamska}, {van Velzen},
  {Greene}  \& {Strauss}}{{Alexandroff} et~al.}{2016}]{alex16}
{Alexandroff} R.~M.,  {Zakamska} N.~L.,  {van Velzen} S.,  {Greene} J.~E.,
  {Strauss} M.~A.,  2016, \mn@doi [\mnras] {10.1093/mnras/stw2124}, \href
  {http://adsabs.harvard.edu/abs/2016MNRAS.463.3056A} {463, 3056}

\bibitem[\protect\citeauthoryear{{Alexandroff} et~al.,}{{Alexandroff}
  et~al.}{2018}]{alex18}
{Alexandroff} R.~M.,  et~al., 2018, \mn@doi [\mnras] {10.1093/mnras/sty1685},
  \href {https://ui.adsabs.harvard.edu/#abs/2018MNRAS.479.4936A} {479, 4936}

\bibitem[\protect\citeauthoryear{{Antonucci}}{{Antonucci}}{1993}]{anto93}
{Antonucci} R.,  1993, \mn@doi [\araa] {10.1146/annurev.aa.31.090193.002353},
  \href {http://adsabs.harvard.edu/abs/1993ARA%26A..31..473A} {31, 473}

\bibitem[\protect\citeauthoryear{{Antonucci} \& {Miller}}{{Antonucci} \&
  {Miller}}{1985}]{anto85}
{Antonucci} R.~R.~J.,  {Miller} J.~S.,  1985, \mn@doi [\apj] {10.1086/163559},
  \href {http://adsabs.harvard.edu/abs/1985ApJ...297..621A} {297, 621}

\bibitem[\protect\citeauthoryear{{Assef} et~al.,}{{Assef}
  et~al.}{2015}]{asse15}
{Assef} R.~J.,  et~al., 2015, \mn@doi [\apj] {10.1088/0004-637X/804/1/27},
  \href {http://adsabs.harvard.edu/abs/2015ApJ...804...27A} {804, 27}

\bibitem[\protect\citeauthoryear{{Assef} et~al.,}{{Assef}
  et~al.}{2016}]{asse16}
{Assef} R.~J.,  et~al., 2016, \mn@doi [\apj] {10.3847/0004-637X/819/2/111},
  \href {http://adsabs.harvard.edu/abs/2016ApJ...819..111A} {819, 111}

\bibitem[\protect\citeauthoryear{{Bahcall}, {Kirhakos}, {Saxe}  \&
  {Schneider}}{{Bahcall} et~al.}{1997}]{bahc97}
{Bahcall} J.~N.,  {Kirhakos} S.,  {Saxe} D.~H.,   {Schneider} D.~P.,  1997,
  \mn@doi [\apj] {10.1086/303926}, \href
  {http://adsabs.harvard.edu/abs/1997ApJ...479..642B} {479, 642}

\bibitem[\protect\citeauthoryear{{Becker}, {White}  \& {Helfand}}{{Becker}
  et~al.}{1995}]{beck95}
{Becker} R.~H.,  {White} R.~L.,   {Helfand} D.~J.,  1995, \mn@doi [\apj]
  {10.1086/176166}, \href {http://adsabs.harvard.edu/abs/1995ApJ...450..559B}
  {450, 559}

\bibitem[\protect\citeauthoryear{{Beelen} et~al.,}{{Beelen}
  et~al.}{2004}]{beel04}
{Beelen} A.,  et~al., 2004, \mn@doi [\aap] {10.1051/0004-6361:20040318}, \href
  {http://adsabs.harvard.edu/abs/2004A%26A...423..441B} {423, 441}

\bibitem[\protect\citeauthoryear{{Begelman} \& {Meier}}{{Begelman} \&
  {Meier}}{1982}]{bege82}
{Begelman} M.~C.,  {Meier} D.~L.,  1982, \mn@doi [\apj] {10.1086/159688}, \href
  {https://ui.adsabs.harvard.edu/abs/1982ApJ...253..873B} {253, 873}

\bibitem[\protect\citeauthoryear{{Bell}}{{Bell}}{2003}]{bell03}
{Bell} E.~F.,  2003, \mn@doi [\apj] {10.1086/367829}, \href
  {http://adsabs.harvard.edu/abs/2003ApJ...586..794B} {586, 794}

\bibitem[\protect\citeauthoryear{{Bischetti} et~al.,}{{Bischetti}
  et~al.}{2018}]{bisc18}
{Bischetti} M.,  et~al., 2018, \mn@doi [\aap] {10.1051/0004-6361/201833249},
  \href {https://ui.adsabs.harvard.edu/#abs/2018A&A...617A..82B} {617, A82}

\bibitem[\protect\citeauthoryear{{Blain}, {Smail}, {Ivison}, {Kneib}  \&
  {Frayer}}{{Blain} et~al.}{2002}]{blai02}
{Blain} A.~W.,  {Smail} I.,  {Ivison} R.~J.,  {Kneib} J.~P.,   {Frayer} D.~T.,
  2002, \mn@doi [\physrep] {10.1016/S0370-1573(02)00134-5}, \href
  {https://ui.adsabs.harvard.edu/#abs/2002PhR...369..111B} {369, 111}

\bibitem[\protect\citeauthoryear{{Boyle} \& {Terlevich}}{{Boyle} \&
  {Terlevich}}{1998}]{boyl98}
{Boyle} B.~J.,  {Terlevich} R.~J.,  1998, \mn@doi [\mnras]
  {10.1046/j.1365-8711.1998.01264.x}, \href
  {http://adsabs.harvard.edu/abs/1998MNRAS.293L..49B} {293, L49}

\bibitem[\protect\citeauthoryear{{Canalizo} \& {Stockton}}{{Canalizo} \&
  {Stockton}}{2001}]{cana01}
{Canalizo} G.,  {Stockton} A.,  2001, \mn@doi [\apj] {10.1086/321520}, \href
  {http://esoads.eso.org/abs/2001ApJ...555..719C} {555, 719}

\bibitem[\protect\citeauthoryear{{Chapman}, {Blain}, {Smail}  \&
  {Ivison}}{{Chapman} et~al.}{2005}]{chap05}
{Chapman} S.~C.,  {Blain} A.~W.,  {Smail} I.,   {Ivison} R.~J.,  2005, \mn@doi
  [\apj] {10.1086/428082}, \href
  {https://ui.adsabs.harvard.edu/#abs/2005ApJ...622..772C} {622, 772}

\bibitem[\protect\citeauthoryear{{Chiaberge}, {Gilli}, {Lotz}  \&
  {Norman}}{{Chiaberge} et~al.}{2015}]{chia15}
{Chiaberge} M.,  {Gilli} R.,  {Lotz} J.~M.,   {Norman} C.,  2015, \mn@doi
  [\apj] {10.1088/0004-637X/806/2/147}, \href
  {http://esoads.eso.org/abs/2015ApJ...806..147C} {806, 147}

\bibitem[\protect\citeauthoryear{{Chiaberge et al.}}{{Chiaberge et
  al.}}{2018}]{chia18}
{Chiaberge et al.} M.,  2018, in prep.

\bibitem[\protect\citeauthoryear{{Croom}, {Schade}, {Boyle}, {Shanks}, {Miller}
   \& {Smith}}{{Croom} et~al.}{2004}]{croo04}
{Croom} S.~M.,  {Schade} D.,  {Boyle} B.~J.,  {Shanks} T.,  {Miller} L.,
  {Smith} R.~J.,  2004, \mn@doi [\apj] {10.1086/382747}, \href
  {http://adsabs.harvard.edu/abs/2004ApJ...606..126C} {606, 126}

\bibitem[\protect\citeauthoryear{{Croton} et~al.,}{{Croton}
  et~al.}{2006}]{crot06}
{Croton} D.~J.,  et~al., 2006, \mn@doi [\mnras]
  {10.1111/j.1365-2966.2005.09675.x}, \href
  {http://adsabs.harvard.edu/abs/2006MNRAS.365...11C} {365, 11}

\bibitem[\protect\citeauthoryear{{Dai}, {McKinney}, {Roth}, {Ramirez-Ruiz}  \&
  {Miller}}{{Dai} et~al.}{2018}]{dai18}
{Dai} L.,  {McKinney} J.~C.,  {Roth} N.,  {Ramirez-Ruiz} E.,   {Miller} M.~C.,
  2018, \mn@doi [\apj] {10.3847/2041-8213/aab429}, \href
  {https://ui.adsabs.harvard.edu/#abs/2018ApJ...859L..20D} {859, L20}

\bibitem[\protect\citeauthoryear{{Dawson}, {Schlegel}  \& et al.}{{Dawson}
  et~al.}{2013}]{daws13}
{Dawson} K.~S.,  {Schlegel} D.~J.,   et al. 2013, \mn@doi [\aj]
  {10.1088/0004-6256/145/1/10}, \href
  {http://adsabs.harvard.edu/abs/2013AJ....145...10D} {145, 10}

\bibitem[\protect\citeauthoryear{{Decarli} et~al.,}{{Decarli}
  et~al.}{2017}]{deca17}
{Decarli} R.,  et~al., 2017, \mn@doi [\nat] {10.1038/nature22358}, \href
  {https://ui.adsabs.harvard.edu/#abs/2017Natur.545..457D} {545, 457}

\bibitem[\protect\citeauthoryear{{Diaz-Santos et al.}}{{Diaz-Santos et
  al.}}{2018}]{diaz18}
{Diaz-Santos et al.} T.,  2018, submitted

\bibitem[\protect\citeauthoryear{{Done}, {Gierli{\'n}ski}  \& {Kubota}}{{Done}
  et~al.}{2007}]{done07}
{Done} C.,  {Gierli{\'n}ski} M.,   {Kubota} A.,  2007, \mn@doi [\aapr]
  {10.1007/s00159-007-0006-1}, \href
  {https://ui.adsabs.harvard.edu/abs/2007A&ARv..15....1D} {15, 1}

\bibitem[\protect\citeauthoryear{{Dunlop}, {McLure}, {Kukula}, {Baum}, {O'Dea}
  \& {Hughes}}{{Dunlop} et~al.}{2003}]{dunl03}
{Dunlop} J.~S.,  {McLure} R.~J.,  {Kukula} M.~J.,  {Baum} S.~A.,  {O'Dea}
  C.~P.,   {Hughes} D.~H.,  2003, \mn@doi [\mnras]
  {10.1046/j.1365-8711.2003.06333.x}, \href
  {http://adsabs.harvard.edu/abs/2003MNRAS.340.1095D} {340, 1095}

\bibitem[\protect\citeauthoryear{{Eisenhardt} et~al.,}{{Eisenhardt}
  et~al.}{2012}]{eise12}
{Eisenhardt} P.~R.~M.,  et~al., 2012, \mn@doi [\apj]
  {10.1088/0004-637X/755/2/173}, \href
  {http://adsabs.harvard.edu/abs/2012ApJ...755..173E} {755, 173}

\bibitem[\protect\citeauthoryear{{Eisenstein} et~al.,}{{Eisenstein}
  et~al.}{2011}]{eise11}
{Eisenstein} D.~J.,  et~al., 2011, \mn@doi [\aj] {10.1088/0004-6256/142/3/72},
  \href {http://adsabs.harvard.edu/abs/2011AJ....142...72E} {142, 72}

\bibitem[\protect\citeauthoryear{{Elmegreen}, {Elmegreen}, {Ravindranath}  \&
  {Coe}}{{Elmegreen} et~al.}{2007}]{elme07}
{Elmegreen} D.~M.,  {Elmegreen} B.~G.,  {Ravindranath} S.,   {Coe} D.~A.,
  2007, \mn@doi [\apj] {10.1086/511667}, \href
  {https://ui.adsabs.harvard.edu/abs/2007ApJ...658..763E} {658, 763}

\bibitem[\protect\citeauthoryear{{Fan} et~al.,}{{Fan} et~al.}{2016}]{fan16}
{Fan} L.,  et~al., 2016, \mn@doi [\apjl] {10.3847/2041-8205/822/2/L32}, \href
  {http://adsabs.harvard.edu/abs/2016ApJ...822L..32F} {822, L32}

\bibitem[\protect\citeauthoryear{{Farrah} et~al.,}{{Farrah}
  et~al.}{2017}]{farr17}
{Farrah} D.,  et~al., 2017, \mn@doi [\apj] {10.3847/1538-4357/aa78f2}, \href
  {https://ui.adsabs.harvard.edu/#abs/2017ApJ...844..106F} {844, 106}

\bibitem[\protect\citeauthoryear{{Floyd}, {Kukula}, {Dunlop}, {McLure},
  {Miller}, {Percival}, {Baum}  \& {O'Dea}}{{Floyd} et~al.}{2004}]{floy04}
{Floyd} D.~J.~E.,  {Kukula} M.~J.,  {Dunlop} J.~S.,  {McLure} R.~J.,  {Miller}
  L.,  {Percival} W.~J.,  {Baum} S.~A.,   {O'Dea} C.~P.,  2004, \mn@doi
  [\mnras] {10.1111/j.1365-2966.2004.08315.x}, \href
  {http://adsabs.harvard.edu/abs/2004MNRAS.355..196F} {355, 196}

\bibitem[\protect\citeauthoryear{{F{\"o}rster Schreiber} et~al.,}{{F{\"o}rster
  Schreiber} et~al.}{2009}]{fors09}
{F{\"o}rster Schreiber} N.~M.,  et~al., 2009, \mn@doi [\apj]
  {10.1088/0004-637X/706/2/1364}, \href
  {https://ui.adsabs.harvard.edu/abs/2009ApJ...706.1364F} {706, 1364}

\bibitem[\protect\citeauthoryear{{Fruchter} \& {Hook}}{{Fruchter} \&
  {Hook}}{2002}]{fruc02}
{Fruchter} A.~S.,  {Hook} R.~N.,  2002, \mn@doi [Publications of the
  Astronomical Society of the Pacific] {10.1086/338393}, \href
  {https://ui.adsabs.harvard.edu/#abs/2002PASP..114..144F} {114, 144}

\bibitem[\protect\citeauthoryear{{Giallongo}, {Salimbeni}, {Menci}, {Zamorani},
  {Fontana}, {Dickinson}, {Cristiani}  \& {Pozzetti}}{{Giallongo}
  et~al.}{2005}]{gial05}
{Giallongo} E.,  {Salimbeni} S.,  {Menci} N.,  {Zamorani} G.,  {Fontana} A.,
  {Dickinson} M.,  {Cristiani} S.,   {Pozzetti} L.,  2005, \mn@doi [\apj]
  {10.1086/427819}, \href
  {https://ui.adsabs.harvard.edu/#abs/2005ApJ...622..116G} {622, 116}

\bibitem[\protect\citeauthoryear{{Glikman}, {Simmons}, {Mailly}, {Schawinski},
  {Urry}  \& {Lacy}}{{Glikman} et~al.}{2015}]{glik15}
{Glikman} E.,  {Simmons} B.,  {Mailly} M.,  {Schawinski} K.,  {Urry} C.~M.,
  {Lacy} M.,  2015, \mn@doi [\apj] {10.1088/0004-637X/806/2/218}, \href
  {http://adsabs.harvard.edu/abs/2015ApJ...806..218G} {806, 218}

\bibitem[\protect\citeauthoryear{{Glikman}, {LaMassa}, {Piconcelli}, {Urry}  \&
  {Lacy}}{{Glikman} et~al.}{2017}]{glik17}
{Glikman} E.,  {LaMassa} S.,  {Piconcelli} E.,  {Urry} M.,   {Lacy} M.,  2017,
  \mn@doi [\apj] {10.3847/1538-4357/aa88ac}, \href
  {http://esoads.eso.org/abs/2017ApJ...847..116G} {847, 116}

\bibitem[\protect\citeauthoryear{{Goulding} et~al.,}{{Goulding}
  et~al.}{2018a}]{goul18b}
{Goulding} A.~D.,  et~al., 2018a, \mn@doi [Publications of the Astronomical
  Society of Japan] {10.1093/pasj/psx135}, \href
  {https://ui.adsabs.harvard.edu/#abs/2018PASJ...70S..37G} {70, S37}

\bibitem[\protect\citeauthoryear{{Goulding} et~al.,}{{Goulding}
  et~al.}{2018b}]{goul18a}
{Goulding} A.~D.,  et~al., 2018b, \mn@doi [\apj] {10.3847/1538-4357/aab040},
  \href {http://esoads.eso.org/abs/2018ApJ...856....4G} {856, 4}

\bibitem[\protect\citeauthoryear{{Greene}, {Zakamska}  \& {Smith}}{{Greene}
  et~al.}{2012}]{gree12}
{Greene} J.~E.,  {Zakamska} N.~L.,   {Smith} P.~S.,  2012, \mn@doi [\apj]
  {10.1088/0004-637X/746/1/86}, \href
  {http://adsabs.harvard.edu/abs/2012ApJ...746...86G} {746, 86}

\bibitem[\protect\citeauthoryear{{Greene} et~al.,}{{Greene}
  et~al.}{2014}]{gree14b}
{Greene} J.~E.,  et~al., 2014, \mn@doi [\apj] {10.1088/0004-637X/788/1/91},
  \href {http://adsabs.harvard.edu/abs/2014ApJ...788...91G} {788, 91}

\bibitem[\protect\citeauthoryear{{Grogin} et~al.,}{{Grogin}
  et~al.}{2005}]{grog05}
{Grogin} N.~A.,  et~al., 2005, \mn@doi [\apj] {10.1086/432256}, \href
  {https://ui.adsabs.harvard.edu/#abs/2005ApJ...627L..97G} {627, L97}

\bibitem[\protect\citeauthoryear{{Grogin} et~al.,}{{Grogin}
  et~al.}{2011}]{grog11}
{Grogin} N.~A.,  et~al., 2011, \mn@doi [\apjs] {10.1088/0067-0049/197/2/35},
  \href {http://adsabs.harvard.edu/abs/2011ApJS..197...35G} {197, 35}

\bibitem[\protect\citeauthoryear{{Guo} et~al.,}{{Guo} et~al.}{2013}]{guo13}
{Guo} Y.,  et~al., 2013, \mn@doi [\apjs] {10.1088/0067-0049/207/2/24}, \href
  {https://ui.adsabs.harvard.edu/abs/2013ApJS..207...24G} {207, 24}

\bibitem[\protect\citeauthoryear{{Hamann}, {Kanekar}, {Prochaska}, {Murphy},
  {Ellison}, {Malec}, {Milutinovic}  \& {Ubachs}}{{Hamann}
  et~al.}{2011}]{hama11}
{Hamann} F.,  {Kanekar} N.,  {Prochaska} J.~X.,  {Murphy} M.~T.,  {Ellison} S.,
   {Malec} A.~L.,  {Milutinovic} N.,   {Ubachs} W.,  2011, \mn@doi [\mnras]
  {10.1111/j.1365-2966.2010.17575.x}, \href
  {http://adsabs.harvard.edu/abs/2011MNRAS.410.1957H} {410, 1957}

\bibitem[\protect\citeauthoryear{{Hamann} et~al.,}{{Hamann}
  et~al.}{2017}]{hama17}
{Hamann} F.,  et~al., 2017, \mn@doi [\mnras] {10.1093/mnras/stw2387}, \href
  {http://adsabs.harvard.edu/abs/2017MNRAS.464.3431H} {464, 3431}

\bibitem[\protect\citeauthoryear{{Hickox}, {Mullaney}, {Alexander}, {Chen},
  {Civano}, {Goulding}  \& {Hainline}}{{Hickox} et~al.}{2014}]{hick14}
{Hickox} R.~C.,  {Mullaney} J.~R.,  {Alexander} D.~M.,  {Chen} C.-T.~J.,
  {Civano} F.~M.,  {Goulding} A.~D.,   {Hainline} K.~N.,  2014, \mn@doi [\apj]
  {10.1088/0004-637X/782/1/9}, \href
  {http://adsabs.harvard.edu/abs/2014ApJ...782....9H} {782, 9}

\bibitem[\protect\citeauthoryear{{Hilbert} et~al.,}{{Hilbert}
  et~al.}{2016}]{hilb16}
{Hilbert} B.,  et~al., 2016, \mn@doi [The Astrophysical Journal Supplement
  Series] {10.3847/0067-0049/225/1/12}, \href
  {https://ui.adsabs.harvard.edu/#abs/2016ApJS..225...12H} {225, 12}

\bibitem[\protect\citeauthoryear{{Hines} \& {Wills}}{{Hines} \&
  {Wills}}{1993}]{hine93}
{Hines} D.~C.,  {Wills} B.~J.,  1993, \mn@doi [\apj] {10.1086/173145}, \href
  {http://adsabs.harvard.edu/abs/1993ApJ...415...82H} {415, 82}

\bibitem[\protect\citeauthoryear{{Hook}, {Stoehr}  \& {Krist}}{{Hook}
  et~al.}{2008}]{hook08}
{Hook} R.,  {Stoehr} F.,   {Krist} J.,  2008, Space Telescope European
  Coordinating Facility Newsletter, \href
  {https://ui.adsabs.harvard.edu/#abs/2008STECF..44...11H} {44, 11}

\bibitem[\protect\citeauthoryear{{Hopkins} \& {Hernquist}}{{Hopkins} \&
  {Hernquist}}{2009}]{hopk09b}
{Hopkins} P.~F.,  {Hernquist} L.,  2009, \mn@doi [\apj]
  {10.1088/0004-637X/694/1/599}, \href
  {http://esoads.eso.org/abs/2009ApJ...694..599H} {694, 599}

\bibitem[\protect\citeauthoryear{{Hopkins}, {Hernquist}, {Cox}, {Di Matteo},
  {Robertson}  \& {Springel}}{{Hopkins} et~al.}{2006}]{hopk06}
{Hopkins} P.~F.,  {Hernquist} L.,  {Cox} T.~J.,  {Di Matteo} T.,  {Robertson}
  B.,   {Springel} V.,  2006, \mn@doi [\apjs] {10.1086/499298}, \href
  {http://adsabs.harvard.edu/abs/2006ApJS..163....1H} {163, 1}

\bibitem[\protect\citeauthoryear{{Hwang}, {Zakamska}, {Alexandroff}, {Hamann},
  {Greene}, {Perrotta}  \& {Richards}}{{Hwang} et~al.}{2018}]{hwan18}
{Hwang} H.-C.,  {Zakamska} N.~L.,  {Alexandroff} R.~M.,  {Hamann} F.,  {Greene}
  J.~E.,  {Perrotta} S.,   {Richards} G.~T.,  2018, \mn@doi [\mnras]
  {10.1093/mnras/sty742}, \href {http://esoads.eso.org/abs/2018MNRAS.477..830H}
  {477, 830}

\bibitem[\protect\citeauthoryear{{Ivison} et~al.,}{{Ivison}
  et~al.}{2012}]{ivis12}
{Ivison} R.~J.,  et~al., 2012, \mn@doi [\mnras]
  {10.1111/j.1365-2966.2012.21544.x}, \href
  {https://ui.adsabs.harvard.edu/#abs/2012MNRAS.425.1320I} {425, 1320}

\bibitem[\protect\citeauthoryear{{Khachikian} \& {Weedman}}{{Khachikian} \&
  {Weedman}}{1974}]{khac74}
{Khachikian} E.~Y.,  {Weedman} D.~W.,  1974, \mn@doi [\apj] {10.1086/153093},
  \href {http://adsabs.harvard.edu/abs/1974ApJ...192..581K} {192, 581}

\bibitem[\protect\citeauthoryear{{Kim}, {Veilleux}  \& {Sanders}}{{Kim}
  et~al.}{2002}]{kim02}
{Kim} D.-C.,  {Veilleux} S.,   {Sanders} D.~B.,  2002, \mn@doi [\apjs]
  {10.1086/343843}, \href {http://adsabs.harvard.edu/abs/2002ApJS..143..277K}
  {143, 277}

\bibitem[\protect\citeauthoryear{{Kirhakos}, {Bahcall}, {Schneider}  \&
  {Kristian}}{{Kirhakos} et~al.}{1999}]{kirh99}
{Kirhakos} S.,  {Bahcall} J.~N.,  {Schneider} D.~P.,   {Kristian} J.,  1999,
  \mn@doi [\apj] {10.1086/307430}, \href
  {http://esoads.eso.org/abs/1999ApJ...520...67K} {520, 67}

\bibitem[\protect\citeauthoryear{{Kocevski} et~al.,}{{Kocevski}
  et~al.}{2012}]{koce12}
{Kocevski} D.~D.,  et~al., 2012, \mn@doi [\apj] {10.1088/0004-637X/744/2/148},
  \href {http://adsabs.harvard.edu/abs/2012ApJ...744..148K} {744, 148}

\bibitem[\protect\citeauthoryear{{Kormendy} \& {Richstone}}{{Kormendy} \&
  {Richstone}}{1995}]{korm95}
{Kormendy} J.,  {Richstone} D.,  1995, \mn@doi [\araa]
  {10.1146/annurev.aa.33.090195.003053}, \href
  {http://adsabs.harvard.edu/abs/1995ARA%26A..33..581K} {33, 581}

\bibitem[\protect\citeauthoryear{{Krist}, {Hook}  \& {Stoehr}}{{Krist}
  et~al.}{2011}]{kris11}
{Krist} J.~E.,  {Hook} R.~N.,   {Stoehr} F.,  2011, in Optical Modeling and
  Performance Predictions V. p. 81270J, \mn@doi{10.1117/12.892762}

\bibitem[\protect\citeauthoryear{{Kukula}, {Dunlop}, {McLure}, {Miller},
  {Percival}, {Baum}  \& {O'Dea}}{{Kukula} et~al.}{2001}]{kuku01}
{Kukula} M.~J.,  {Dunlop} J.~S.,  {McLure} R.~J.,  {Miller} L.,  {Percival}
  W.~J.,  {Baum} S.~A.,   {O'Dea} C.~P.,  2001, \mn@doi [\mnras]
  {10.1111/j.1365-2966.2001.04692.x}, \href
  {http://adsabs.harvard.edu/abs/2001MNRAS.326.1533K} {326, 1533}

\bibitem[\protect\citeauthoryear{{Lang}, {Hogg}  \& {Schlegel}}{{Lang}
  et~al.}{2016}]{lang16}
{Lang} D.,  {Hogg} D.~W.,   {Schlegel} D.~J.,  2016, \mn@doi [\aj]
  {10.3847/0004-6256/151/2/36}, \href
  {https://ui.adsabs.harvard.edu/#abs/2016AJ....151...36L} {151, 36}

\bibitem[\protect\citeauthoryear{{Liu}, {Zakamska}, {Greene}, {Nesvadba}  \&
  {Liu}}{{Liu} et~al.}{2013a}]{liu13a}
{Liu} G.,  {Zakamska} N.~L.,  {Greene} J.~E.,  {Nesvadba} N.~P.~H.,   {Liu} X.,
   2013a, \mn@doi [\mnras] {10.1093/mnras/stt051}, \href
  {http://adsabs.harvard.edu/abs/2013MNRAS.430.2327L} {430, 2327}

\bibitem[\protect\citeauthoryear{{Liu}, {Zakamska}, {Greene}, {Nesvadba}  \&
  {Liu}}{{Liu} et~al.}{2013b}]{liu13b}
{Liu} G.,  {Zakamska} N.~L.,  {Greene} J.~E.,  {Nesvadba} N.~P.~H.,   {Liu} X.,
   2013b, \mn@doi [\mnras] {10.1093/mnras/stt1755}, \href
  {http://adsabs.harvard.edu/abs/2013MNRAS.436.2576L} {436, 2576}

\bibitem[\protect\citeauthoryear{{Lotz}, {Primack}  \& {Madau}}{{Lotz}
  et~al.}{2004}]{lotz04}
{Lotz} J.~M.,  {Primack} J.,   {Madau} P.,  2004, \mn@doi [\aj]
  {10.1086/421849}, \href {http://esoads.eso.org/abs/2004AJ....128..163L} {128,
  163}

\bibitem[\protect\citeauthoryear{{Maraston}}{{Maraston}}{2005}]{mara05}
{Maraston} C.,  2005, \mn@doi [\mnras] {10.1111/j.1365-2966.2005.09270.x},
  \href {https://ui.adsabs.harvard.edu/abs/2005MNRAS.362..799M} {362, 799}

\bibitem[\protect\citeauthoryear{{McAlpine}, {Bower}, {Rosario}, {Crain},
  {Schaye}  \& {Theuns}}{{McAlpine} et~al.}{2018}]{mcal18}
{McAlpine} S.,  {Bower} R.~G.,  {Rosario} D.~J.,  {Crain} R.~A.,  {Schaye} J.,
   {Theuns} T.,  2018, \mn@doi [\mnras] {10.1093/mnras/sty2489}, \href
  {https://ui.adsabs.harvard.edu/#abs/2018MNRAS.481.3118M} {481, 3118}

\bibitem[\protect\citeauthoryear{{McConnell}, {Ma}, {Gebhardt}, {Wright},
  {Murphy}, {Lauer}, {Graham}  \& {Richstone}}{{McConnell}
  et~al.}{2011}]{mcco11}
{McConnell} N.~J.,  {Ma} C.-P.,  {Gebhardt} K.,  {Wright} S.~A.,  {Murphy}
  J.~D.,  {Lauer} T.~R.,  {Graham} J.~R.,   {Richstone} D.~O.,  2011, \mn@doi
  [\nat] {10.1038/nature10636}, \href
  {http://adsabs.harvard.edu/abs/2011Natur.480..215M} {480, 215}

\bibitem[\protect\citeauthoryear{{McLeod} \& {Bechtold}}{{McLeod} \&
  {Bechtold}}{2009}]{mcle09}
{McLeod} K.~K.,  {Bechtold} J.,  2009, \mn@doi [\apj]
  {10.1088/0004-637X/704/1/415}, \href
  {http://adsabs.harvard.edu/abs/2009ApJ...704..415M} {704, 415}

\bibitem[\protect\citeauthoryear{{Mechtley} et~al.,}{{Mechtley}
  et~al.}{2016}]{mech16}
{Mechtley} M.,  et~al., 2016, \mn@doi [\apj] {10.3847/0004-637X/830/2/156},
  \href {http://esoads.eso.org/abs/2016ApJ...830..156M} {830, 156}

\bibitem[\protect\citeauthoryear{{Moustakas}, {Kennicutt}  \&
  {Tremonti}}{{Moustakas} et~al.}{2006}]{mous06}
{Moustakas} J.,  {Kennicutt} Jr. R.~C.,   {Tremonti} C.~A.,  2006, \mn@doi
  [\apj] {10.1086/500964}, \href
  {http://adsabs.harvard.edu/abs/2006ApJ...642..775M} {642, 775}

\bibitem[\protect\citeauthoryear{{Nesvadba}, {De Breuck}, {Lehnert}, {Best}  \&
  {Collet}}{{Nesvadba} et~al.}{2017}]{nesv17}
{Nesvadba} N.~P.~H.,  {De Breuck} C.,  {Lehnert} M.~D.,  {Best} P.~N.,
  {Collet} C.,  2017, \mn@doi [\aap] {10.1051/0004-6361/201528040}, \href
  {https://ui.adsabs.harvard.edu/#abs/2017A&A...599A.123N} {599, A123}

\bibitem[\protect\citeauthoryear{{Ni} et~al.,}{{Ni} et~al.}{2018}]{ni18}
{Ni} Q.,  et~al., 2018, \mn@doi [\mnras] {10.1093/mnras/sty1989}, \href
  {https://ui.adsabs.harvard.edu/abs/2018MNRAS.480.5184N} {480, 5184}

\bibitem[\protect\citeauthoryear{{Noirot} et~al.,}{{Noirot}
  et~al.}{2018}]{noir18}
{Noirot} G.,  et~al., 2018, \mn@doi [\apj] {10.3847/1538-4357/aabadb}, \href
  {https://ui.adsabs.harvard.edu/#abs/2018ApJ...859...38N} {859, 38}

\bibitem[\protect\citeauthoryear{{Obied}, {Zakamska}, {Wylezalek}  \&
  {Liu}}{{Obied} et~al.}{2016}]{obie16}
{Obied} G.,  {Zakamska} N.~L.,  {Wylezalek} D.,   {Liu} G.,  2016, \mn@doi
  [\mnras] {10.1093/mnras/stv2850}, \href
  {http://adsabs.harvard.edu/abs/2016MNRAS.456.2861O} {456, 2861}

\bibitem[\protect\citeauthoryear{{P{\^a}ris} et~al.,}{{P{\^a}ris}
  et~al.}{2014}]{pari14}
{P{\^a}ris} I.,  et~al., 2014, \mn@doi [\aap] {10.1051/0004-6361/201322691},
  \href {http://adsabs.harvard.edu/abs/2014A%26A...563A..54P} {563, A54}

\bibitem[\protect\citeauthoryear{{P{\^a}ris} et~al.,}{{P{\^a}ris}
  et~al.}{2017}]{pari17}
{P{\^a}ris} I.,  et~al., 2017, \mn@doi [\aap] {10.1051/0004-6361/201527999},
  \href {http://adsabs.harvard.edu/abs/2017A%26A...597A..79P} {597, A79}

\bibitem[\protect\citeauthoryear{{Patterson}}{{Patterson}}{1940}]{patt40}
{Patterson} F.~S.,  1940, Harvard College Observatory Bulletin, \href
  {https://ui.adsabs.harvard.edu/#abs/1940BHarO.914....9P} {914, 9}

\bibitem[\protect\citeauthoryear{{Peng}, {Impey}, {Rix}, {Kochanek}, {Keeton},
  {Falco}, {Leh{\'a}r}  \& {McLeod}}{{Peng} et~al.}{2006}]{peng06b}
{Peng} C.~Y.,  {Impey} C.~D.,  {Rix} H.-W.,  {Kochanek} C.~S.,  {Keeton} C.~R.,
   {Falco} E.~E.,  {Leh{\'a}r} J.,   {McLeod} B.~A.,  2006, \mn@doi [\apj]
  {10.1086/506266}, \href {http://adsabs.harvard.edu/abs/2006ApJ...649..616P}
  {649, 616}

\bibitem[\protect\citeauthoryear{{Perrotta et al.}}{{Perrotta et
  al.}}{2019}]{perr19}
{Perrotta et al.} S.,  2019, MNRAS

\bibitem[\protect\citeauthoryear{{Petrosian}}{{Petrosian}}{1976}]{petr76}
{Petrosian} V.,  1976, \mn@doi [\apj] {10.1086/182253}, \href
  {http://adsabs.harvard.edu/abs/1976ApJ...209L...1P} {209, L1}

\bibitem[\protect\citeauthoryear{{Pickles}}{{Pickles}}{1998}]{pick98}
{Pickles} A.~J.,  1998, \mn@doi [\pasp] {10.1086/316197}, \href
  {http://esoads.eso.org/abs/1998PASP..110..863P} {110, 863}

\bibitem[\protect\citeauthoryear{{Pitchford} et~al.,}{{Pitchford}
  et~al.}{2019}]{pitc19}
{Pitchford} L.~K.,  et~al., 2019, \mn@doi [\mnras] {10.1093/mnras/stz1471},
  \href {https://ui.adsabs.harvard.edu/abs/2019MNRAS.487.3130P} {487, 3130}

\bibitem[\protect\citeauthoryear{{Polletta} et~al.,}{{Polletta}
  et~al.}{2007}]{poll07}
{Polletta} M.,  et~al., 2007, \mn@doi [\apj] {10.1086/518113}, \href
  {http://adsabs.harvard.edu/abs/2007ApJ...663...81P} {663, 81}

\bibitem[\protect\citeauthoryear{{Polletta} et~al.,}{{Polletta}
  et~al.}{2008}]{poll08}
{Polletta} M.,  et~al., 2008, \mn@doi [\aap] {10.1051/0004-6361:200810345},
  \href {http://adsabs.harvard.edu/abs/2008A%26A...492...81P} {492, 81}

\bibitem[\protect\citeauthoryear{{Reyes} et~al.,}{{Reyes}
  et~al.}{2008}]{reye08}
{Reyes} R.,  et~al., 2008, \mn@doi [\aj] {10.1088/0004-6256/136/6/2373}, \href
  {http://adsabs.harvard.edu/abs/2008AJ....136.2373R} {136, 2373}

\bibitem[\protect\citeauthoryear{{Ricci} et~al.,}{{Ricci}
  et~al.}{2017}]{ricc17}
{Ricci} C.,  et~al., 2017, \mn@doi [\apj] {10.3847/1538-4357/835/1/105}, \href
  {http://adsabs.harvard.edu/abs/2017ApJ...835..105R} {835, 105}

\bibitem[\protect\citeauthoryear{{Richards} et~al.,}{{Richards}
  et~al.}{2006a}]{rich06b}
{Richards} G.~T.,  et~al., 2006a, \mn@doi [\aj] {10.1086/503559}, \href
  {http://adsabs.harvard.edu/abs/2006AJ....131.2766R} {131, 2766}

\bibitem[\protect\citeauthoryear{{Richards} et~al.,}{{Richards}
  et~al.}{2006b}]{rich06}
{Richards} G.~T.,  et~al., 2006b, \mn@doi [\apjs] {10.1086/506525}, \href
  {http://adsabs.harvard.edu/abs/2006ApJS..166..470R} {166, 470}

\bibitem[\protect\citeauthoryear{{Ross} et~al.,}{{Ross} et~al.}{2015}]{ross15}
{Ross} N.~P.,  et~al., 2015, \mn@doi [\mnras] {10.1093/mnras/stv1710}, \href
  {http://adsabs.harvard.edu/abs/2015MNRAS.453.3932R} {453, 3932}

\bibitem[\protect\citeauthoryear{{Rowan-Robinson} et~al.,}{{Rowan-Robinson}
  et~al.}{2018}]{rowa18}
{Rowan-Robinson} M.,  et~al., 2018, \mn@doi [\aap]
  {10.1051/0004-6361/201832671}, \href
  {https://ui.adsabs.harvard.edu/abs/2018A&A...619A.169R} {619, A169}

\bibitem[\protect\citeauthoryear{{Rupke}, {G{\"u}ltekin}  \&
  {Veilleux}}{{Rupke} et~al.}{2017}]{rupk17}
{Rupke} D.~S.~N.,  {G{\"u}ltekin} K.,   {Veilleux} S.,  2017, \mn@doi [\apj]
  {10.3847/1538-4357/aa94d1}, \href
  {http://esoads.eso.org/abs/2017ApJ...850...40R} {850, 40}

\bibitem[\protect\citeauthoryear{{Sanders}, {Soifer}, {Elias}, {Madore},
  {Matthews}, {Neugebauer}  \& {Scoville}}{{Sanders} et~al.}{1988}]{sand88}
{Sanders} D.~B.,  {Soifer} B.~T.,  {Elias} J.~H.,  {Madore} B.~F.,  {Matthews}
  K.,  {Neugebauer} G.,   {Scoville} N.~Z.,  1988, \mn@doi [\apj]
  {10.1086/165983}, \href {http://adsabs.harvard.edu/abs/1988ApJ...325...74S}
  {325, 74}

\bibitem[\protect\citeauthoryear{{Savorgnan}, {Graham}, {Marconi}, {Sani},
  {Hunt}, {Vika}  \& {Driver}}{{Savorgnan} et~al.}{2013}]{savo13}
{Savorgnan} G.,  {Graham} A.~W.,  {Marconi} A.,  {Sani} E.,  {Hunt} L.~K.,
  {Vika} M.,   {Driver} S.~P.,  2013, \mn@doi [\mnras] {10.1093/mnras/stt1027},
  \href {https://ui.adsabs.harvard.edu/#abs/2013MNRAS.434..387S} {434, 387}

\bibitem[\protect\citeauthoryear{{Schawinski}, {Treister}, {Urry}, {Cardamone},
  {Simmons}  \& {Yi}}{{Schawinski} et~al.}{2011}]{scha11}
{Schawinski} K.,  {Treister} E.,  {Urry} C.~M.,  {Cardamone} C.~N.,  {Simmons}
  B.,   {Yi} S.~K.,  2011, \mn@doi [\apjl] {10.1088/2041-8205/727/2/L31}, \href
  {http://adsabs.harvard.edu/abs/2011ApJ...727L..31S} {727, L31}

\bibitem[\protect\citeauthoryear{{S{\'e}rsic}}{{S{\'e}rsic}}{1968}]{sers68}
{S{\'e}rsic} J.~L.,  1968, {Atlas de galaxias australes}.
Cordoba, Argentina: Observatorio Astronomico, 1968

\bibitem[\protect\citeauthoryear{{Seymour} et~al.,}{{Seymour}
  et~al.}{2007}]{seym07}
{Seymour} N.,  et~al., 2007, \mn@doi [\apjs] {10.1086/517887}, \href
  {http://adsabs.harvard.edu/abs/2007ApJS..171..353S} {171, 353}

\bibitem[\protect\citeauthoryear{{Shakura} \& {Sunyaev}}{{Shakura} \&
  {Sunyaev}}{1973}]{shak73}
{Shakura} N.~I.,  {Sunyaev} R.~A.,  1973, \aap, \href
  {https://ui.adsabs.harvard.edu/abs/1973A&A....24..337S} {500, 33}

\bibitem[\protect\citeauthoryear{{Shlosman}, {Frank}  \& {Begelman}}{{Shlosman}
  et~al.}{1989}]{shlo89}
{Shlosman} I.,  {Frank} J.,   {Begelman} M.~C.,  1989, \mn@doi [\nat]
  {10.1038/338045a0}, \href
  {https://ui.adsabs.harvard.edu/#abs/1989Natur.338...45S} {338, 45}

\bibitem[\protect\citeauthoryear{{Silk} \& {Rees}}{{Silk} \&
  {Rees}}{1998}]{silk98}
{Silk} J.,  {Rees} M.~J.,  1998, \aap, \href
  {http://adsabs.harvard.edu/abs/1998A%26A...331L...1S} {331, L1}

\bibitem[\protect\citeauthoryear{{S{\k{a}}dowski} \&
  {Narayan}}{{S{\k{a}}dowski} \& {Narayan}}{2016}]{sado16}
{S{\k{a}}dowski} A.,  {Narayan} R.,  2016, \mn@doi [\mnras]
  {10.1093/mnras/stv2941}, \href
  {https://ui.adsabs.harvard.edu/#abs/2016MNRAS.456.3929S} {456, 3929}

\bibitem[\protect\citeauthoryear{{Smith}, {Schmidt}, {Hines}, {Cutri}  \&
  {Nelson}}{{Smith} et~al.}{2002}]{smit02a}
{Smith} P.~S.,  {Schmidt} G.~D.,  {Hines} D.~C.,  {Cutri} R.~M.,   {Nelson}
  B.~O.,  2002, \mn@doi [\apj] {10.1086/339208}, \href
  {http://adsabs.harvard.edu/abs/2002ApJ...569...23S} {569, 23}

\bibitem[\protect\citeauthoryear{{Steinborn}, {Hirschmann}, {Dolag}, {Shankar},
  {Juneau}, {Krumpe}, {Remus}  \& {Teklu}}{{Steinborn} et~al.}{2018}]{stei18}
{Steinborn} L.~K.,  {Hirschmann} M.,  {Dolag} K.,  {Shankar} F.,  {Juneau} S.,
  {Krumpe} M.,  {Remus} R.-S.,   {Teklu} A.~F.,  2018, preprint, \href
  {http://adsabs.harvard.edu/abs/2018arXiv180506902S} {} (\mn@eprint {arXiv}
  {1805.06902})

\bibitem[\protect\citeauthoryear{{Thomas}, {Ma}, {McConnell}, {Greene},
  {Blakeslee}  \& {Janish}}{{Thomas} et~al.}{2016}]{thom16}
{Thomas} J.,  {Ma} C.-P.,  {McConnell} N.~J.,  {Greene} J.~E.,  {Blakeslee}
  J.~P.,   {Janish} R.,  2016, \mn@doi [\nat] {10.1038/nature17197}, \href
  {https://ui.adsabs.harvard.edu/#abs/2016Natur.532..340T} {532, 340}

\bibitem[\protect\citeauthoryear{{Treister}, {Schawinski}, {Urry}  \&
  {Simmons}}{{Treister} et~al.}{2012}]{trei12}
{Treister} E.,  {Schawinski} K.,  {Urry} C.~M.,   {Simmons} B.~D.,  2012,
  \mn@doi [\apjl] {10.1088/2041-8205/758/2/L39}, \href
  {http://adsabs.harvard.edu/abs/2012ApJ...758L..39T} {758, L39}

\bibitem[\protect\citeauthoryear{{Tsai} et~al.,}{{Tsai} et~al.}{2015}]{tsai15}
{Tsai} C.-W.,  et~al., 2015, \mn@doi [\apj] {10.1088/0004-637X/805/2/90}, \href
  {http://adsabs.harvard.edu/abs/2015ApJ...805...90T} {805, 90}

\bibitem[\protect\citeauthoryear{{Villforth}, {Sarajedini}  \&
  {Koekemoer}}{{Villforth} et~al.}{2012}]{vill12}
{Villforth} C.,  {Sarajedini} V.,   {Koekemoer} A.,  2012, \mn@doi [\mnras]
  {10.1111/j.1365-2966.2012.21732.x}, \href
  {https://ui.adsabs.harvard.edu/#abs/2012MNRAS.426..360V} {426, 360}

\bibitem[\protect\citeauthoryear{{Villforth} et~al.,}{{Villforth}
  et~al.}{2017}]{vill17}
{Villforth} C.,  et~al., 2017, \mn@doi [\mnras] {10.1093/mnras/stw3037}, \href
  {http://adsabs.harvard.edu/abs/2017MNRAS.466..812V} {466, 812}

\bibitem[\protect\citeauthoryear{{Vito} et~al.,}{{Vito} et~al.}{2018}]{vito18}
{Vito} F.,  et~al., 2018, \mn@doi [\mnras] {10.1093/mnras/stx3120}, \href
  {http://esoads.eso.org/abs/2018MNRAS.474.4528V} {474, 4528}

\bibitem[\protect\citeauthoryear{{Wethers} et~al.,}{{Wethers}
  et~al.}{2018}]{weth18}
{Wethers} C.~F.,  et~al., 2018, \mn@doi [\mnras] {10.1093/mnras/stx3332}, \href
  {https://ui.adsabs.harvard.edu/#abs/2018MNRAS.475.3682W} {475, 3682}

\bibitem[\protect\citeauthoryear{{Wright} et~al.,}{{Wright}
  et~al.}{2010}]{wrig10}
{Wright} E.~L.,  et~al., 2010, \mn@doi [\aj] {10.1088/0004-6256/140/6/1868},
  \href {http://adsabs.harvard.edu/abs/2010AJ....140.1868W} {140, 1868}

\bibitem[\protect\citeauthoryear{{Wu} et~al.,}{{Wu} et~al.}{2018}]{wu18}
{Wu} J.,  et~al., 2018, \mn@doi [\apj] {10.3847/1538-4357/aa9ff3}, \href
  {http://adsabs.harvard.edu/abs/2018ApJ...852...96W} {852, 96}

\bibitem[\protect\citeauthoryear{{Wylezalek} et~al.,}{{Wylezalek}
  et~al.}{2013}]{wyle13}
{Wylezalek} D.,  et~al., 2013, \mn@doi [\apj] {10.1088/0004-637X/769/1/79},
  \href {https://ui.adsabs.harvard.edu/abs/2013ApJ...769...79W} {769, 79}

\bibitem[\protect\citeauthoryear{{Wylezalek}, {Zakamska}, {Liu}  \&
  {Obied}}{{Wylezalek} et~al.}{2016}]{wyle16a}
{Wylezalek} D.,  {Zakamska} N.~L.,  {Liu} G.,   {Obied} G.,  2016, \mn@doi
  [\mnras] {10.1093/mnras/stv3022}, \href
  {http://adsabs.harvard.edu/abs/2016MNRAS.457..745W} {457, 745}

\bibitem[\protect\citeauthoryear{{Yang} et~al.,}{{Yang} et~al.}{2017}]{yang17}
{Yang} G.,  et~al., 2017, \mn@doi [\apj] {10.3847/1538-4357/aa7564}, \href
  {http://adsabs.harvard.edu/abs/2017ApJ...842...72Y} {842, 72}

\bibitem[\protect\citeauthoryear{{Yuan}, {Strauss}  \& {Zakamska}}{{Yuan}
  et~al.}{2016}]{yuan16}
{Yuan} S.,  {Strauss} M.~A.,   {Zakamska} N.~L.,  2016, \mn@doi [\mnras]
  {10.1093/mnras/stw1747}, \href
  {http://adsabs.harvard.edu/abs/2016MNRAS.462.1603Y} {462, 1603}

\bibitem[\protect\citeauthoryear{{Zakamska} \& {Greene}}{{Zakamska} \&
  {Greene}}{2014}]{zaka14}
{Zakamska} N.~L.,  {Greene} J.~E.,  2014, \mn@doi [\mnras]
  {10.1093/mnras/stu842}, \href
  {http://adsabs.harvard.edu/abs/2014MNRAS.442..784Z} {442, 784}

\bibitem[\protect\citeauthoryear{{Zakamska} et~al.,}{{Zakamska}
  et~al.}{2003}]{zaka03}
{Zakamska} N.~L.,  et~al., 2003, \mn@doi [\aj] {10.1086/378610}, \href
  {http://adsabs.harvard.edu/abs/2003AJ....126.2125Z} {126, 2125}

\bibitem[\protect\citeauthoryear{{Zakamska} et~al.,}{{Zakamska}
  et~al.}{2005}]{zaka05}
{Zakamska} N.~L.,  et~al., 2005, \mn@doi [\aj] {10.1086/427543}, \href
  {http://adsabs.harvard.edu/abs/2005AJ....129.1212Z} {129, 1212}

\bibitem[\protect\citeauthoryear{{Zakamska} et~al.,}{{Zakamska}
  et~al.}{2006}]{zaka06}
{Zakamska} N.~L.,  et~al., 2006, \mn@doi [\aj] {10.1086/506986}, \href
  {http://adsabs.harvard.edu/abs/2006AJ....132.1496Z} {132, 1496}

\bibitem[\protect\citeauthoryear{{Zakamska} et~al.,}{{Zakamska}
  et~al.}{2016}]{zaka16b}
{Zakamska} N.~L.,  et~al., 2016, \mn@doi [\mnras] {10.1093/mnras/stw718}, \href
  {http://adsabs.harvard.edu/abs/2016MNRAS.459.3144Z} {459, 3144}

\bibitem[\protect\citeauthoryear{{de Vaucouleurs}}{{de
  Vaucouleurs}}{1953}]{deva53}
{de Vaucouleurs} G.,  1953, \mn@doi [\mnras] {10.1093/mnras/113.2.134}, \href
  {https://ui.adsabs.harvard.edu/#abs/1953MNRAS.113..134D} {113, 134}

\bibitem[\protect\citeauthoryear{{van Dokkum} et~al.,}{{van Dokkum}
  et~al.}{2010}]{vand10}
{van Dokkum} P.~G.,  et~al., 2010, \mn@doi [\apj]
  {10.1088/0004-637X/709/2/1018}, \href
  {http://esoads.eso.org/abs/2010ApJ...709.1018V} {709, 1018}

\makeatother
\end{thebibliography}
\input{astroph.bbl}



\clearpage
\begin{table*}
\caption{Aperture measurements of host galaxies of red and obscured quasars\label{tab:measure}}
\begin{threeparttable}
\begin{tabular}{r|r|r|c|c|c|r|r|r|r|r}
\hline
ID & R.A. & declination & $z$ & object & WFC3 & host lum. & host lum. & 0.3$-$0.6\arcsec & 0.6$-$1\arcsec & 1$-$2\arcsec \\
 & deg. & deg. &  & type & filter & $\log(\nu L_{\nu}/L_{\odot})$ & $\log(L_B/L_{\odot})$ & flux density & flux density & flux density \\
\hline
SDSSJ0832$+$1615 &  128.00083 &  16.25011 & 2.431 & ERQ & F140W & 10.90 & 11.03 &  8.54 $ \pm $  1.54 &  3.18 $ \pm $  0.34 &  0.63 $ \pm $  0.09 \\ 
SDSSJ0834$+$0159 &  128.70200 &   1.98919 & 2.594 & ERQ & F160W & 10.89 & 10.92 &  8.66 $ \pm $  0.57 &  2.82 $ \pm $  0.20 &  0.61 $ \pm $  0.08 \\ 
SDSSJ0913$+$2344 &  138.26625 &  23.74311 & 2.436 & ERQ & F140W & 10.92 & 11.07 & 17.12 $ \pm $  2.63 &  2.57 $ \pm $  0.38 &  0.39 $ \pm $  0.10 \\ 
SDSSJ1013$+$3427 &  153.35221 &  34.45072 & 2.473 & ERQ & F140W & $<$11.02 & $<$11.18 & 12.82 $ \pm $  7.35 &  4.36 $ \pm $  1.02 &  1.21 $ \pm $  0.26 \\ 
SDSSJ1217$+$0234 &  184.26958 &   2.57142 & 2.426 & ERQ & F140W & 11.41 & 11.55 & 21.08 $ \pm $  5.58 &  7.66 $ \pm $  0.98 &  1.91 $ \pm $  0.25 \\ 
SDSSJ1232$+$0912 &  188.17387 &   9.20258 & 2.374 & ERQ & F140W & 11.20 & 11.31 & 13.79 $ \pm $  1.14 &  5.14 $ \pm $  0.23 &  2.50 $ \pm $  0.09 \\ 
SDSSJ1356$+$0730 &  209.03467 &   7.50478 & 2.279 & ERQ & F140W & 11.17 & 11.22 & 16.05 $ \pm $  4.68 &  2.85 $ \pm $  0.40 &  0.98 $ \pm $  0.10 \\ 
SDSSJ1652$+$1728 &  253.01100 &  17.48119 & 2.942 & ERQ & F160W & 12.03 & 12.26 & 53.85 $ \pm $ 12.63 & 15.14 $ \pm $  3.71 & 14.61 $ \pm $  0.82 \\ 
SDSSJ2215$-$0056 &  333.85000 &  -0.94550 & 2.493 & ERQ & F160W & 11.14 & 11.13 & 20.45 $ \pm $  1.77 &  5.39 $ \pm $  0.35 &  1.03 $ \pm $  0.11 \\ 
SDSSJ2323$-$0100 &  350.85904 &  -1.00919 & 2.352 & ERQ & F140W & 11.58 & 11.68 & 22.07 $ \pm $  1.16 & 15.30 $ \pm $  0.25 &  9.24 $ \pm $  0.08 \\ 
\hline
SDSSJ1001$+$4151 &  150.39106 &  41.85220 & 2.364 & T2C & F160W & 11.06 & 11.00 & 13.11 $ \pm $  2.50 &  5.42 $ \pm $  0.43 &  2.01 $ \pm $  0.13 \\ 
SDSSJ1122$+$3415 &  170.62645 &  34.26056 & 2.398 & T2C & F160W & 10.58 & 10.53 &  6.46 $ \pm $  0.38 &  1.46 $ \pm $  0.18 &  0.14 $ \pm $  0.08 \\ 
SDSSJ1444$-$0013 &  221.17104 &  -0.22873 & 2.548 & T2C & F160W & 10.36 & 10.37 &  3.66 $ \pm $  0.39 &  1.53 $ \pm $  0.19 & -0.06 $ \pm $  0.09 \\ 
SDSSJ1515$+$1757 &  228.93337 &  17.96474 & 2.402 & T2C & F160W & 10.90 & 10.85 & 15.99 $ \pm $  2.82 &  3.50 $ \pm $  0.30 &  0.34 $ \pm $  0.10 \\ 
SDSSJ2201$+$0012 &  330.35878 &   0.20876 & 2.636 & T2C & F160W & $<$11.11 & $<$11.17 & 24.50 $ \pm $ 14.53 &  5.42 $ \pm $  4.92 &  0.36 $ \pm $  1.21 \\ 
SDSSJ2233$+$0249 &  338.45039 &   2.82579 & 2.587 & T2C & F160W & 10.63 & 10.66 &  7.80 $ \pm $  0.36 &  3.12 $ \pm $  0.21 &  0.39 $ \pm $  0.09 \\ 
\end{tabular}
\begin{tablenotes}
\item[a] `Object type' describes whether the target is drawn from the extremely red quasar (ERQ) parent sample or the type 2 quasar candidate (T2C) parent sample. `WFC3 filter' is the filter used for the near-infrared HST observations, with the approximate effective wavelength of 1.4\micron\ and 1.6\micron\ for F140W and F160W, respectively. All objects were also observed with the ACS F814W filter. Host luminosities from aperture photometry are tabulated at the rest wavelength given by the wavelength of the WFC3 filter divided by $(1+z)$, as well as k-corrected to the rest-frame $B$-band. Luminosity uncertainties are dominated by PSF subtraction systematics and are estimated to be 0.25 dex. The last three columns present some of the aperture photometry of PSF-subtracted WFC3 images in units of $10^{-30}$ erg s$^{-1}$ cm$^{-2}$ Hz$^{-1}$, with error bars combining Poisson noise, background subtraction error and PSF subtraction uncertainty. 
\end{tablenotes}
\end{threeparttable}
\end{table*}

\clearpage
\begin{table}
\caption{Results of Sersic fits\label{tab:sersic}}
\begin{threeparttable}
\begin{tabular}{r|r|r|r|r|r}
\hline
ID & $n_s$ & $R_{50}$, & $\log$ & $\log$ & PSF-to \\
& & kpc & $(\nu L_{\nu}/L_{\odot})$ & $(L_B/L_{\odot})$ & -host\\
\hline
SDSSJ0832$+$1615 & 3.2 & 4.5 & 11.07 & 11.21 &  4.71 \\ 
SDSSJ0834$+$0159 & 2.6 & 5.1 & 10.66 & 10.69 &  9.54 \\ 
SDSSJ0913$+$2344 & 1.1 & 2.6 & 11.12 & 11.26 &  1.29 \\ 
SDSSJ1013$+$3427 &     &     & $<$10.70 & $<$10.86 & $>$10.02 \\ 
SDSSJ1217$+$0234 &     &     & 11.33 & 11.46 &  8.70 \\ 
SDSSJ1232$+$0912 & 2.7 & 6.5 & 11.14 & 11.24 &  3.01 \\ 
SDSSJ1356$+$0730 &     &     & 10.97 & 11.02 &  7.77 \\ 
SDSSJ1652$+$1728 & 3.4 & 3.0 & 11.76 & 11.99 &  2.07 \\ 
SDSSJ2215$-$0056 & 2.5 & 5.2 & 11.05 & 11.03 &  2.14 \\ 
SDSSJ2323$-$0100 &     &     & 11.64 & 11.73 &  0.94 \\ 
\hline
SDSSJ1001$+$4151 & 3.1 & 7.3 & 10.63 & 10.57 &  4.80 \\ 
SDSSJ1122$+$3415 & 2.8 & 4.3 & 10.49 & 10.44 &  7.66 \\ 
SDSSJ1444$-$0013 & 2.4 & 4.4 & 10.23 & 10.24 &  7.76 \\ 
SDSSJ1515$+$1757 & 2.7 & 4.9 & 10.71 & 10.66 &  7.24 \\ 
SDSSJ2201$+$0012 &     &     & $<$11.32 & $<$11.38 & $>$ 4.09 \\ 
SDSSJ2233$+$0249 & 2.4 & 5.2 & 10.60 & 10.63 &  4.44 \\ 
\end{tabular}
\begin{tablenotes}
\item[a] We omit structural information on the two objects that are at best marginally resolved and on the three objects that are not well-fit with single Sersic profiles or where fitting was not successful. While we list structural parameters $n_s$ and $R_{50}$ we caution that the uncertainties on these measurements are quite large, with estimated $\sigma(n_s)\simeq 1$ and $\sigma(\log(R_{50}))\simeq 0.2$ dex at $R_{50}>4.5$ kpc and $0.3$ dex otherwise. Luminosities are listed both at the rest wavelength corresponding to the filter coverage and k-corrected to the rest-frame $B$ band. Individual uncertainties in luminosities are 0.25 dex. PSF-to-host ratios are given for Sersic fits. Their individual uncertainty from comparison with aperture-based values is 0.25 dex.
\end{tablenotes}
\end{threeparttable}
\end{table}






\bsp	
\label{lastpage}
\end{document}